# Internet of Things
## Technology, Applications and Standardization

*Edited by Jaydip Sen*

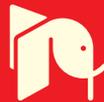

# INTERNET OF THINGS - TECHNOLOGY, APPLICATIONS AND STANDARDIZATION

Edited by **Jaydip Sen**

**Contributors**

Moonkun Lee, Sunghyeon Lee, Yeongbok Choe, Menachem Domb, Arpan Pal, Hemant Kumar Rath, Samar Shailendra, Abhijan Bhattacharyya, Albena Mihovska, Mahasweta Sarkar, Hyun Jung Lee, Myungho Kim, Alexandru Averian

**Notice**

# Meet the editor

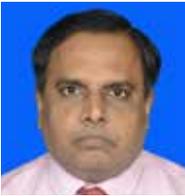

Professor Jaydip Sen has around 25 years of experience in the field of communication networks, protocol design, network analysis, cryptography, network security, and data analytics in reputable organizations such as the Oil and Natural Gas Corporation Ltd., India, Oracle India Pvt. Ltd., Akamai Technology Pvt. Ltd., Tata Consultancy Services Ltd., National Institute of Science and Technology, India, and the Calcutta Business School. Currently, he is associated with Praxis Business School, Kolkata, India, as a professor. His research areas include security and privacy issues in computing and communications, intrusion detection systems, secure routing protocols in wireless ad hoc and sensor networks, and privacy issues in the Internet of Things. Professor Sen obtained a Bachelor of Engineering in Mechanical Engineering with honors from Jadavpur University, Kolkata, India, in 1988 and a Master of Technology in Computer Science with honors from the Indian Statistical Institute, Kolkata, India, in 2001.

# Contents

# Preface

During the last decade, a novel paradigm named the "Internet of Things" (IoT) has evolved and it is slowly becoming an integrated part of our day-to-day life. The concept of the IoT was first introduced by Kevin Ashton in 1998, and over the years it has gained increasing importance and focus from both the academic world and industry. The essential objective of the IoT is to embed short-range and power-constrained mobile transceivers into a whole gamut of gadgets in our daily use and to enable communication between humans and things and between things themselves. It is not surprising that IoT applications are increasingly finding more deployments in the real world, thereby heralding a new paradigm in the world of information and communication.

It is easy to visualize that the main impact of the IoT will be on several aspects of everyday life and the way it will change the lives of its potential users. Naturally, from the perspective of a private user, the most striking effect of the IoT will be visible in both working and domestic fields. Some of the examples where the IoT will find increasing applications in this regard are assisted living, smart home, smart office, smart car, smart city, e-health, enhanced learning, etc. From the perspective of business users, automation and industrial manufacturing, control systems, and intelligent transportation systems will be some of the important applications of the IoT in the future.

However, several challenges and problems need to be addressed before IoT applications find large-scale deployment in the real world. These challenges include both technological and social issues. The most critical issues are ensuring interoperability between disparate interconnected objects, providing the objects with a high degree of smartness by autonomous and adaptable computing, and enforcing trust, security, and privacy of users and their data. Efficient utilization of computational power and memory space in tiny and resource-constrained devices and objects is also an important requirement in the IoT.

Looking at the current state-of-the-art technologies and the current deployment of IoT applications, it is not difficult to visualize how the IoT will be implemented on a universal level in the coming years. It is clearly evident that an urgent need exists for designing a framework for IoT governance. Standardization of communication protocols, semantic and protocol interoperability, and security-, privacy-, and trust-related issues will have to be done at a rapid pace to avoid possible situations in which one may witness proliferation of architectures, identification schemes, protocols, and frameworks for a particular and specific use case. Such fragmented deployment of the IoT can potentially hamper its large-scale adoption leading to the creation of a major barrier in its rollout.

About the book: The purpose of the book is to discuss and critically analyze some of the important challenges in design and deployment of real-world applications of the IoT. Some of the issues that have been discussed in the chapters in the book are standardization of various communication protocols for smart objects, ensuring and enforcing security and privacy requirements, establishing interoperability among various disparate protocols and devices, and optimizing the computational power and memory requirements in tiny objects. For effectively addressing these challenges, the book presents a collection of theoretical and practical research work done by experts in the field of the IoT.

The organization of this book is as follows. The book contains six chapters dealing with different aspects of the IoT, e.g., architecture, applications, communication protocols, and standardization.

In Chapter 1, entitled "dT-Calculus: A Formal Method to Specify Distributed Mobile Real-Time IoT Systems," Lee et al. propose a process algebra-based approach—known as dT calculus—for modeling distributed real-time mobile applications for deployment in IoT systems. The authors have demonstrated the feasibility of their proposition by conducting several experiments using a tool called SAVE. In Chapter 2, entitled "An Adaptive Lightweight Security Framework Suited for IoT," Domb presents three mechanisms for establishing a high level of security in IoT applications while optimizing on the memory space and computational power requirements. The author claims that optimization of computational and space overhead has been possible to achieve by eliminating the frequent use of the classification data and by using a random forest machine learning approach in a parallel and distributed environment. In Chapter 3, entitled "IoT Standardization—the Road Ahead," Pal et al. discuss various aspects of the IoT, including deployment and standardization. The authors have also identified a number of broad areas in the IoT on which current standardization efforts are going on, e.g., security and privacy, interoperability, reliability, agility, and scalability. In Chapter 4, entitled "Cooperative Human-Centric Sensing Connectivity," Mihovska and Sarkar present a "human-centric sensing" approach in the IoT. The authors discuss various issues in the state of the art of "human-centric sensing" and also identify several challenges in the deployment of the concept in real-world applications. Several solutions have also been proposed to address those challenges. In Chapter 5, entitled "The Internet of Things in a Smart Connected World ," Lee and Kim present a survey of various issues in IoT applications and their deployment challenges in the real world with a particular focus on the smart city use case. Several challenges in IoT technology, especially security and privacy-related threats, have also been highlighted. Finally, in Chapter 6, entitled "A Reference Architecture for Digital Ecosystems," Averian presents the concept of a "digital ecosystem" that consists of digital entities communicating with each other and achieving a goal in a collaborative and distributed way. The author proposes a reference architecture for such a "digital ecosystem" and identifies a path for standardizing such an architecture.

Judging by the high-quality technical contents in an area that is of extremely high interest in the current academic and professional world, I am confident that the book will be very useful to researchers, engineers, graduate and doctoral students, and faculty members of graduate schools and universities, who work in the broad areas of the IoT, especially on its applications, standardization, and communication protocols.

I express my sincere thanks to the authors who have contributed their valuable work in this volume. Without their rich contributions, the book would not have been able to attain the high level of quality. The authors have been extremely cooperative during the submission, editing, and publication process. I would like to express my special thanks to Mr. Julian Vir-ag, Publishing Process Manager of IntechOpen Ltd., London, for his constant support, encouragement, patience, and cooperation during the period of the publication of the volume. My sincere thanks also go to Ms. Ana Pantar, Senior Commissioning Editor of IntechOpen Ltd., London, for having faith in me and delegating me with the critical responsibility of editorship of such a prestigious academic volume. I would be failing in my duty if I did not acknowledge the motivation and encouragement that I received from my faculty colleagues in Praxis Business School, Kolkata, India. Prof. Charanpreet Singh and Prof. Prithwis Mukherjee deserve special mention for being my wonderful academic colleagues and for being sources of motivation for me always. Last but not the least, I would like to thank my mother Ms. Krishna Sen, my wife Ms. Nalanda Sen, and my daughter Ms. Ritabrata Sen for being my pillars of strength and the major sources of inspiration.

**Professor Jaydip Sen**
Department of Analytics and Information Technology
Praxis Business School
Kolkata, India

# dT-Calculus: A Formal Method to Specify Distributed Mobile Real-Time IoT Systems

Sunghyeon Lee, Yeongbok Choe and Moonkun Lee

Additional information is available at the end of the chapter

**Abstract**

In general, process algebra can be the most suitable formal method to specify IoT systems due to the equivalent notion of processes as things. However there are some limitations for distributed mobile real-time IoT systems. For example, *Timed pi-Calculus* has capability of specifying time property, but is lack of direct specifying both execution time of action and mobility of process at the same time. And *d-Calculus* has capability of specifying mobility of process itself, but is lack of specifying various time properties of both action and process, such as, *ready time, timeout, execution time, deadline*, as well as priority and repetition. In order to overcome the limitations, this paper presents a process algebra, called, *dT-Calculus*, extended from *d-Calculus*, by providing with capability of specifying the set of time properties, as well as priority and repetition. Further the method is implemented as a tool, called *SAVE*, on ADOxx meta-modeling platform. It can be considered one of the most practical and innovative approaches to specify distributed mobile real-time IoT systems.

**Keywords:** dT-Calculus, process algebra, mobility, time, SAVE, ADOxx

## 1. Introduction

The main characteristics of distributed mobile real-time IoT systems can be movement of things on some geographical space and real-time communication among them with deadlines [1]. Therefore it is necessary to specify these characteristics with formal methods during design phase of the system development process, and process algebra is known to be best suitable for the specification of the systems since things can be considered as processes and the characteristics can be depicted as both the movements of processes and the timed communications among them [2]. For example, the most suitable process algebras for IoT systems can be as follows:

1. *Timed pi-Calculus* [3]: It is the timed version of the existing *pi-Calculus* [4], which expresses process movements indirectly by using the notion of *value passing*. It allows *time-stamp* and *clock* to be passed additionally during value passing, with which the temporal requirements of the process movements can be specified.

2. *Timed Mobile Ambient* [5]: It is the timed version of the existing *Mobile Ambient* [6], where process can move by ambient with *in, out,* and *open* capabilities. In contrast to pi-Calculus, it is based on the semantics of autonomous movement, and makes timed specification possible by adding time property to the movement.

3. *d-Calculus* [7]: This is a process algebra that can express direct process movements into or out of other processes by using the four types of synchronous movements with simple temporal conditions: a bound of the minimum and maximum limits. It naturally allows process nesting by the resulting inclusion relations among processes.

However it is noticed that there are fundamental limitations in the above process algebra to specify the main characteristics of distributed mobile real-time IoT systems due to lack of both full description power of mobile and temporal properties, as follows:

1. Timed pi-Calculus: It allows various types of temporal requirements to be specified, but it is not possible to specify directly both the actual execution time of action itself and the type of its movement in the same requirements.

2. Timed Mobile Ambient: It is possible to specify temporal requirements by adding temporal property to ambient, but it is difficult to understand intuitively process synchronization since the synchronization is represented by the movement of the ambient.

3. d-Calculus: It allows various types of temporal requirements to be specified, but only simple types of temporal requirements for process movements are possible. For example, a temporal bound of the minimum and maximum limits. It results in limited specification of the temporal requirements of the movements as well as analysis of the requirements.

In order to overcome the limitations, this paper proposes process algebra, namely, *dT-Calculus*, which is the timed version of d-Calculus, extended for more specific temporal specification and analysis of the requirements of the IoT systems. More specifically, dT-Calculus allows the temporal properties of the actions of processes to be expressed as follows:

- *Ready time*: The time needed before execution of an action or a process.

- *Timeout*: The maximum waiting time up to the actual execution of an action or a process, after the execution will be ready with *ready time*.

- *Execution Time*: The actual execution time of an action or a process.

- *Deadline*: The time that the execution of action is to be terminated.

- *Period*: Period for repetition of an action or a process.

These specific temporal properties allow various types of temporal requirements of process movements and communications over the IoT environment to be specified and analyzed, without modifying any types of the process movements and communications from d-Calculus.

This paper is organized as follows. Section II introduces some of the existing process algebras with temporal properties. Section III introduces the basic algebra for dT-Calculus, that is, d-Calculus. Section VI describes syntax and semantics of dT-Calculus, focusing on its temporal properties. Section V demonstrates usability of dT-Calculus with a simple IoT example. Section VI shows some comparison of dT-Calculus with other process algebras. Section VII introduces a tool, called SAVE [8, 9], which is developed on ADOxx meta-modeling platform, to specify and analyze the temporal requirements of the process movements with dT-Calculus. Finally conclusions will be made and some of future researches will be discussed.

## 2. Related research

### 2.1. Timed pi-Calculus

One of the best known process algebra to specify the temporal properties is Timed pi-Calculus. It is the timed version of pi-Calculus, adding the temporal properties to process movements. **Figure 1** shows the syntax of Timed pi-Calculus.

In the *send* and *receive* actions of the calculus, $t_c$ and $c$ represent *time-stamp* and *clock* used for creating of the time-stamp, respectively. Further $\delta$ and $\gamma$ represent *temporal restriction condition* and *clock reset*, respectively. The process specification with temporal restriction condition is to be used as follows:

$$P = (c < 2)\bar{x}\langle y, t_c, c\rangle.P^{'} \tag{1}$$

It implies that, in 2 time units after *clock c* is reset, *name y* can be transmitted through *channel x* in $t_c$.

The notion of clock in Timed pi-Calculus is based on local clock concept, which allows various kinds of temporal restriction conditions. For example,

| | | | |
|---|---|---|---|
| $P ::= M$ | message | $M ::= \delta\gamma\bar{x}\langle y, t_c, c\rangle.P$ | send |
| $\mid (P \mid P')$ | composition | $\mid \delta\gamma x((y, t_c, c)).P$ | receive |
| $\mid !P$ | replication | $\mid \delta\gamma\tau.P$ | internal action |
| $\mid (z)P$ | restriction | $\mid 0$ | inactive process |
| $\mid [x = y]P$ | match(name) | $\mid M + M'$ | choice |
| $\mid [c = d]P$ | match(clock) | | |

**Figure 1.** Syntax of Timed pi-Calculus.

$$Q = (e > 5)(d - t_z \leq 3)x(\langle z, t_z, d \rangle).Q' \tag{2}$$

It specifies two temporal conditions with clock: $(e > 5)$ represents a condition for a local clock $e$, and $(d - t_z \leq 3)$ represents a temporal condition related to a receiving message. $d$ and $t_z$ are the temporal conditions on the clock for the receiving message and its time-stamp, but, since the clock ticks continuously, $(d - t_z \leq 3)$ implies the temporal condition that the message should be transmitted in 3 time units.

The mobile property of Timed pi-Calculus is represented indirectly by changing the state of channel connection among processes through passing the connecting channel names. For example,

$$\overline{y}x.P' | y(z).Q' | R \xrightarrow{\tau} P' | Q' \{x/z\} | R \tag{3}$$

As shown in **Figure 2**, it represents the state of $P$ and $R$, connected by $x$, to be changed to the state of $Q$ and $R$, newly connected by $x$, after passing the name $x$ to $Q$ by $P$ through the channel $y$. Obviously the connection between of $P$ and $R$ is invalid since there is no $x$ in $P$.

## 2.2. Timed Mobile Ambient

Timed Mobile Ambient is another process algebra to specify process movements and temporal properties. It is the timed version of Mobile Ambient. **Figure 3** shows the syntax of Timed Mobile Ambient.

In Timed Mobile Ambient, 0 represents the process with no action. $n$ in $n^{\triangle t}[P]^{\mu}$ implies the location where Process $P$ executes, and $\triangle t$ does that P should terminate its execution in $t$.

If $t > 0$, then ambient $n^{\triangle t}[P]^{\mu}$ is equal to $n[P]$. If a timer becomes 0 by $t = 0$, then $n^{\triangle t}[P]^{\mu}$ can be represented as a pair of $(n^{\triangle t}[P]^{\mu}, Q)$, where $Q$ is a safe process, implying that, in case that $n^{\triangle t}[P]^{\mu}$ is not completed in time or timed out, a safe process $Q$ can be activated in order to handler the time-out case of $n^{\triangle t}[P]^{\mu}$. For example, if the *open n* capability does not occur in the time $t$, ambient $n^{\triangle t}[P]^{\mu}$ is deactivated, and a safe process $Q$ is activated instead as a handler. If $Q = 0$, then $n^{\triangle t}[P]^{\mu}$ can be simple enough to represent $(n^{\triangle t}[P]^{\mu}, Q)$.

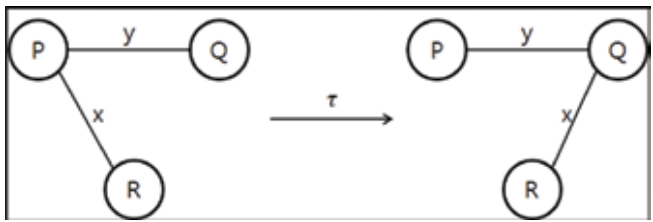

**Figure 2.** Movement in Timed pi-Calculus.

| $a, p,$ | ambient tags | $P, Q ::=$ | processes |
| $c$ | channel name | $\mid 0$ | inactivity |
| $n, m$ | ambient names | $\mid M^{\triangle t}.(P, Q)$ | movement |
| $x$ | variable | $\mid (n^{\triangle t}[P]^{\mu}, Q)$ | ambient |
| | | $\mid P \mid Q$ | composition |
| $M ::=$ | capabilities | $\mid (vn : Amb[\Gamma])P$ | restriction |
| $\mid in\ n$ | can enter n | $\mid c^{\triangle t}! < m : Amb[\Gamma] >.(P, Q)$ | output action |
| $\mid out\ n$ | can exit n | $\mid c^{\triangle t}?(x : Amb[\Gamma]).(P, Q)$ | input action |
| $\mid open\ n$ | can open n | $\mid * P$ | replication |

**Figure 3.** Syntax of Timed Mobile Ambient.

| (R-GTProcess) | $\dfrac{P \nrightarrow}{P \rightarrow \phi\Delta(P)}$ | (R-Res) | $\dfrac{P \rightarrow Q}{(vn : Amb[\Gamma])P \rightarrow (vn : Amb[\Gamma])Q}$ |
| (R-In) | $n^{\triangle t'}[in^{\triangle t}m.(P, P')\mid Q]^{\hat{a}}\mid(m^{\triangle t''}[R]^{\mu}, S'') \rightarrow$ $(m^{\triangle t'}[(n^{\triangle t'}[P\mid Q]P.S')\mid R]^{\hat{a}}, S'')$ | (R-Amb) | $\dfrac{P \rightarrow Q}{(n^{\triangle t}[P]^{\mu}, R) \rightarrow (n^{\triangle t}[Q]^{\mu}, R)}$ |
| (R-Out) | $\left(m^{\triangle t'}\left[(n^{\triangle t''}[out^{\triangle t}m.(P, P')\mid Q]^{\hat{a}}, S'')\mid R\right]^{\mu}, S'\right) \rightarrow$ $(n^{\triangle t'}[P\mid Q]P.S'')\mid(m^{\triangle t'}[R]^{\mu}, S')$ | (R-Par1) | $\dfrac{P \rightarrow Q}{R\mid P \rightarrow R\mid Q}$ |
| (R-Com) | $c^{\triangle t}! < m : Amb[\Gamma] >.(P, Q)\mid c^{\triangle t}?(x : Amb[\Gamma]).(P', Q') \rightarrow$ $P\mid P'\{m/x\}$ | (R-Par2) | $\dfrac{P \rightarrow P', Q \rightarrow Q'}{P\mid Q \rightarrow P'\mid Q'}$ |
| (R-Open) | $\dfrac{n : Amb[\Gamma'], m : Amb[\Gamma], \Gamma < :\Gamma'}{(m^{\triangle t'}[open^{\triangle t}n.(P, P')](n^{\triangle t'}[Q]\hat{a}, S'')], S') \rightarrow (m^{\triangle t'}[P\mid Q]\mu, S')}$ | (R-Struct) | $\dfrac{P' \equiv P, P \rightarrow Q, Q \equiv Q'}{P' \rightarrow Q'}$ |

**Figure 4.** Reduction rules of Timed Mobile Ambient.

Tags are related to reductions, which are similar to execution rules, and are classified into active and passive ones. And $\mu$ is a neutral tag to represent whether a tag is active or passive. An active tag performs a reduction in a time unit by consuming capability, and a passive tag performs a series of reductions in time units. The reduction rule is defined in **Figure 4**.

The movement $M^{\triangle t}.(P, Q)$ is provided by the capability $M$, and followed by the execution of Process $P$. If the time becomes 0 as in $t = 0$, the safe process $Q$ is executed instead of $P$.

An output action implies that Process $P$ releases a name $m$ on Channel $c$. An input action implies that that Process $P$ brings a name from Channel $c$ and binds it to a name $n$ within the scope of $P$. Restriction does that a new unique name $n$ is declared within the scope of $P$.

Since the communication method used in Timed Mobile Ambient is not direct, it is possible to define appropriate types for receivers in communication. The $Amb[\Gamma]$ in the restriction and the output and input actions is used to define such types.

**Figure 5** shows a part of the Cab Protocol in Timed Mobile Ambient [5]. The basic scenario of the protocol is that *cab* takes on a *client* sending the signal *call* from the place *from*. If the call

$$load\ client = loading^{\Delta t_1} \left[ out^{\Delta t_2} cab.in^{\Delta t_3} client \right]^{\mu}$$
$$call = call^{\Delta t_7} \left[ out^{\Delta t_8} client.out^{\Delta t_9} from.in^{\Delta t_{10}} cab.in^{\Delta t_{11}} from.load\ client \right]^{\mu}$$
$$recall = recall^{\Delta t_{12}} \left[ out^{\Delta t_{13}} cab.in^{\Delta t_{14}} from.in^{\Delta t_{15}} client \right]^{\mu}$$
$$call\ from\ client = (call, recall)$$

**Figure 5.** Timed Mobile Ambient example.

from the client is not replied, the client should *recall*. The cab can be absent or full of customers, the client can be waiting for a cab at the specific place while sending signals or be on a cab. In order to specify the scenario, four processes are defined: *load client, call, recall,* and *call from client.*

In specification, Ambient *client* must enter *cab*, and *cab* can release Ambient *load client*. After Ambient *client* gets off *cab*, Ambient *from* looks for *cab* for another *client*'s transportation. If Ambient *from* finds *cab*, *client* gets on *cab* by the *R-In* reduction.

$$\left( call^{\triangle t_7} \left[ in^{\triangle t_{10}} cab.in^{\triangle t_{11}} from.\ \dots \right]^a, recall \right) | cab^{\infty} [\ ]^{\mu} \rightarrow cab^{\infty} \left[ call^{\triangle t_7} \left[ in^{\triangle t_{11}} from.\ \dots \right]^p, recall \right]^{\mu} \quad (4)$$

If the timer $\triangle t_7$ of Ambient *call* is terminated before getting-on *cab*, Ambient *call* is released automatically. This kind of specification allows for Ambient *cab* and Ambient *call* not to contact each other in $\triangle t_7$. After releasing Ambient *call*, a safe process can be executed by the *R-GTProgress* reduction.

$$\left( call^{\triangle t_7} \left[ in^{\triangle t_{10}} cab.in^{\triangle t_{11}} from.\ \dots \right]^a, recall \right) \rightarrow recall \quad (5)$$

Once Ambient *recall* enters Ambient *client*, other *call*s will be informed for execution. The *recall* process will repeat itself until *load client* is released.

# 3. Preliminary research

d-Calculus is the process algebra developed to specify and analyze the process movements directly on geographical space. There are four types of movements in d-Calculus, all of which are synchronously defined.

## 3.1. Syntax

The syntax of d-Calculus is shown in **Figure 6** and is defined as follows:

1. **Action:** Actions performed by a process.

2. **Priority:** The priority of the process $P$ represented by a natural number $n \geq 0$. The higher number represents the higher priority. Exceptionally, 0 represents the highest priority.

| $P ::= A$ | action | $M ::= m_r^p(k)\,P$ | request |
| $\mid P_{(n)}$ | priority | $\mid P\,m(k)_t$ | permission |
| $\mid P[Q]$ | nesting | | |
| $\mid P\langle r_t\rangle$ | channel | $m ::= in \mid out \mid get \mid put$ | movement types |
| $\mid P + Q$ | choice | | |
| $\mid P \parallel Q$ | parallel | $C ::= new\,P$ | create process |
| $\mid P\backslash_t F$ | exception | $\mid kill\,P$ | kill process |
| $\mid A \cdot P$ | sequence | $\mid exit$ | exit process |
| | | | |
| $A ::= \emptyset$ | empty | | |
| $\mid r_t(\overline{msg})$ | send | | |
| $\mid r_t(msg)$ | receive | | |
| $\mid M$ | movement | | |
| $\mid C$ | control | | |

**Figure 6.** Syntax of d-Calculus.

3. Nesting: $P$ contains $Q$. The internal process is controlled by its external process. If the internal process has a higher priority than that of its external, it can move out of the external without the permission of the external.

4. Channel: A channel $r$ of $P$ to communicate with other processes. $t$ implies the time needed for the communication through the channel.

5. Choice: Only one of $P$ and $Q$ will be selected non-deterministically for execution.

6. Parallel: Both $P$ and $Q$ are running concurrently.

7. Exception: Execution of $P$, but $F$ in case of violation of the deadline $t$.

8. Sequence: $P$ follows after action $A$.

9. Empty: No action.

10. Send/Receive: Communication between processes, exchanging a message by a channel $r$. $t$ represents deadline of the communication.

11. Request: Requests for movement. $t$, $p$ and $k$ represent deadline, priority and key, respectively.

12. Permission: Permissions for movement. $t$ represents deadline.

13. Create process: Creation of a new internal process. The new process cannot have a higher priority than its creator.

14. Kill process: Termination of other processes. The terminator should have the higher priority than that of the terminate.

15. Exit process: Termination of its own process. All internal processes will be terminated at the same time.

Generally all the movements are synchronous. In order for a process to move in or out of another process, the moving process (*mover*) needs permission from the target process.

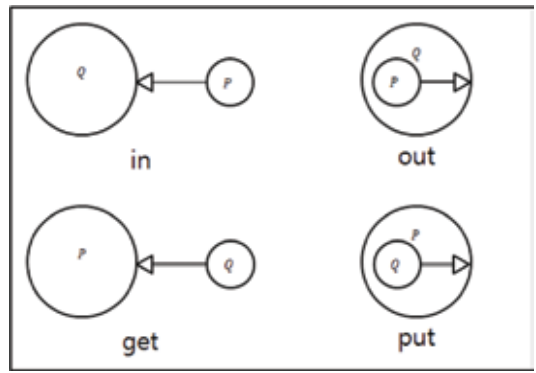

**Figure 7.** Pictorial view of d-Calculus movements.

Reversely, in order for a process to be moved in or out of another process forcefully, the moving process needs permission from the being-moved process (*movee*).

By means of the strict method of synchrony, the movements of processes can be controlled, and further the security and safety of the IoT systems can be guaranteed by pre-cautiously preventing insecure or unsafe movements.

### 3.2. Mobility

As stated, the process movement in d-Calculus occurs synchronously between the requesting process and the permitting process. It implies that the movement cannot be allowed without permission. It prevents any unplanned movement from occurring unexpectedly, and clarifies control of the movement explicitly. There are four types of such movements in d-Calculus as follows:

- *in*: A process moves into another process directly.

- *out*: A process moves out of another process directly.

- *get*: A process makes another process move into itself.

- *put*: A process makes another process move out of itself.

The types of movements can be pictorial depicted as shown in **Figure 7**.

# 4. dT-Calculus

dT-Calculus is the process algebra developed to specify and analyze the movements of things in the IoT systems with temporal restrictions directly on geographical space. In order to represent precise temporal properties explicitly, it extended the basic temporal property of the movements in d-Calculus to specify the different types of temporal properties for period and sporadic actions or processes, with the additional syntax and semantics accordingly.

## 4.1. Temporal properties

As shown in **Figure 8**, there are five temporal properties in dT-Calculus: *ready time*, *timeout*, *execution time*, *deadline*, and *period*. The first four properties are used to specify the temporal properties of sporadic actions or processes, and the last one is used to specify the temporal properties of periodic actions and processes inclusively. The definition of each property is as follows:

1. *Ready time*: It represents the waiting time for an action. At the point of the action in a process, the process was to wait in *ready time* before executing the action. No other or synchronous actions are possible during *ready time*.

2. *Timeout*: It represents the maximum waiting time for the actual execution of an action to be started after the action is ready for execution. If the waiting time in *ready time* is over and the partner for its synchronous action is not ready, the action cannot be executed. If the partner is ready for the action in *timeout*, the action can be executed. If not, the action will be in the state of *timeout*, the process will be in some fault state unless some proper handling action is not specified.

3. Execution Time: The time needed to execute an action. In case that the action can be performed in *timeout* after *ready time*, the action will be executed in *execution time* and be terminated. And then the next action will be available.

4. *Deadline*: The termination time for the execution of an action. All actions must be terminated in *deadline*. *Deadline* starts as *ready time* does. If the action is terminated in *deadline*, the process will be in some fault state. In order to prevent the process from being in the fault state, an exceptional handling must be specified accordingly.

5. *Period*: The duration of period for the execution of an action or process in repetition. The action will repeat itself after period of executing the action or process. This is an additional temporal property to specify the periodic action or process, different from the previous four temporal properties. The periodic action or process can be put into some fault state due to failure or *timeout* and *deadline*.

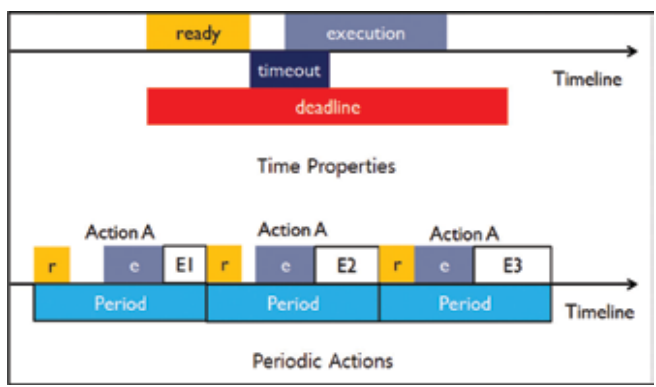

**Figure 8.** Time properties of dT-Calculus.

All actions and processes are defined or specified with these temporal properties. However the properties cannot be applied to some actions and processes. For example, *empty* action, no-time action, timed process, etc.

## 4.2. Syntax

The syntax of dT-Calculus is shown in **Figure 9**, and the extended notions from d-Calculus for temporality are as follows:

1. Timed action: The execution of an action with temporal restrictions. The temporal properties of [*r, to, e, d*] represent *ready time*, *timeout*, *execution time*, and *deadline*, respectively. *p* and *n* are properties for periodic action or processes: *p* for period and *n* for the number of repetition.

2. Timed process: Process with temporal properties.

3. Exception: *P* will be executed. But *F* will be executed in case that *P* is out of timeout or deadline.

The biggest difference of dT-Calculus with d-Calculus is the notion of *timed* action and processes. In d-Calculus, the temporal property is simple, defined with a time interval in action or process: the boundary of the lower and upper time limits. However, in dT-Calculus, the property is divided into more specific properties, as described. In addition, the exceptions caused by the violation of the temporal properties are more specifically divided into the one by *deadline* and the one by *timeout*.

Consequently the separate notions for temporal properties for action and process in d-Calculus can be represented in one single notion and form of the properties in dT-Calculus.

If there is no temporal properties to be specified in an action, it will be considered to be [0,-,1,-] by default. That it, there is no waiting time so that the action can be executed immediately, and infinite waiting for the synchronous co-action is possible without *timeout* and *deadline*.

$$
\begin{array}{llll}
P ::= A & \text{action} & A ::= \emptyset & \text{empty} \\
\quad | \ A^{p,n}_{[r,to,e,d]} & \text{timed action} & \quad | \ r(\overline{msg}) & \text{send} \\
\quad | \ P^{p,n}_{[r,to,e,d]} & \text{timed process} & \quad | \ r(msg) & \text{receive} \\
\quad | \ P_{(n)} & \text{priority} & \quad | \ M & \text{movement} \\
\quad | \ P[Q] & \text{nesting} & \quad | \ C & \text{control} \\
\quad | \ P(r) & \text{channel} & M ::= m^p(k) \ P & \text{request} \\
\quad | \ P + Q & \text{choice} & \quad | \ P \ m(k) & \text{permission} \\
\quad | \ P \parallel Q & \text{parallel} & m ::= in \ | \ out \ | get \ | \ put & \text{movement types} \\
\quad | \ P \backslash F & \text{exception} & C ::= new \ P & \text{create process} \\
\quad | \ A \cdot P & \text{sequence} & \quad | \ kill \ P & \text{kill process} \\
& & \quad | \ exit & \text{exit process}
\end{array}
$$

**Figure 9.** Syntax of dT-Calculus.

### 4.3. Semantics

The semantics of dT-Calculus for the temporal properties in action and process are defined as transition rules as shown in **Table 1**.

Each rule in the table is defined as follows:

1. *Tick-Time R*: The rule for *ready time* of an action. As time passes in *ready time*, the values of $r$ and $d$ decrease accordingly.

2. *Tick-Time TO*: The rule for *timeout* of an action. The action, not executing, but in waiting, decreases its *timeout* time accordingly as time passes.

3. *Tick-Time End*: The rule for termination of an action. After the execution of the action started and the value of $e$ becomes 0, the next action can start.

4. *Tick-Time SyncE*: The rule for execution of an action. When an action and its partner co-action are executed synchronously, the values of $e$ and $d$ decrease accordingly as time passes.

5. *Tick-Time AsyncE*: The rule for *execution time* of an asynchronous action. In case of asynchronous action, there is no need for timeout: it goes into its own execution immediately just after *ready time* and the values of $e$ and $d$ decrease accordingly as time passes.

| | |
|---|---|
| Tick-Time R | $\dfrac{-}{A_{[r,to,e,d]} \xrightarrow{\;\rhd_1\;} A_{[r-1,to,e,d-1]}}\,(r \geq 1)$ |
| Tick-Time TO | $\dfrac{-}{A_{[0,to,e,d]} \xrightarrow{\;\rhd_1\;} A_{[0,to-1,e,d-1]}}\,(to \geq 1)$ |
| Tick-Time End | $\dfrac{-}{A_{[0,0,0,d]} \cdot A' \xrightarrow{\;\rhd_1\;} A'}$ |
| Tick-Time SyncE | $\dfrac{A\|A' \xrightarrow{(\tau \vee \delta) \wedge \rhd_1} A''\|A'''}{A_{[0,to_1,e_1,d_1]}\|A'_{[0,to_2,e_2,d_2]} \xrightarrow{(\tau \vee \delta) \wedge \rhd_1} A_{[0,to_1,e_1-1,d_1-1]}\|A'_{[0,to_2,e_2-1,d_2-1]}}\,(e_1 \geq 1 \wedge e_2 \geq 1)$ |
| Tick-Time AsyncE | $\dfrac{-}{A_{[0,to,e,d]} \xrightarrow{\;\rhd_1\;} A_{[0,to,e-1,d-1]}}$ |
| Tick-Time P | $\dfrac{-}{P_{[r,to,e,d]} \xrightarrow{\;\rhd_1\;} P_{[r,to,e,d-1]}}$ |
| Timeout | $\dfrac{-}{A_{[0,0,e,d]} \setminus P \xrightarrow{\;\rhd_1\;} P}$ |
| Deadline | $\dfrac{-}{A_{[r,to,e,0]} \setminus P \xrightarrow{\;\rhd_1\;} P}$ |
| Period | $\dfrac{-}{A^{p,n}_{[r,to,e,d]} \xrightarrow{\;\rhd_p\;} A^{p,n-1}_{[r,to,e,d]}}\,(n > 1)$ |
| Period End | $\dfrac{-}{A^{p,1}_{[r,to,e,d]} \cdot A' \xrightarrow{\;\rhd_p\;} A'}$ |

**Table 1.** Temporal semantics of dT-Calculus.

6. *Tick-Time P*: The rule for passage of time in process. Since the temporal property for a process uses only deadline in its temporal requirements, the value of *e* decreases accordingly as time passes.

7. *Timeout*: The rule for *timeout* to occur. When the value of *to* becomes 0, its timeout error will occur. However, when an exception for the *timeout* defines, its exception handling will be activated accordingly.

8. *Deadline*: The rule for violation of *deadline*. When the value of *d* becomes 0, its deadline error will occur. However, when an exception for the *deadline* defines, its exception handling will be activated accordingly.

9. *Period*: The rule for execution of a periodic action. The action will be executed again after the period passes, and the value of *n* will be decremented by 1.

10. *Period End*: The rule for termination of a periodic action. In case that the value of *n* is 1, no action will be repeated after the period passed over.

## 4.4. Laws

The laws for the additional temporal properties in dT-Calculus are shown in **Table 2**. The laws represent the notion and restrictions of temporal properties in dT-Calculus as follows:

1. *Timed Process*: Only applicable temporal property for a process is *deadline*.

2. *Non-time Action*: The action with no temporal properties is same as the one with the temporal properties of [0,-,1,-].

3. *Empty*: Only applicable temporal property for the *Empty* action is *execution time*.

## 4.5. Characteristics

The temporal properties are directly specified to each action and process in dT-Calculus. The specification of the temporal properties for both actions and processes allows the temporal requirements for both actions in a processes and the process itself to be specified and analyzed at the same time.

The introduction of the periodic temporal property has many advantages than other process algebras in specification of different types of repeating processes. Generally, the starting time of each synchronous action depends on the ready time of its partner action so that the same actions may require different total execution or termination time of their synchronous actions.

| | |
|---|---|
| $P_{[r,to,e,d]} = P_{[-,-,-,d]}$ | Timed Process |
| $A = A_{[0,-,1,-]}$ | Non-time Action |
| $\varnothing_{[r,to,e,d]} = \varnothing_{[-,-,e,-]}$ | Empty |

**Table 2.** Temporal Laws of dT-Calculus.

That is, there is some problem of not being able to specify explicitly and precisely the temporal properties of periodic actions in the following form:

$$A \cdot \varnothing_{[-,-,e,-]} \cdot A \cdot \varnothing_{[-,-,e,-]} \cdot A \cdot \varnothing_{[-,-,e,-]} \cdot \ldots \qquad (6)$$

It is intended to specify the above periodic actions with empty actions, but the empty actions with fixed execution time are not appropriate because their interaction times for synchronization can be different from each other. However, there is an advantage that there is no need to consider such time for synchronous interactions if the periodic temporal property is used. The specification of the periodic requirements becomes very simple since the next execution of an action will be performed after elapsing the periodic temporal duration without calculating the temporal length left over up to the next re-execution of the action following the immediate execution of the action.

### 4.6. Graphical representation

There are two graphical representations for dT-Calculus: system view and process view. *System view* represents graphical relationships among processes in a system: containment and interactions. *Process view* represents graphical relationships among actions in a process: precedency and control flow. These views show in-the-large (ITL) view of a system and in-the-small (ITS) views for its processes, respectively. And they provide better understanding of the system and the processes in the visual representation. **Tables 3** and **4** show the icons for the views, respectively.

| Icon | |
|---|---|
| Process | 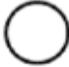 |
| Channel | 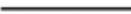 |
| Movement | 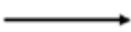 |

**Table 3.** Icon for system view.

| Icon | | | | | | | | | |
|---|---|---|---|---|---|---|---|---|---|
| Process Lane | 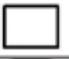 | Start | 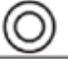 | End | 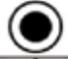 | Other Process | 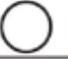 | | |
| Exit | 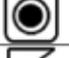 | Choice | 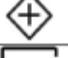 | Parallel | 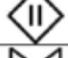 | Send | 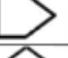 | | |
| Receive | 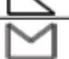 | Empty | 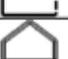 | InR | 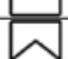 | OutR | 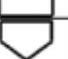 | | |
| GetR | 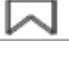 | PutR | 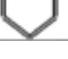 | InP | 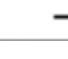 | OutP | 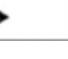 | | |
| GetP | 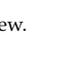 | PutP | 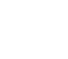 | Sequence | 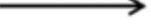 | | | | |

**Table 4.** Icon for process view.

## 5. Example

This section describes the specification of a distributed mobile real-time IoT system in dT-Calculus with a *Smart Emergency Evacuation System* (SEES) example.

SEES is a system that activates evacuation plan with supporting devices in buildings or facilities, in case of fire or threat, by detecting the source of fire or threat as well as the people and their movements in the building, and guiding them safely out of the building until all of them move out of the building safely in both active or passive manner [10].

### 5.1. Requirements

SEES needs a set of secure requirements since it guarantees safe evacuation of people in a building in case of fire or threat. The requirements include, as stated, provision of the evacuation plan, detection of the source of fire or threat as well as the people and their movements in the building, automatic notification of the fir and threat to police and 911, and safe guidance of the residents out of the building. It can be summarized as follows:

1. Req 1: Sensors must confirm occurrence of fire or threat continuously.

2. Req 2: Controller must send fire or threat alarm to all the people in case of fire or threat.

3. Req 3: Controller must guide all the people to the safe areas without fire in both present and near future.

4. Req 4: The evacuation of all the people must be completed in 25 time units.

5. Req 5: 911 must evacuate the people who are not escaped from the fire.

In case that these requirements are not satisfied, it is possible for people not to escape from fire or to escape through insecure paths, causing loss of human lives. Therefore it is very important to specify these requirements formally and to verify their satisfiability.

### 5.2. Specification

As shown in **Figure 10** in dT-Calculus, the SEES in the example operates as follows:

1. A fire is detected by sensor(s), and is informed to the controller by the sensor(s).

2. The controller informs the people in the building of the fire or threat, and, at the same time, shows the evacuation paths as planned.

$Sys \coloneqq Building[Control\ System|StairA[SensorA]|StairB[SensorB]\ |1st\ floor|2nd\ floor|P1|P2]]|\ 911;$

$Control\ System \coloneqq (CS(FireA) \cdot P1(\overline{StairB}) \cdot P2(\overline{StairB}) + CS(FireB) \cdot P1(\overline{StairA}) \cdot P2(\overline{StairA}))$

$\qquad\qquad \cdot CE(\overline{Call}) \cdot CS(P1)_{[0,0,1,7]} \backslash CE(\overline{P1}) \cdot CS(P2)_{[0,0,1,14]} \backslash CE(\overline{P2});$

$SensorA \coloneqq \left((SA(\overline{Fire})_{[0,3,1,0]} \cdot CS(\overline{FireA})) \backslash \emptyset_3\right)^{6,\infty};$

$SensorB \coloneqq \left((SB(\overline{Fire})_{[0,3,1,0]} \cdot CS(\overline{FireB})) \backslash \emptyset_3\right)^{6,\infty};$

$P1 = (P1(StairB) \cdot (\emptyset.RC(P1).out\ 2nd.out\ Building + out\ 2nd.in\ StairB.out\ StairB.in\ 1st.out\ 1st.out\ Building))$

$\qquad + P1(StairA) \cdot (\emptyset.RC(P1).out\ 2nd.out\ Building + out\ 2nd.in\ StairA.out\ StairA.in\ 1st.out\ 1st.out\ Building));$

$P2 = (P2(StairB) \cdot (\emptyset.RC(P2).out\ 2nd.out\ Building + out\ 2nd.in\ StairB.out\ StairB.in\ 1st.out\ 1st.out\ Building))$

$\qquad + P2(StairA) \cdot (\emptyset.RC(P2).out\ 2nd.out\ Building + out\ 2nd.in\ StairA.out\ StairA.in\ 1st.out\ 1st.out\ Building));$

$StairA \coloneqq (P1\ in_{[0,0,1,10]} \cdot P1\ out) \backslash \emptyset \cdot (P2\ in_{[0,0,1,10]} \cdot P2\ out) \backslash \emptyset;$

$StairB \coloneqq (P1\ in_{[0,0,1,10]} \cdot P1\ out) \backslash \emptyset \cdot (P2\ in_{[0,0,1,10]} \cdot P2\ out) \backslash \emptyset;$

$1st\ floor \coloneqq (P1\ in_{[0,0,1,10]} \cdot P1\ out) \backslash \emptyset \cdot (P2\ in_{[0,0,1,10]} \cdot P2\ out) \backslash \emptyset;$

$2nd\ floor \coloneqq P1\ out_{[0,0,1,10]} \cdot P2\ out_{[0,0,1,10]} \backslash \emptyset \cdot (911\ in_{[0,0,1,20]} \cdot P1\ out_{[0,0,1,5]} \backslash \emptyset \cdot P2\ out_{[0,0,1,5]} \backslash \emptyset \cdot 911\ out) \backslash \emptyset;$

$Building \coloneqq (SA(\overline{Fire}) + SB(\overline{Fire})) \cdot \left(P1\ out_{[0,0,1,13]} \cdot CS(\overline{P1})\right) \backslash \emptyset \cdot \left(P2\ out_{[0,0,1,13]} \cdot CS(\overline{P2})\right) \backslash \emptyset$

$\qquad\qquad \cdot (911\ in_{[0,0,1,10]} \cdot P1\ out_{[0,0,1,5]} \backslash \emptyset \cdot P2\ out_{[0,0,1,5]} \cdot 911\ out) \backslash \emptyset;$

$911 \coloneqq Ce(Call) \cdot (CE(P1) \cdot in\ Building \cdot in\ 2nd \cdot RC(\overline{P1}) \cdot out\ 2nd \cdot out\ Building$

$\qquad + CE(P2) \cdot in\ Building \cdot in\ 2nd \cdot RC(\overline{P2}) \cdot out\ 2nd \cdot out\ Building$

$\qquad + CE(P1) \cdot CE(P2) \cdot in\ Building \cdot in\ 2nd \cdot RC(\overline{P1}) \cdot RC(\overline{P2}) \cdot out\ 2nd \cdot out\ Building)_{[0,10,0,0]} \backslash \emptyset;$

**Figure 10.** The SEES example in dT-Calculus.

**3.** The controller tracks all the people in the building while they are evacuating, and informs the current status of the evacuation to 911 in real-time, so that the people trapped in the building can be monitored in real-time as planned.

**4.** 911 rescues the people trapped in the building in order, based on the status of the fire or threat in the building and the availability of the rescue facilities and devices.

In the specification, the following actions have been declared in Process *Building* and Process *Control System* to detect the case that the people cannot be evacuated from building autonomously:

$$Building \coloneqq \cdots P1\ out_{[0,0,1,14]} \cdot CS(\overline{P1}) \cdots \qquad (7)$$

$$Control\ System \coloneqq \cdots CS(P1)_{[0,0,1,7]} \backslash CE(\overline{P1}) \qquad (8)$$

The above code implies that, when *P1* moves out of the building, it sends *CS* a signal of its safe evacuation, and that, if not, that is, if the signal is not received in the deadline of 7 time units of [0,0,1,7] by *CS*, the non-evacuation situation of *P1* is informed to 911 by the exception handler process *CE* of *CS*.

In the specification from **Figure 10**, sensors, *SensorA*, and *SensorB*, are defined to perform their actions in repetition by the period properties of dT-Calculus: normally their fire alarm actions do not occur by timeout in normal case of no fire, however, in case of fire, they have to occur in order to inform *Control System* of the fire.

There are two people in the building and there are two choices for them in case of fire: one for evacuation safely from the building, and another for non-evacuation.

## 5.3. Graphical representation

The textual specification in dT-Calculus can be represented graphically in two views: *in-the-large* (ITL) and *in-the-small* (ITS). The ITL view can be considered as *system view* consisting of processes interacting together with communication and movements. The ITL view can be considered as *process view* with the detailed actions. **Figure 11** shows the ITL view of the SEES example, and **Figures 12** and **13** show the ITS views of the processes in the example.

In order to construct the ITL view for the example, it is necessary to understand main processes and their containment relations from the example, which is textually specified with dT-Calculus in Section 5.2 as follows:

$$Sys:=Building[Control\ System|StairA[SensorA]|StairB[SensorB]|1stfloor|2nd\ floor[P1|P2]]|911; \quad (9)$$

In **Figure 11**, *P1* and *P2* are placed in *2nd floor* since they are defined as contained processes of *2nd floor* in Eq. 9. Similarly, *SensorA* and *SensorB* are placed in *StairA* and *StairB*, respectively, in the figure, since they are defined as contained process of *StairA* and *StairB*, respectively, in the equation. Further *1st floor*, *2nd floor*, *StairA* and *StairB* are placed in *Building* in the figure, since they are defined as contained processes of *Building* in the equation. However 911 is placed outside of *Building* in the figure since it is defined as a parallel process of *Building* in the equation. In addition, the edges in the view are the channels for communication among the processes in the example.

In order to construct the ITS view of each process as shown in **Figures 12** and **13**, it is necessary to understand the types of actions in each process and their order of execution. For example, **Figure 14** shows the ITS view of *Building* from **Figure 13**. The figure shows actions as nodes and their execution order as directed edges for *Building*, which is textually specified with dT-Calculus in Section 5.2 as follows:

$$
\begin{aligned}
Building := &\left( SA\left(\overline{Fire}\right) + SB\left(\overline{Fire}\right) \right) \cdot \left( P1\ out_{[0,0,1,13]} \cdot CS\left(\overline{P1}\right) \right) \backslash \varnothing \cdot \left( P2\ out_{[0,0,1,13]} \cdot CS\left(\overline{P2}\right) \right) \backslash \varnothing \\
&\cdot \left( 911\ in_{[0,0,1,10]} \cdot P1\ out_{[0,0,1,5]} \backslash \varnothing \cdot P2\ out_{[0,0,1,5]} \backslash \varnothing \cdot 911\ out \right) \backslash \varnothing;
\end{aligned}
$$

$$(10)$$

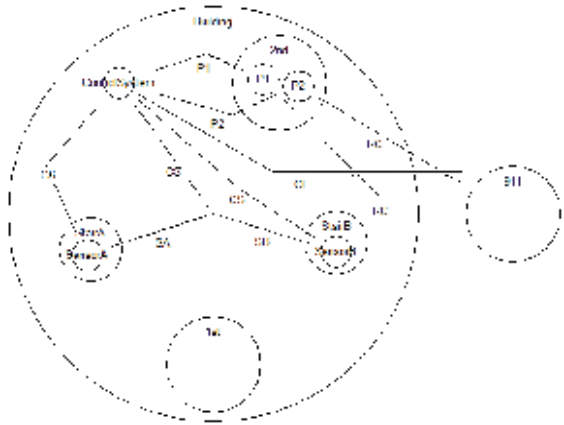

**Figure 11.** ITL view of the SEES example.

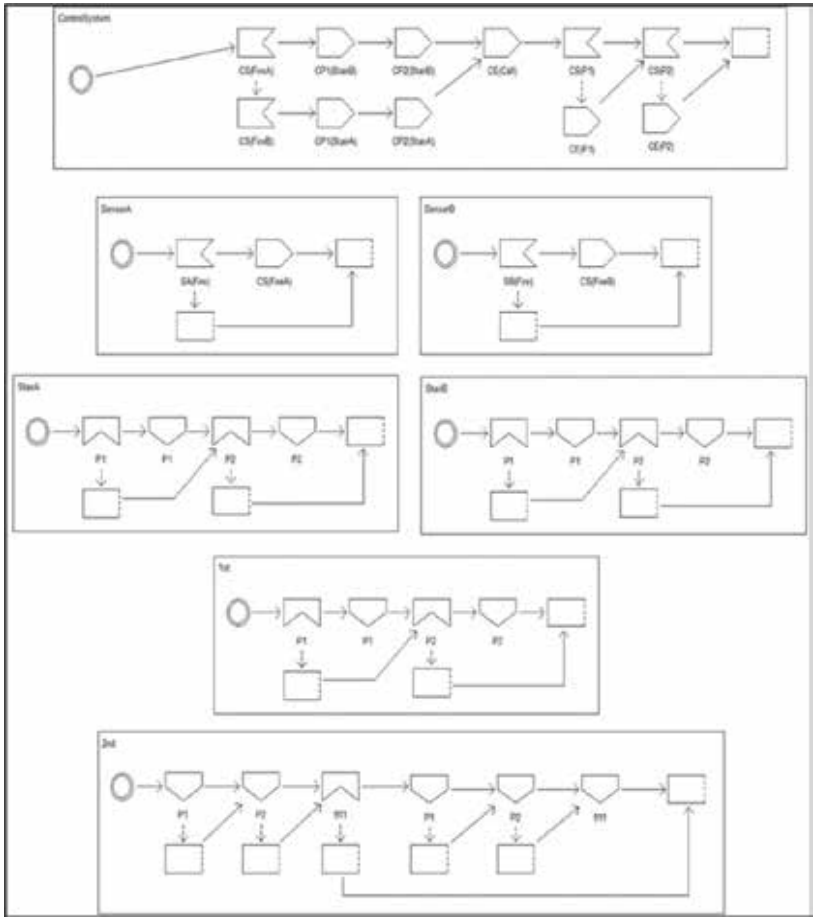

**Figure 12.** ITS views of the SEES e2xample (1).

*Building* performs the $SA\left(\overline{Fire}\right) + SB\left(\overline{Fire}\right)$ first. The *Choice* operation in the action is graphically represented with its *Choice* icon in the figure, including its two independent execution paths. And it is followed by a sequence of timed actions with exception, represented by their graphical icons. Firstly, $\left( P1\ out_{[0,0,1,13]} \cdot CS\left(\overline{P1}\right) \right) \backslash \varnothing$ is graphically represented by a pair of ordered action of $P1\ out_{[0,0,1,13]}$ and $CS\left(\overline{P1}\right)$ with its exception, that is, $\varnothing$, in the figure. Other timed actions are similarly represented in the same graphical pattern.

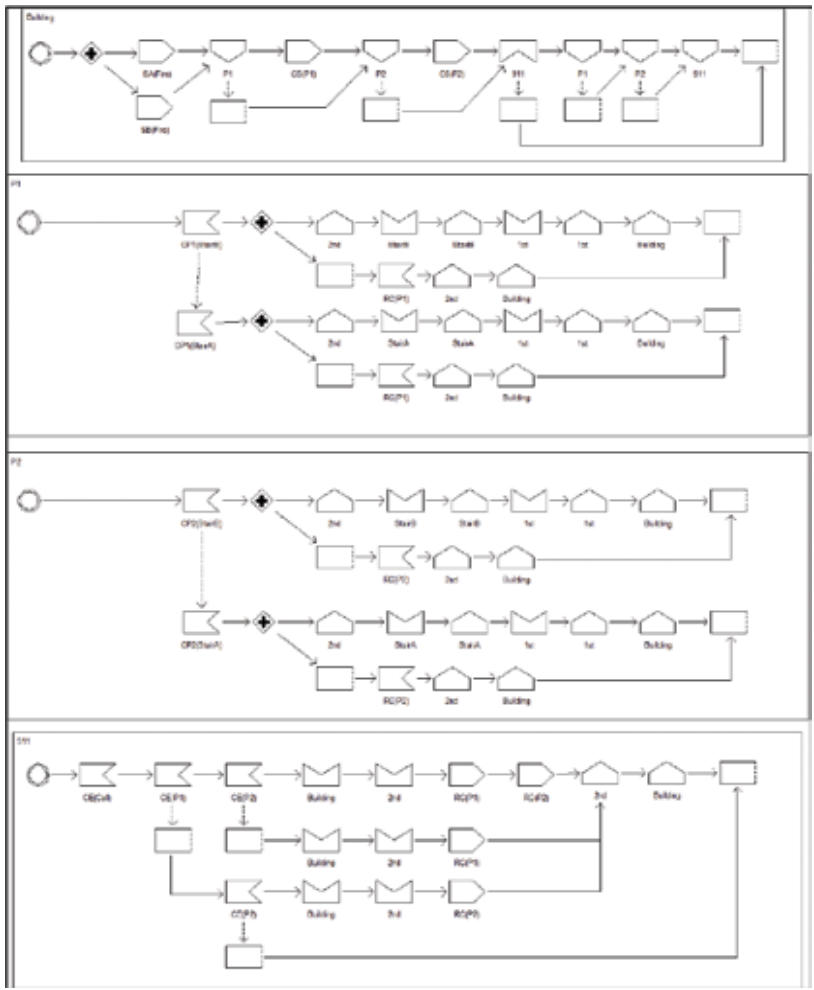

**Figure 13.** ITS views of the SEES example (2).

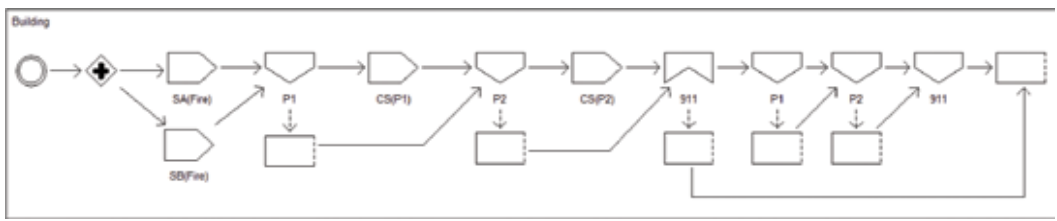

**Figure 14.** ITS view of the building process.

## 5.4. Execution

**Figure 15** shows the execution model for the SEES example. It consists of total 8 execution paths. Note that an execution path implies each independent case of execution by the example.

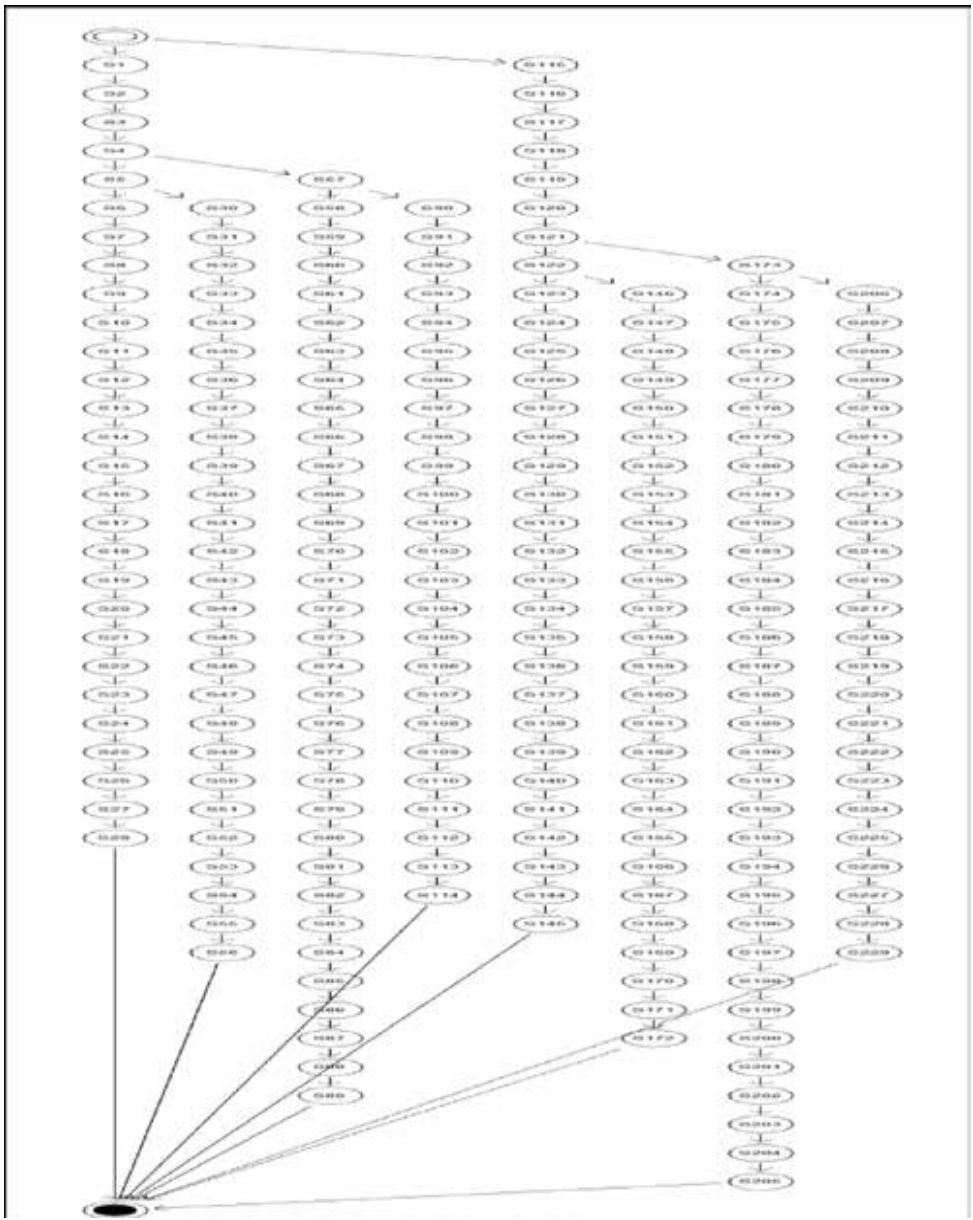

**Figure 15.** Execution paths of the SEES system.

It can be obtained by analyzing all the possible synchronization cases in the example. The icons in the model are defined in **Table 5**.

As shown in **Figure 15**, there are total eight paths: two locations for fire, two cases of evacuation for two persons, and consequently eight cases in total.

Firstly, for each execution path of successful evacuation, it is possible to perform analysis of their temporal properties as follows. Let us consider the case that a fire occurs at Stairs *A*:

| Icon | |
|---|---|
| Start | ◎ |
| End | ◉ |
| State | ○ |
| Deadlock | ◎ |
| Sequence | → |

**Table 5.** Icon for execution model.

1. T1: A fire occurs at Stairs *A*.

2. T2: A sensor detects the fire and informs a controller of the fire with a signal.

3. T3: The controller informs *P*1 on Floor 2 of an evacuation path through Stairs *B*.

4. T4: The controller informs *P*2 on Floor 2 of an evacuation path through Stairs *B*.

5. T5: The controller informs 911 of the fire, and *P*1 enters Stairs *B*.

6. T6: *P*2 enters Stairs *B*.

7. T7: *P*1 enters Floor 1.

8. T8: *P*2 enters Floor 1.

9. T9: *P*1 moves out of the building.

10. T10: *P*2 moves out of the building, and the controller detects that *P*1 moved out of the building.

11. T11: The controller detects that *P*2 moved out of the building.

All the people moved out of the building in 10 time units. And the controller detected their evacuation in 11 time units. Since there are more actions left to be performed by 911, it takes more time units for the system to terminate its own mission.

Secondly, for each execution path of failed evacuation, it is also possible to perform analysis of their temporal properties as follows. Let us consider the case that a fire occurs at Stairs *A*:

1. T1: A fire occurs at Stairs *A*.

2. T2: A sensor detects the fire and informs a controller of the fire with a signal.

3. T3: The controller informs *P*1 of an evacuation path through Stairs *B*.

4.  T4: The controller informs *P*2 of an evacuation path through Stairs *B*, and *P*1 is not able to move of out Floor 2.

5.  T5: The controller informs 911 of the fire.

6.  T6: *P*2 enters Stairs *B*.

7.  T8: *P*2 enters Floor 1.

8.  T10: *P*2 moves out of the building.

9.  T12: The controller detects that *P*1 is still on Floor 2.

10. T13: The controller informs 911 of the non-evacuation of *P*1.

11. T14: The controller detects that *P*2 moved out of the building, and 911 moves into the building to rescue *P*1.

12. T16: 911 finds *P*1 and provides the first treatment.

13. T18: *P*1 moves out of the building.

14. T19: 911 moves out of the building.

For evacuation, *P*2 takes 10 time units, but p1 takes 18 time units due to rescue time required for 911 to handle *P*1's non-evacuation situation. Once all the people are safely evacuated, the system will terminate its mission. However it will takes little more time due to some left-over actions by 911.

As a result of analysis, it can be confirmed that, in case of the fire at Stairs A, all the people were evacuated safely in 20 time units. Similar to the case of the fire at Stairs A, it can be confirmed that, in case of the fire at Stairs B, all the people were evacuated safely in 20 time units. Consequently it can be concluded that all the people in the building will be safely evacuated in time in any case of fires.

### 5.5. Analysis

In order to assure the safety of SEES, it is necessary to verify if the safety requirements, specified in dT-Calculus, in Section 5.1, are satisfied or not. All the five requirements specified in the section must be verified in order to prevent loss of lives from happening by fire as follows:

1.  Req 1: Sensors must confirm occurrence of fire continuously.

    It is specified in the SEES specification for *SensorA* and *SensorB* as follows. They are detecting fires in the same actions in different locations, that is, *A* and *B*:

$$SensorA := \left( \left( SA(\overline{Fire})_{[0,3,1,0]} \cdot CS(\overline{FireA}) \right) \setminus \varnothing_3 \right)^{6,\infty} \tag{11}$$

Each sensor performs a fire-detecting action for 3 time units. In case of no fire, it terminates its action immediately, but it repeats its fire-detecting action repeatedly as the following periodic actions with the specifier of its "$6, \infty$." However, in case of fire, it notifies the fire to the controller, and similarly, it repeats its fire-detecting action repeatedly as the following periodic actions.

2. Req 2: Controller must send a fire alarm to all the people in case of fire or threat.

The controller performs the following actions in case of fire:

$$\left(CS(FireA) \cdot P1\left(\overline{StairB}\right) \cdot P2\left(\overline{StairB}\right) + CS(FireB) \cdot P1\left(\overline{StairA}\right) \cdot P2\left(\overline{StairA}\right)\right) \qquad (12)$$

No matter where the fire occurs, it can be verified that the alarm is sent to all the people in the building: *P1* and *P2*.

3. Req 3: Controller must guide all the people to the safe areas without fire in both present and near future.

In the actions in 2), it can be varified that the people receiving the *FireA* by *CS* get the *StairB* signal for evacuation and, similarly, that the people receiving the *FireB* by *CS* get the *StairA* signal for evacuation. It guarantees that the people in the fire areas are evacuating through the non-fire areas.

4. Req 4: The evacuation of all the people must be completed in 25 time units.

As shown in Section 5.4, the autonomous evacuaiton, that is, the evacuation of the people without 911, takes 10 time units. However the heteronomous evacuaiton, that is, the evacuation of the people by 911, takes little longer that the autonomous case, since it requires the time that 911 arrives at the site. In this case, the controller has to recognize the situation of non-evacuation of the people at T12 ans T17, and 911 has to evacuate the people at T21 and T22. Finally, P1 is evacuated at P1, and P2 is evacuated at T24. In both cases, it can be verified that all the people are evacuated in 25 time units.

5. Req 5: 911 must evacuate all the people who are not escaped from the fire.

911 performs the following actions after the call:

$$(CE(P1) \cdot \cdots + CE(P2) \cdot \cdots + CE(P1) \cdot CE(P2) \cdot \cdots) \qquad (13)$$

It shows that the evacuations are performed by the signals from the controller, as the following calls for the signals of the controller show:

$$\cdots \cdot CS(P1)_{[0,0,1,7]} \setminus CE\left(\overline{P1}\right) \cdot CS(P2)_{[0,0,1,4]} \setminus CE\left(\overline{P2}\right) \qquad (14)$$

CS(P1) and CS(P2) are the signals from the people when they are evacuating from the building. In case that the signals are not transmitted to the controller in the certain period of time, the controller sends 911 the non-escaping signal to indicate the non-evacuation situation of the people. It is be verified that SEES guarantees that the controller recognizes all the non-evacuated people in the building and informs 911 of the situations, and that 911 evacuates them in time.

# 6. Comparative analysis

## 6.1. Main characteristics of IoT systems

The main characteristics of the IoT-based systems are shown in many literatures [11–13]. These can be summarized as follows with respect to process algebra:

1. Mobility: A number of devices in the systems are able to move their positions in geographical space. The devices should be able to get IoT services at any place and environment.

2. Real-time: The IoT devices in the systems should be able to get IoT services in real-time.

3. Interactivity: The interactions among the IoT devices in the systems must be possible, i.e., communication among the electronic devices in the smart home.

Especially, distributed mobile real-time IoT systems must have the above characteristics in order to operate properly in real-time without faults over geographical space with temporal restrictions.

## 6.2. Timed pi-Calculus

Timed pi-Calculus is a process algebra that is designed to specify and analyze mobile services. Timed pi-Calculus is the timed version of pi-Calculus, which allows time-stamp and clock be passed additionally during value passing: the temporal requirements of the process movements can be specified. However there is a limitation that the execution time of an action cannot be specified directly on the action. Further it is difficult to analyze the execution time, the deadline, and others of an action, since such temporal properties are represented by the passing time-stamp and clock. Similarly the movement in the algebra is inappropriate to represent a real movement of a process since it is represented by value passing. Consequently such indirect representation of a movement may result in distortion of the patterns of real movements since the representation reduces the scope of the possible movements in expression.

## 6.3. Timed Mobile Ambient

Timed Mobile Ambient is a process algebra that allows specification of temporal requiements by adding temporla properties on the existing movments of processes from Mobile Ambient. Temporal properties are added to process movements controlled by capabilities, and the process with the properties performs as follows: if the process performs an action within the valid time, it performs normally as in Mobile Ambient. If not, the existing process is intentionally terminated and a safe process is executed instead, in order to handle this abnormal situation. Timed Mobile Ambient solves the incapability of temporal specification of Mobile Ambient, but it is difficult to reason about starting time of processes since there is no other temporal properties except deadline. In addition, it is difficult to understand intuitively process synchronization since the synchronization is represented by the movements of ambients.

## 6.4. d-Calculus

d-Calculus is a process algebra that is designed to express direct process movements into or out of other processes both autonomously and heteronomously. It allows various types of mobile requirements to be specified, but only a simple type of temporal requirements for process movements is possible: a temporal bound of the minimum and maximum limits. It results in limited specification of the temporal requirements of the movements as well as analysis of the requirements. In addition, specification can be represented in both text and graph in order to increase visibility of the specification as well as comprehensibility. However there are limitations in specification of temporal properties: the execution time is only possible for an action and deadline is specified only by exception. It implies that only simple temporal specification is possible, but complex temporal specification for the smart EMS example is not allowed.

## 6.5. dT-Calculus

However, dT-Calculus overcomes these limitations of these algebras. Since it is an extension version of d-Calculus, it can utilize all different types of direct movements of processes. Besides, it is possible to specify complex temporal requirements of the smart mobile service by supplying a variety of additional temporal properties. Further, the analysis of the temporal properties is relatively easy since the properties are directly specified on actions and processes. And it is possible to specify exceptional handling to solve errors or faults caused by any violation of timeout and deadline.

## 6.6. IoT-based comparison

The first three process algebras can be analyzed with dT-Calculus with respect to the IoT characteristics stated in Section 6.1, as follows:

1. Mobility: A number of IoT devices are moving around in the IoT systems in a various manners. For example, a device containing other devices can move in and out of other devices, autonomously or heteronomously. In Timed pi-Calculus, the movements of processes can be expressed with value passing only. Consequently there are limitations to express various kinds of direct movements. In Timed Mobile Ambient, there are three in, out, open movement actions. However there is no movement action for passive or heteronomous movement. In d-Calculus and dT-Calculus, it is possible to express both autonomous and heteronomous movements of processes since they provide both the active actions of in, out and the passive actions of get, put.

2. Real-time: The IoT systems should provide their services in real-time. It means that the process algebras for the systems must have capability to express real-time properties of the services. In Timed pi-Calculus, it is possible to specify the temporal properties of processes by providing time-stamp and clock through value passing. But it is not possible to specify execution time of its actions. In Timed Mobile Ambient, it is possible to specify only temporal property of deadline for process movement with capability, but other properties are not possible. In d-Calculus, only execution time and deadline properties are possible, but others are not possible. However, in dT-Calculus, other properties, such as, ready time and time out, are possible, beside execution time and deadline properties of d-Calculus.

| Process Algebra | IoT Characteristic | | |
| --- | --- | --- | --- |
| | **Types of movement Properties** | **Types of temporal properties** | **Interactions** |
| Timed pi-Calculus | Indirect movements | Deadline | Communication |
| Timed Mobile Ambient | in, out, open | Deadline | Communication Movements |
| d-Calculus | in, out, get, put | Execution time Deadline | Communication Movements |
| dT-Calculus | in, out, get, put | Ready time Time out Execution time deadline period | Communication Movements |

**Table 6.** Comparison of dT-Calculus with other algebras by the IoT characteristics.

3. Interactions: All the devices in the IoT system should interact together. It implies that the process algebras for the systems must have capability to express the interactions. All the above algebras are able to express interactions among processes, but there are differences in the types of the interactions. In Timed pi-Calculus, the interactions are based on of synchronized communication. In Time Mobile Ambient, the interactions are based on capability-based movements, besides communication. In d-Calculus and dT-Calculus, the interactions are based on both communication and movements by synchronization.

**Table 6** shows the summary of the analysis with respect to the IoT characteristics.

## 7. SAVE

In order to demonstrate the feasibility of the approach in the paper, a tool, called SAVE (Specification, Analysis, Verification and Evaluation) [14], has been developed on the ADOxx meta-modeling platform [15]. As shown in **Figure 16**, it consists of four basic components as follows:

- Modeler: It provides capability to specify system and process views.

- EM Generator: It generates an execution model (EM) for the views and makes each path of the model to be selected for simulation.

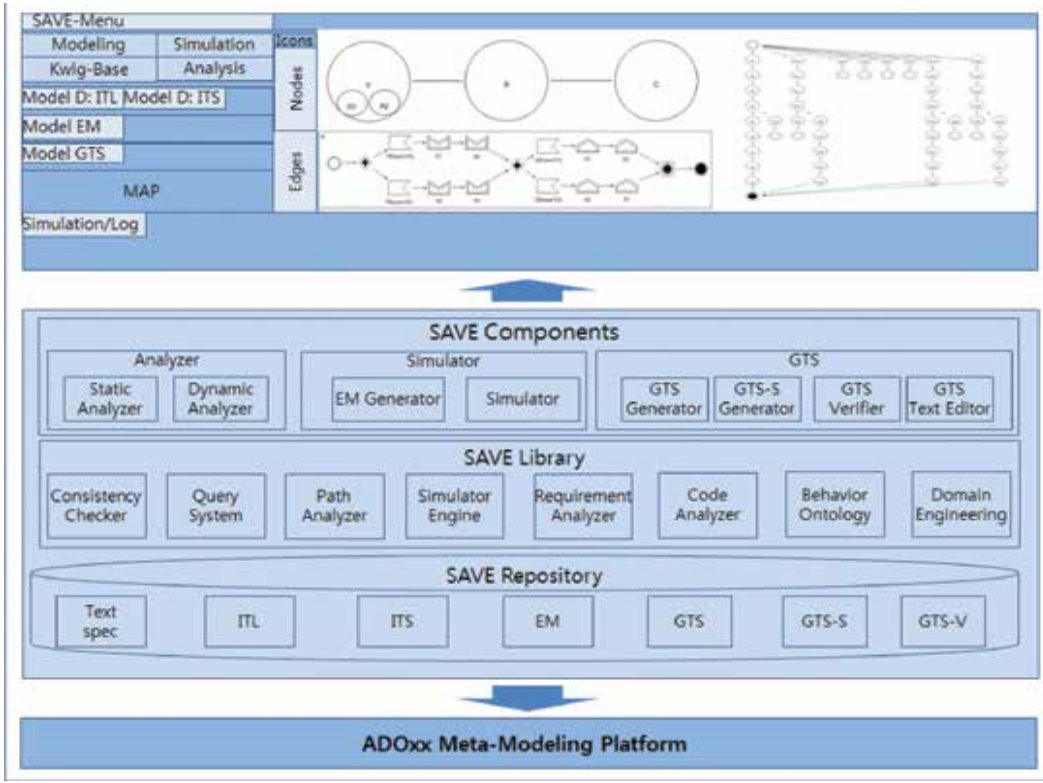

**Figure 16.** SAVE architecture.

- Simulator: It generates a model for the selected simulation, in a Geo-Temporal Space (GTS) diagram.

- Analyzer: It analyzes the secure requirements of the system by model-checking on the diagrams.

The graphical models in SAVE are designed by the ADOxx Development Tool, and the procedures of the SAVE components are built from the ADOxx libraries. The detailed logics of the procedures are programmed in the ADOScript language.

The first step to use SAVE for analysis is to specify systems in dT-Calculus. There are two specification models in SAVE, as shown in **Figures 17** and **18** for the SEES example: ITL (In-The-Large) and ITS (In-The-Small). From specification, the execution model can be automatically generated by the execution model generator. The execution model reveals all possible execution paths and determines whether each path is of normal or deadlock. **Figure 19** shows the execution model for the SEES example.

After generating an execution model, the simulation model for each execution path is automatically generated. The simulation model is represented in GTS (*Geo-Temporal Space*), where all the execution and movements resulted in the path are described in the model in detail.

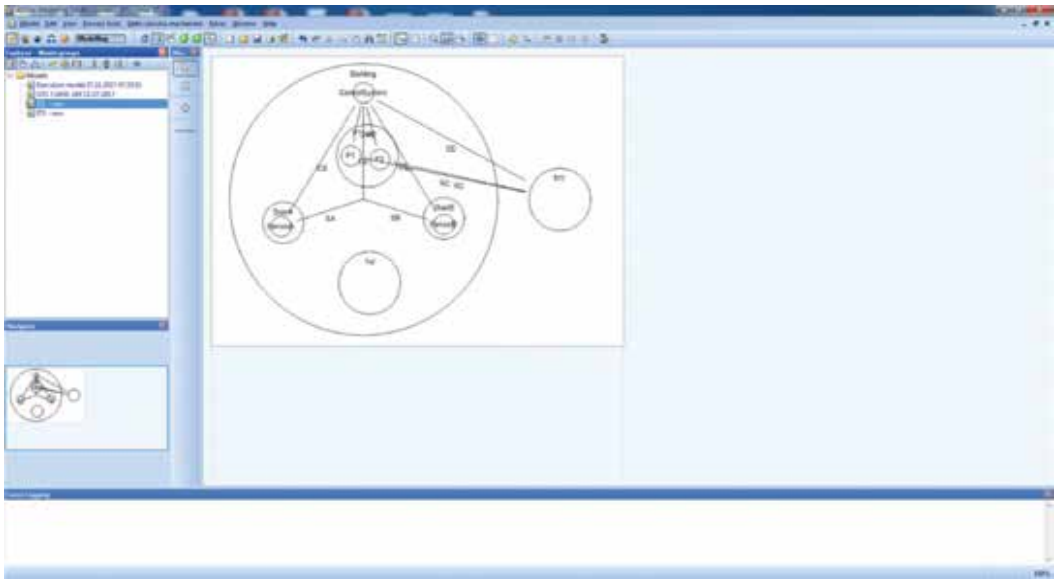

**Figure 17.** ITL model in SAVE.

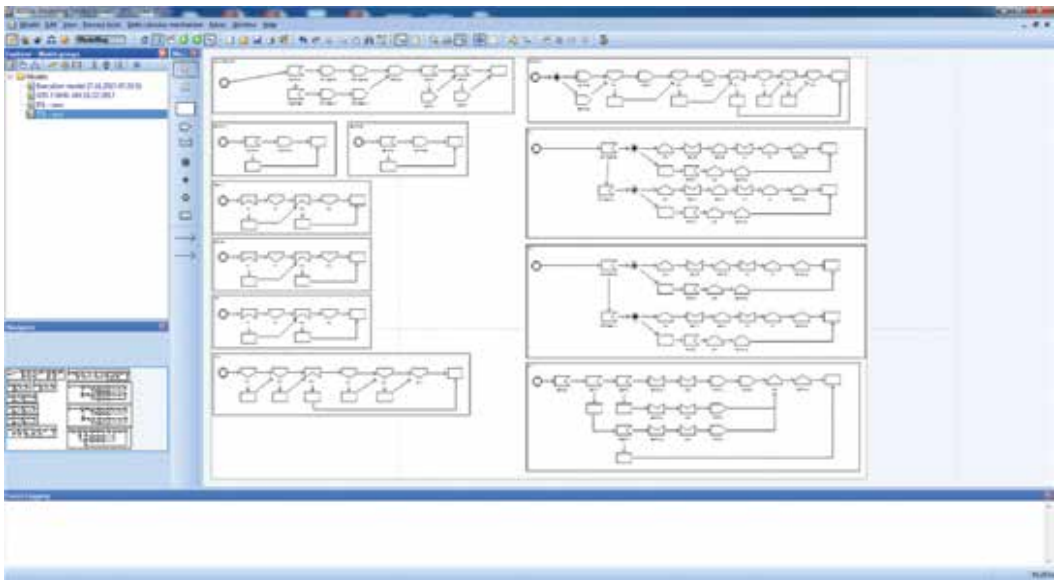

**Figure 18.** ITS models in SAVE.

Based on the simulation model, it is possible to analyze and verify the temporal requirements of IoT systems. **Figure 20** shows the simulation model for the first path of the SEES example in **Figure 19**.

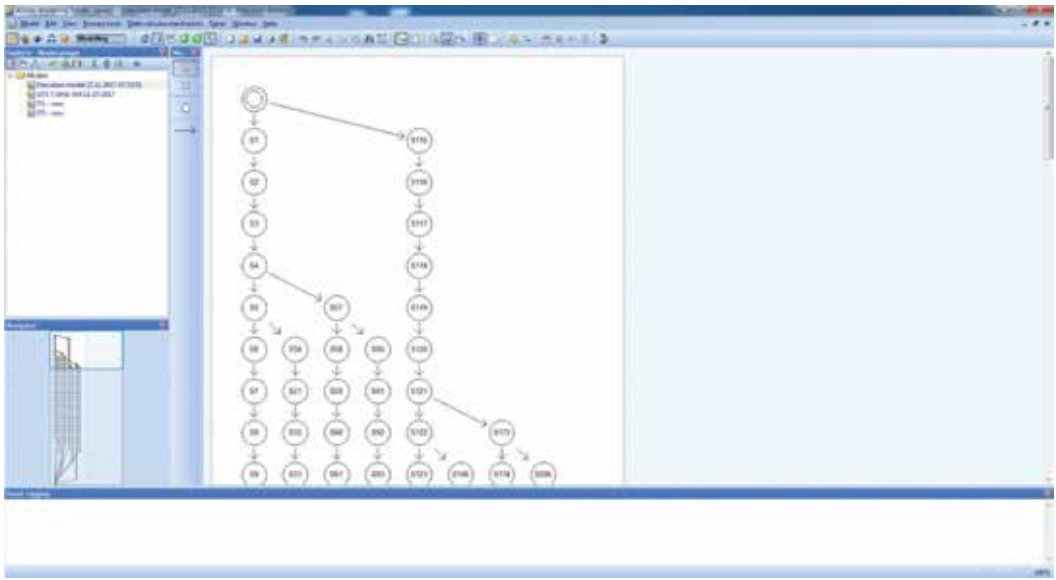

**Figure 19.** Execution model in SAVE.

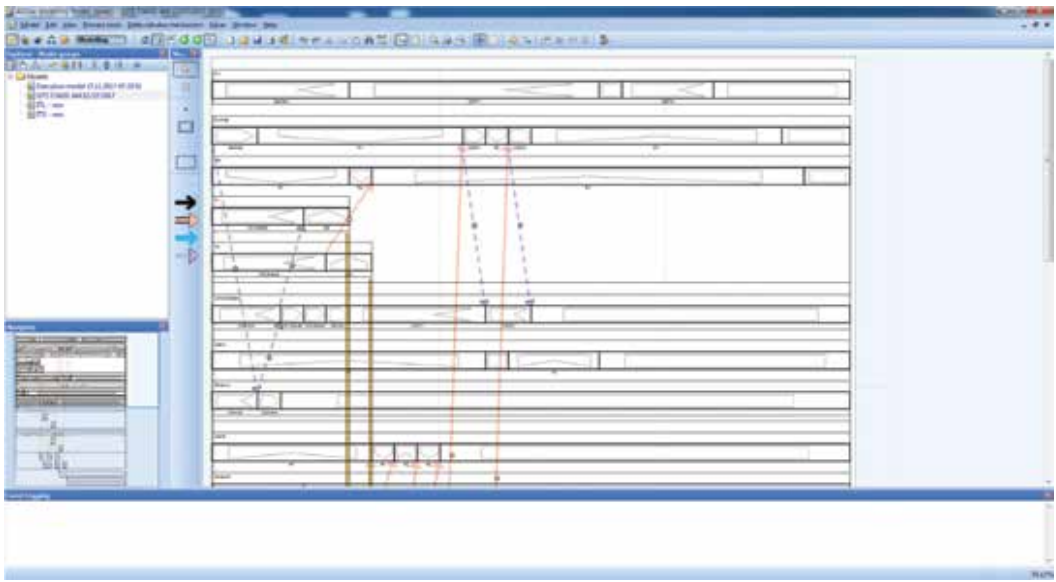

**Figure 20.** Simulation model in SAVE.

## 8. Conclusions and future research

This paper proposed dT-Calculus for mobile and temporal specification of the distributed mobile real-time IoT systems. The algebra extended d-Calculus for specification and analysis of a variety of different types of temporal properties at the direct movement actions and the mobile processes. Further a tool, called SAVE, has been developed to demonstrate the feasibility of the approach with the algebra.

In the paper, the process algebra for specification with temporal properties is presented. In the future research, the different types of verification methods are developed to demonstrate the usability of dT-Calculus, based on a logic system, including SAVE with a verification model.

## Acknowledgements

This work was supported by Basic Science Research Programs through the National Research Foundation of Korea (NRF) funded by the Ministry of Education (2010-0023787), and Space Core Technology Development Program through the NRF (National Research Foundation of Korea) funded by the Ministry of Science, ICT and Future Planning (NRF-2014M1A3A3A020 34792), and Basic Science Research Program through the National Research Foundation of Korea (NRF) funded by the Ministry of Education (NRF-2015R1D1A3A01019282).

## Author details

Sunghyeon Lee, Yeongbok Choe and Moonkun Lee*

*Address all correspondence to: moonkun@jbnu.ac.kr

Chonbuk National University, Jeonju-si Jeonbuk, Republic of Korea

## References

[1] Chen C-Y, Hasan M, Mohan S. Securing Real-Time Internet-of-Things. arXiv preprint arXiv. 2017;**1705**:08489

[2] Choe Y, Lee M. Algebraic method to model secure IoT. In: Karagiannis D, Mayr HC, Mylopoulos J, editors. Domain-Specific Conceptual Modeling. Switzerland: Springer International Publishing; 2016. pp. 335-355. Ch. 15

[3] Saeedloei N, Gupta G. Timed π-Calculus. Trustworthy Global Computing. Lecture Notes in Computer Science. Vol. 8358. Cham: Springer. 2014

[4] Milner R, Parrow J, Walker D. A calculus of mobile processes (i–ii). Information and Computation. 1992:1-77

[5] Aman B, Ciobanu G. Mobile ambients with timer and types. In: International Colloquium on Theoretical Aspects of Computing. Berlin Heidelberg: Springer; 2007

[6] Cardelli L, Gordon A. Mobile ambients. In: Nivat M, editors. ETAPS 1998 and FOSSACS 1998. LNCS. Vol. 1378. Heidelberg: Springer; 1998. pp. 140-155

[7] Choe Y, Lee M. δ-Calculus: Process algebra to model secure movements of distributed mobile processes in real-time business application. In: 23rd European Conference on Information Systems; 2015

[8] Choe Y, Choi W, Jeon G, Lee M. A tool for visual specification and verification for secure process movements. In: eChallenges e-2015; 2015

[9] SAVE tool. Available from: http://austria.omilab.org/psm/content/save/info

[10] Snoonian D. Smart buildings. IEEE Spectrum. 2003;**40**(8):18-23

[11] Chung S-m, Choi J-h, Park J-w. Design of software quality evaluation model for IoT. Journal of the Korea Institute of Information and Communication Engineering. 2016;**20**(7):1342-1354

[12] Liu Y, Zhou G. Key technologies and applications of internet of things. 2012 Fifth International Conference on Intelligent Computation Technology and Automation (ICICTA); IEEE; 2012

[13] Sarkar C, et al. A scalable distributed architecture towards unifying IoT applications. 2014 IEEE World Forum on Internet of Things (WF-IoT); IEEE; 2014

[14] Choe Y, Lee S, Lee M. SAVE: An environment for visual specification and verification of IoT. 2016 IEEE 20th International on Enterprise Distributed Object Computing Workshop (EDOCW); IEEE; 2016

[15] Fill H, Karagiannis D. On the conceptualisation of modeling methods using the ADOxx meta modeling platform. Enterprise Modeling and Information Systems Architectures. 2013;**8**(1):4-25

# An Adaptive Lightweight Security Framework Suited for IoT

Menachem Domb

Additional information is available at the end of the chapter

**Abstract**

Standard security systems are widely implemented in the industry. These systems consume considerable computational resources. Devices in the Internet of Things [IoT] are very limited with processing capacity, memory and storage. Therefore, existing security systems are not applicable for IoT. To cope with it, we propose downsizing of existing security processes. In this chapter, we describe three areas, where we reduce the required storage space and processing power. The first is the classification process required for ongoing anomaly detection, whereby values accepted or generated by a sensor are classified as valid or abnormal. We collect historic data and analyze it using machine learning techniques to draw a contour, where all streaming values are expected to fall within the contour space. Hence, the detailed collected data from the sensors are no longer required for real-time anomaly detection. The second area involves the implementation of the Random Forest algorithm to apply distributed and parallel processing for anomaly discovery. The third area is downsizing cryptography calculations, to fit IoT limitations without compromising security. For each area, we present experimental results supporting our approach and implementation.

**Keywords:** IoT, anomaly detection, entropy, machine learning, random forest, cryptography, RSA

## 1. Introduction

The area of the Internet of Things [IoT] is rapidly growing, raising severe security concerns to the entire network. Due to its high traffic volume and real-time operation, a security framework is essential. The system should timely predict possible attacks and react accordingly. Standard security systems are widely implemented in the industry. These systems consume considerable computational resources and cannot operate in IoT devices (i.e., sensors) due to their very limited memory and computation power. To cope with these limitations, two alternatives come to mind, i.e., the development of novel security measures tailored to IoT [1] or downsizing existing security processes to enable properly operation in IoT devices. We apply the latter option as it is highly recommended to use proven algorithms, which have been extensively analyzed and tested, while new algorithms exposes the user to vulnerability.

We introduce lightweight versions of several known security processes. We analyze each relevant process and its corresponding limitations, and then we divide each complex and large process into a collection of smaller processes. These small processes are distributed and executed by sensors connected to the same network, based on its available capacity. Once all small processes are completed, we collect the partial results and input them into a complementary process that integrates the partial results to compose the desired result. The final result is the same as if the original process was generated. In this chapter, we describe three areas, where we minimize the required storage space and processing power. The first is the classification process required for ongoing anomaly detection, whereby values accepted or generated by a sensor are classified as valid or abnormal. We collect historic data and analyze it using machine learning techniques to draw a contour, and all streaming values are expected to fall within the contour space. The detailed collected data are no longer required, thereby considerably reducing the storage space. The second area involves the implementation of the Random Forest algorithm to apply distributed and parallel processing for anomaly discovery, resulting in the use of limited processing power. The third area is downsizing cryptography calculations, such as RSA, a public-key cryptosystem, to fit IoT limitations. The rest of this chapter is divided into three sections, one dedicated to each downsized area. In the last section, we conclude this chapter.

The rest of this chapter is organized as follows: In Section 2, we describe the preparation stage of the classification process, which minimizes the need for the entire historic data and then the anomaly detection processes using the outcome of the previous stage. In Section 3, we describe the use of the Random Forest algorithm for distributed and parallel processing of automatic classification and anomaly detection. In Section 4, we present an improved implementation of RSA to allow high class cryptography that runs in an IoT configuration. In Section 5, we conclude this chapter and discuss our ongoing and future work.

## 2. Classification framework for data streaming anomaly detection

To predict the behavior of a system, we usually examine its past data to discover common patterns and other classification issues. This process consumes considerable computational power and data storage. In this section, we describe an approach and a system, which requires much less resources without compromising prediction capabilities and accuracy. It employs three basic methods: a common behavior graph, the contour surrounding the graph, and entropy calculation methods. When the system is about to be implemented for a specific domain, the optimized combination of these three methods is considered, such that it fits the unique nature of the domain and its corresponding type of data. In addition, we present a framework and a process that will assist system designers in finding the optimal methods for the case at hand. We use a case study to demonstrate this approach with meteorological data collected over 15 years to classify and detect anomalies in new data.

This section is organized as follows: We begin by defining the problem, proceed with various solutions proposed in the literature, and then present our adjustable contour approach. We then show how it is applicable for IoT. We proceed with a case study demonstrating the build-up of the contour and how it is used for instant anomaly detection. We conclude with a summary of the section.

## 2.1. Problem definition

The problem we attempt to solve is the optimization of the amount of sampling data collected to maintain a proper balance between the quantity of sampling data and the information extracted from it. The problem statement focuses on extracting concepts, methods, rules, and measurements, so that at the end of the process, the original sampling data become redundant and no longer need to be stored. However, to keep improving and adjusting the extracted items to natural changes in the behavior of the sampled mechanism, we incorporate in the approach an ongoing learning process. In addition, in the study, we concentrate on time-dependent streaming sampling data, divided by fixed periods, so that we can repeat the analysis process for each period/cycle. Thus, while there are many classification algorithms using time series sampling, the aim is not to compare the performance of yet another classifier, but rather present a flexible method to compactly represent the data with several parameters that can be chosen and adjusted. We suggest an independent framework that allows a flexible adaptation of the contour to the nature of the given domain. Indeed, some of the reviewed works, such as Reeves et al. [6], can be revised and adjusted to the problem statement and serve as a valid alternative to the approach we present. We are striving for the best sampling strategy given sequential data, generated from IoT devices.

The input given is a set of time series: $D = \{d^{(1)}, d^{(2)}, \ldots, d^{(n)}\}$, where each time series $d^{(i)}$ contains pairs (timestamp and numeric value). The required output is an optimal set $Dw = \{a_1, a_2, \ldots, a_m\}$, where $ai$ can be any sampling item, such as a minimal data set, trends, graphs, measurements, or rules, which strongly represents and supports the purpose of the original data set $D$.

We consider the set $Dw$ and the full data set $D$ as containing the same information, if they produce the same classifier. That is, if $f(d) = fw(d) \in \{-1, 1\}$ for every new data series $d$, where $f$ is a classifier learned from $D$ and $fw$ is a classifier based on $Dw$. For instance, we can judge whether a series of yearly temperatures represent an El Nino (EN) year or not, or whether a series of sensor data is characteristic of a suspected intrusion or not. Here, we consider two sets $D$ and $Dw$ as containing the same (or similar) information if both can predict the future pattern of an initial series $d$. That is, we can use either $D$ or $Dw$ to predict a future item $dn$ with similar accuracy.

## 2.2. Literature review

Real-world data typically contain repeated and periodic patterns. This suggests that the data can be effectively represented and compressed using only a few coefficients of an appropriate basis. Mairal et al. [2] study modeling data vectors as sparse linear combinations of basic elements generating a generic dictionary and then adapt it to specific data. Jankov et al. [3] present an implementation of a real-time anomaly detection system over data streams and report experimental results and performance tuning strategies. Vlachos et al. [4] formulate the problem of estimating lower/upper distance bounds as an optimization problem and establish the properties of optimal solutions to develop an algorithm which obtains an exact solution to the problem. Sakurada and Yairi [5] use auto-encoders with nonlinear dimensionality reduction for the anomaly detection task. They demonstrate the ability to detect subtle anomalies where linear PCA fails. Reeves et al. [6] present a multi-scale analysis to decompose time series and to obtain sparse representations in various domains. Chilimbi and Hirzel [7] implement a dynamic pre-fetching scheme that operates in several phases. The first is profiling, which gathers a temporal data reference profile from a running program. Next, an algorithm extracts hot data streams, which are data reference sequences that frequently repeat in the same order. Then, a code is dynamically injected into appropriate program points to detect and pre-fetch the hot data streams. Finally, the process enters the hibernation phase where the program continues to execute with the added pre-fetch instructions. At the end, the program is deoptimized to remove the inserted checks and pre-fetch instructions and control returns to the profiling phase. Lane and Brodley [8] claim that features can be extracted from object behavior and a domain heuristic. Experiments show that it detects anomalous conditions, and it is able to identify a profiled user from other users. They present several techniques for reducing 70% of the storage required for user profile. Kasiviswanathan et al. [9] proposed a two-stage approach based on detection and clustering of novel user-generated content to derive a scalable approach by using the alternating directions method to solve the resulting optimization problems. Aldroubi et al. [10] show that for each dataset there is an optimized collection of cells spanning the entire space and so generate the optimized sampling set.

The common underlying idea of the reviewed approaches is the definition of the problem they are aiming to solve. The problem attempted to be solved is optimizing the size of the collected sampling data so that it keeps the proper balance between the quantity of sampling data and the information extracted from it.

## 2.3. Contour-based approach

Briefly, we analyze sampling data collected over several periods. We divide the period into time-units. For example, for a period of a year, we divide it into daily time-units. For each time-unit, we extract one value that represents it. This is done by averaging the samples collected during the time-unit. In the example, we may calculate the average value of all samples of that day. We may also decide to select one of the samples to represent the day, e.g., the first or last sample. We then calculate the average value for each time-unit from the collected values for the same time-unit in all periods, resulting in an average value for a given time-unit. We repeat this process for all time-units in the period and obtain a graph that represents the average values for an average and common period.

Assuming we have the average graph line for an average period, we now calculate the contour around this average. The generated contour represents the standard range of values, such that an unanalyzed period can be compared to this contour. If its graph value is completely within the contour, the period is a standard period. If it is completely out of the contour, then it is purely not standard. If the sections of the graph are within the contour, while others are out of it, we use an entropy measure to calculate the overall "distance" of the given period from the standard contour. Assuming an existing entropy threshold, we can decide whether the period is a standard one or not. We apply the same concept at the unit level and decide whether a specific time-unit in a period is within the standard or not. This specific check is relevant, for example, to anomaly detection of IoT behavior.

In conclusion, the entire process is based on three key elements: the average graph per period, the contour around the average graph, and an entropy value representing the overall distance of a period from the contour. Each of these elements—average, contour, and entropy—can be one of the several possibilities. For the contour, a simplistic choice would be minimum and maximum (min-max) values. Alternatively, the *SD* or confidence interval (CI) could be employed. These three elements affect each other, and every choice of such a triplet—average, contour, and entropy—will produce a different behavior of the compressed classifier. The object is to find the best triplet that will be able to disregard the original data after extracting the representative contour, without compromising the ability to successfully analyze future series. In our work, we consistently use the arithmetic average and classical entropy and focus on finding the best contour.

### 2.3.1. Finding the optimal contour

We begin with a supervised learning approach, for classification, in which each time series is labeled as one of two classes. To demonstrate, using the data set from the experiments, the time series are year-long recordings of temperature samplings, labeled as positive, if the corresponding year was an EN year, or otherwise negative. We now describe in detail the process of building the classifier, with emphasis on finding the optimal contour.

Constructing the best contour is described in **Figure 1**. We begin with raw data collected during $N$ periods, where each record corresponds to a specific time-unit. These cycles have already been classified positive or negative according to some classification criteria. These classified cycles will later be used to determine the best contour.

The process is divided into four stages. In stage one, we use a selected average method and calculate the average graph line representing the $N$ given cycles. This is done horizontally by calculating the average of the values related to the same time-unit across all $N$ cycles. For example, we calculate the average of the values for January 1st across the various years. Doing so for all time-units will generate the average graph line. In stage two, we select several distance calculation methods, and for each method, we construct its associated contour. This is done by calculating the distance value for each distance method, e.g., the min-max difference, *SD*, and CI. Taking the distance value, we add and subtract it from the average line to get the contour around the average. We repeat this process for all distance methods. At this stage, we have constructed several contours around the average line. The goal now is to select the contour, which is most effective in classifying unclassified cycles. This is done in stages three and four. In stage three, we calculate the prediction power for each contour and select the one with the highest prediction power. This is done by summing, for each contour, the number of cases in which its prediction was right and calculating the average entropy of these correctly classified cycles. We do the same for wrong predictions. In stage four, we use one entropy method with an associated threshold value. An unclassified cycle with an entropy value lower than the threshold will be classified positive and otherwise negative. For each contour, we calculate the entropy of the given classified cycles. The result is a set of entropy values, where some are below the threshold and others are above it.

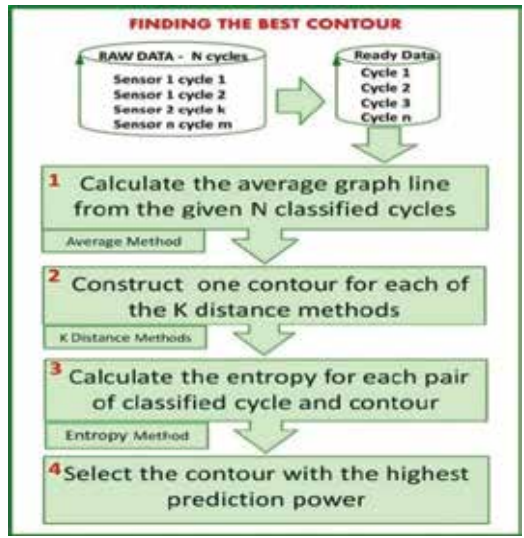

**Figure 1.** Process of finding the optimal contour.

**a.** We repeat this for all classified cycles. We then sum up the number of correct predictions and their total entropies. We do the same for wrong predictions. We then subtract the total wrong numbers from the correct numbers. We repeat this process for all the constructed contours and select the contour with the highest prediction power.

**b.** Calculating the entropy.

The entropy of a period, given a contour, is calculated as follows:

- Marking for every timestamp whether the cycle's value at that timestamp is below, within, or above the contour.

- Calculating the frequency of each of these three possibilities: below ($p_1$), within ($p_2$), and above ($p_3$)

- Using these as a ternary probability distribution, its entropy is calculated according to the formula: $p_1 \log^{(p1)} + p_2 \log^{(p2)} + p_3 \log^{(p3)}$

- The entropy measure is expected to return its minimum value at the two extreme cases: When the cycle graph is entirely contained within the contour and when the cycle graph lies entirely outside of the contour. All other cycles are expected to fall mostly within the contour, and those which diverge enough from the contour, will have a high entropy value which will lead to the right conclusion **c.** Classifying a cycle/period

**Figure 2** describes the process of classifying unlabeled data cycles, as listed below:

**1.** Apply the given data cycle to the contour and match it according to timestamps.

**2.** Noting for each timestamp whether the data point is below the contour, within it, or above it.

**3.** Marking these cases respectively as −1, 0, and +1.

**4.** Calculating the frequencies of each of the three values: −1 ($p1$), 0 ($p2$), and +1($p3$).

**5.** Calculating the entropy of the distribution defined by $p1$, $p2$, and $p3$.

**6.** Classifying as belonging to the contour, if the entropy is below the threshold determined in the learning phase.

*2.3.2. Advantages of the proposed technique*

The proposed technique has several advantages over other methods. The technique is a family of sampling methods and is defined by the three parameters described above. It is reasonable to expect that different datasets will require different parameters for the best sampling. Different combinations can be tested and evaluated to ensure optimal treatment of the data. The technique we propose is therefore flexible and adjustable and thus suits every given data set. Secondly, this technique can be applied not only for classification but also for prediction of time series.

Thirdly, the technique can be used to evaluate reliability of data online. In cases of high fluctuations or sharp changes in the cycle graph, which do not conform to either of the two class contours, suspicion may arise that the reliability of the data has been compromised. This can indicate that the sensor is damaged or that there has been a security breach.

Fourthly, the approach allows self-learning and automatic adjustments in cases of common behavior changes and a new standard has been established. Lastly, occasionally, a post-mortem may be run to check the system's reaction to actual behavior and thereafter adjust the system parameters accordingly.

## 2.4. Anomaly detection for IoT security

IoT devices generate time-related data, i.e., structured records containing a timestamp and one or more numeric values. In many cases, we can identify recurrent time frames where the system behavior has a repetitive format. Hence, IoT data have a structure to which the contour approach is highly applicable.

IoT security utilizes common data patterns and quantitative measurements. Based on the identified patterns and measurements, we can extract logical rules that will be executed once an exception is discovered. An exception may be any violation of predefined patterns, measurements, and other parameters, which represent normal, standard, and permitted behavior.

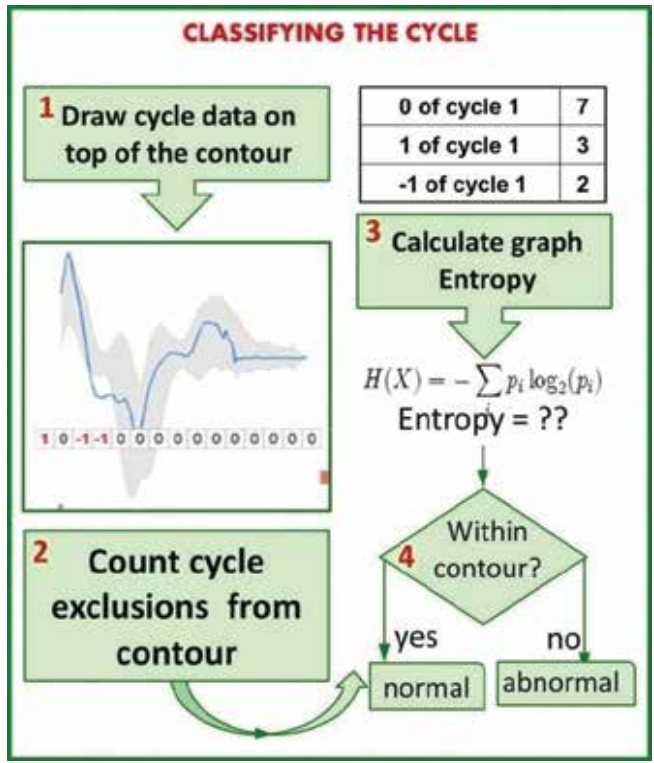

**Figure 2.** Classifying a cycle.

In IoT, there is an abundance of possible patterns, starting with column level patterns up to a super internet controlling several IoT networks. The goal is to find the methods and tools to define standard patterns and how they can be identified. Once this is done, we can apply the contour method. In our work, we show a two-dimensional contour. Using the same concept, we can expand it to be a multi-dimensional contour. This case is common where there is a dependency among several columns within one record and the same applies for the case where there are dependencies among networks of IoT systems.

### 2.5. Case study

In the following case study, we used meteorological data collected on EN years (positive class) and NEN years (negative class) from 1980 to 1998. For the positive contours, we took data from the EN years 1982, 1983, 1987, 1988, 1991, and 1992. All other years in the range were NEN years. We tested three methods for generating contours: (a) max-min over all cycles; (b) average cycle $\pm SD$; and (c) CI.

**Figures 3** and **4** depict the contours for NEN years. **Figure 3** shows the NEN contour in black according to the average $\pm SD$ and depicts how EN years diverge from this contour, as compared to the NEN year—1995. The 1992 and 1988 (EN years) show clear divergence from the contour while 1995 (a NEN) is more contained within the contour. This is nicely captured by the entropy values, which for 1992 was 0.4266 and for 1988 was 0.3857—above the threshold, leading to the conclusion that they are not NEN years—while for 1995, the entropy was 0.3631—significantly lower than those of the EN years, leading to the correct conclusion that 1995 was indeed a NEN year.

**Figure 4** shows two contours: the min-max contour and the average ± *SD* contour. The *Y*-axis in these graphs is the temperature value, and the *X*-axis is the time. Within each contour, the year 1995 (a NEN year) is graphed. Its entropy is 0.3631 for the average *SD* contour and 0.2932 for the min-max contour. Both are the threshold, which leads to the correct conclusion that it should indeed be classified as NEN.

In the case study, we compared the constructed contours, by using the average graph ± *SD* and the average graph ± min-max. For the *SD* contour, we obtained a significant entropy value difference between a classified EN case and a NEN case. In comparison, the min-max contour resulted in close values of entropy for the EN cycle and the NEN cycle. Thus, the ability to differentiate between two extreme situations using entropy depends on the parameter used to build the contour.

### 2.6. Section summary

In this section, we dealt with the classification problem of an unclassified cycle of IoT streaming data. We introduced the contour approach to draw the borders around the standard area representing a specific class. If there was an unclassified cycle, we measured its distance from the contour using an entropy formula. Then, we compared the result to a predefined threshold. If the entropy value is below the threshold, the cycle is of the same class.

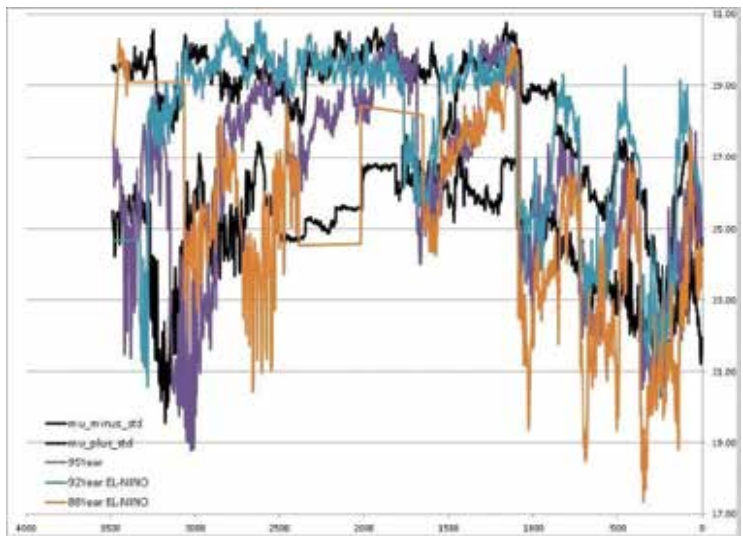

**Figure 3.** EN cycles on NEN average ± *SD* contour.

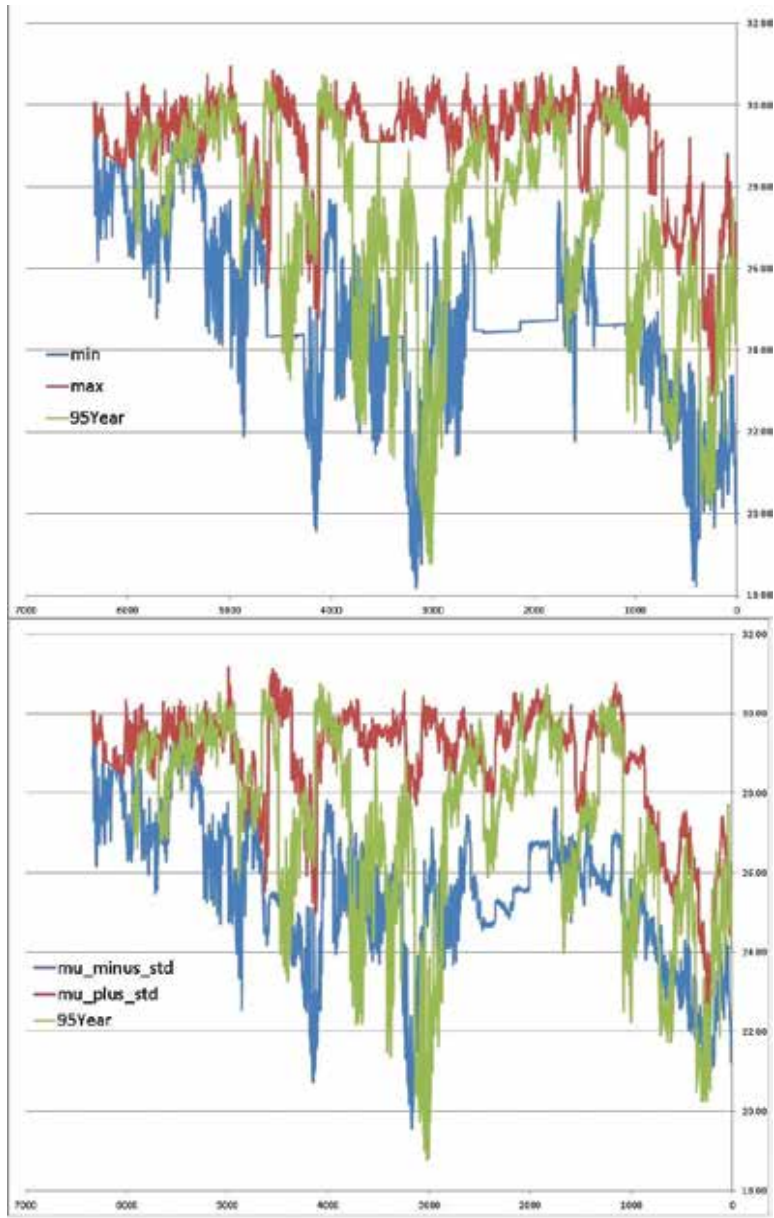

**Figure 4.** NEN contours—min-max and *SD*.

We propose a process for constructing the best contour that will presumably classify the correct underlying class. The process is based on three measurement methods: average, distance, and entropy. For each method, there are several alternate formulas that we may use. Each combination of these three methods may result in different contour and hence different entropy value for the same unclassified cycle. We select the combination with the maximum difference between positive and negative values.

In addition to the initial construction of the class contours from the given data, we suggest ongoing improvements of the initial contours. Namely, we recalculate the class averages and their contours to refine and revise the contours for improved classification performance.

In this manner, we are able to improve the contour approach, in reference to several aspects, such as determining the minimal number of classified cycles required to define the best contour, expanding the use of the contour to discover early trends or discover significant changes in behavior and adjusting the contour accordingly, exploring the possibility of dividing one cycle into several segments, and associating a different contour method to each segment.

## 3. Lightweight adaptive random forest for rule generation and execution

The volume of transmitted data over the various sensors continuously grows. Sensors typically are low in resources of storage, memory, and processing power. Data security and privacy are part of the major concerns and drawbacks of this growing domain. An IoT network intrusion detection system is required to monitor and analyze the traffic and predict possible attacks. Machine leaning techniques can automatically extract normal and abnormal patterns from a large set of training sensors data. Due to the high volume of traffic and the need for real-time reaction, accurate threat discovery is mandatory. This section focuses on designing a lightweight comprehensive IoT rules generation and execution framework. It is composed of three components, a machine learning rule discovery, a threat prediction model builder and tools to ensure timely reaction to rules violation and unstandardized and ongoing changes in traffic behavior. The generated detection model is expected to identify exceptions in real time and notify the system accordingly.

We use random forest (RF) as the machine learning platform for the discovery of rules and real-time anomaly detection. To allow RF adaptation for IoT, we propose several improvements to make it lightweight and propose a process that combines IoT network capabilities, messaging and resource sharing, to build a comprehensive and efficient IoT security framework.

The rest of this section is organized as follows: We begin with an introduction followed by the relevant literature review. We then discuss rules extraction using machine learning techniques. We present random forest as the most suitable ML for IoT. We proceed with various improvements, utilizing RF and IoT attributes. We then outline an experiment that executes RF building and its corresponding classifications using 15 different configurations, each based on a unique combination of the number of processors and the forest size.

## 3.1. Introduction

IoT is a network of objects, consisting of sensors, Internet, software, and exchange of data. This generates critical issues of security, which must be addressed. Since to date there is no standard for sensors, any system under development at this stage must consider the possibility that soon a standard will be defined, and the systems must be able to easily adjust to it. Along with the limited processing power and the fact that the security issues must be dealt with in real time, we realize the immediate need for a flexible and lightweight solution. The solution should be dynamic, open, scalable, distributed and decentralized. The analysis discovers patterns and measurements from the data, which are then translated into anomaly detection rules associated with actions to be executed when a rule is violated. The rules are then deployed in the IoT devices. When data are received from, or transmitted to an IoT device, the rules are executed. If the result is positive, the corresponding action is triggered to cope with the situation.

## 3.2. Literature review

Mansoori et al. [11] proposed a systematic process for retrieving fuzzy rules from a given data set. To improve performance, the retrieved rules are then crystallized based on its effectiveness and applicability. Dubois et al. [12] use Sugeno integrals, which are qualitative criteria aggregations where it is possible to assign weights to groups of criteria. They show how to extract if-then rules that express the selection of situations based on local evaluations and rules to detect bad situations. Sumit-Gulwani, Hart, and Zorn [13] deal with converting data into an appropriate layout, which requires major investment in manual reformatting. The paper introduces a synthesis engine to extract structured relational data. It uses examples to synthesize a program using an extraction language. Bharathidason et al. [23] presented a fast and compact decision rules algorithm. The algorithm works online to learn rule sets. It presents a technique to detect local drifts by taking advantage of the modularity of the rule sets. Each rule monitors the evolution of performance metrics to detect a concept drift. It provides useful information about the dynamics of the process generating data, faster adaptation to changes, and generates more compact rule sets. Jafarzadeh et al. [15] used averaging techniques to propose a method in which a previous algorithm for association rules mining is improved upon to specify minimum support. It uses fuzzy logic to distribute data in different clusters and then tries to provide the user with the most appropriate threshold automatically. Limb et al. [16] used Fuzzy ARTMAP and Q learning to build a data classification and rule mining model. To justify the classification, the model provides a fuzzy conditional rule. Q-values are used to minimize QFAM prototyping. Mashinchi et al. [17] proposed a granular-rules extraction method to simplify a data set into a granular-rule set with unique granular rules. It performs in two stages to construct and prune the granular rules. Yang H. et al [18] proposed an anomaly detection algorithm of Quick Access Recorder (QAR) data, based on attribute support of a rough set. The method retains the time characteristics of QAR data and strengthens the relation between the condition and decision attributes. Tang [19] described an approach of data mining with Excel, using the XLMiner add-in. This is an example of mining association rules to illustrate all the steps of this approach. Tong S and Koller D. [20] introduced an algorithm for choosing which instances to request next, in a setting in which the learner has access to a pool of unlabeled instances and can request the labels for some number of them. The algorithm is based on a theoretical motivation for using support vector machines (SVMs). Osungi et al. [21] proposed an active learning algorithm that balances exploration by dynamically adjusting the probability to explore each step. Lang T et al. [22] proposed an active learning method for multi-class classification. The method selects informative training compounds to optimally support the learning progress. Bharathidason et al. [23] improved the performance and the accuracy by including only uncorrelated high performing trees in a random forest.

The reviewed literature focuses on improvements to known rule discovery mechanisms, such as machine learning, to transform them into lightweight systems that can be executed in limited resources settings. In most cases, the proposed solutions remain for general purposes but can run with less required resources. We are seeking a solution that takes advantage of the unique IoT attributes and utilizes them to build a combined comprehensive framework for IoT security.

### 3.3. Rules generation and deployment process

The process consists of seven stages (see **Figure 1**). Stage 1 collects training data from the IoT network, removes irrelevant records, and complements data in records with missing data. In stage 2, we apply discovery techniques to extract important measurements and patterns. Stage 3 consists of generating a rule for each measurement and pattern. In stage 4, we evaluate the effectiveness of each rule with a set of training data. If the number of times a rule has been executed is below a given threshold, the rule is removed from the rules set. Next, in stage 5, we check the completeness and the integrity of the generated set of rules. Rules that contradict another rule are removed and missing rules are added. Stage 6 runs a simulation with the same training data with the presumption that all the designated rules will be executed. Finally, in stage 7, we deploy the generated rules set. At this point, the system is ready to accept the IoT traffic data in real time and automatically check it against the set of rules.

### 3.4. Extracting rules from training data

A typical sensor record contains the sensor ID, timestamp, and one or more values per feature. The main source for extracting rules is data collected from the concrete processes involved in the explored domain. The significance to IoT is taking the accurate decision in real time and react in real time to security alerts, notifications, automation, and predictive maintenance. To ensure the completeness and the integrity of the generated set of rules, we use a consistent multi-layer process of accumulating rules, starting with the simplest rules up to the most complicated and multi-stage rules. Simple rules are extracted at the single feature level, and then we proceed with rules extracted from a combination of any number of features having a common relation, such as features of sensors sharing the same workflow. The generated rules at this level relate to basic data such as maximum, minimum, average, standard deviation, median, and most frequent value. More complex relations, such as proportions among subsequent values, sequence trends, and significant patterns, require reasoning capabilities and can be reached by machine learning and data mining techniques. The outcomes are measurements, thresholds, and patterns used to draw the corresponding decision trees. These decision trees tend to grow fast, consuming large storage, and memory space along with high runtime when pruning and analyzing it to find the specific rule. The depth of the tree grows linearly with the number of variables, but the number of branches grows exponentially with the number of states. Decision trees are useful when the number of states per variable is limited. It becomes complicated when the state of the variables depends on a threshold or complex computations. Communicating this rationale requires labeling every edge and then tracing the tree path to understand the logic incorporated in it. Complex event processing (CEP) engines are popular in IoT. They support matching time series data patterns that originate from different sources. However, they suffer from the same modeling issues as trees and pipeline processing.

Rule engines have two major drawbacks in the context of IoT, the logic representation is not compact and the use of it requires much processing power and time. We will cope with these drawbacks in two ways. 1. Reduce the number of decision trees and improve the search navigation scope, resulting in a reasonable and acceptable search time. 2. Utilize IoT attributes and functionality to optimize the tree navigation flow and process sharing.

In the following sections, we present the random forest machine learning and propose several improvements where the known drawbacks are removed.

### 3.5. Decision automation using random forest

Random forest employs bootstrap aggregation for training. While the predictions of a single tree are sensitive to noise in its training set, the average of many uncorrelated trees is not. Bootstrap sampling is a way of decorrelating the trees by showing them different training sets. Many trees reduce the depth and width of each tree and so save pruning and analysis time, which suit IoT constraints.

The algorithm has two key parameters: the number of K trees to form a random forest and the number of features F, randomly sampled features for building a decision tree. For large and high dimensional data, a large K should be used. Estimating the performance of random forest for one core is based on the following parameters: # trees [K], # features [F], # rows [R], and maximum depth [D]. The estimated runtime is influenced by the number of features. Hence, keeping only the most important features lowers the number of records and maintains the maximum depth low, which will improve the overall random forest performance.

Random forest performance is better than the classical tree decision algorithm. However, it may still be insufficient for IoT due to the memory space and processing power it requires. Hence, building a lightweight RF process and utilizing IoT networking are required.

In the following section, we describe four proposals that make random-forest lightweight.

### 3.6. Improving RF performance and consumption of resources

**a.** Randomization may cause occurrence of redundant, irrelevant or even contradicting trees, which may lead to redundant searches or even to the wrong decision. Therefore, selection of trees with high classification accuracies leads to improved performance and better decision accuracy. A decision process is effective when the difference among the relevant alternatives is significant. RF contains many decision trees, where each of them may contribute to the final decision. Many such trees generally require wider searches and thus expand the decision process. On the one hand, reducing the number of the searched trees will shorten the process but on the other hand may increase the probability of making the wrong decision. Therefore, a selection criterion for removing the "redundant" trees is required. An initial approach is to remove similar trees as correlated trees hardly contribute to reaching the correct decision. Thus, for effective RF decisions, we strive to remove uncorrelated trees [14]. The correlation between two trees may be defined in various ways, such as:

1. Distance—we transform the tree into a sequence of values, and then we apply a hashing function on this sequence and get a score. Two trees are correlated if the difference between the scores is below a predefined threshold.

2. Common components—count the number of similar components and compare.

3. Empirically by removing the tree and trying a vast number of cases, we will reach the same decisions as we would if the tree was included, which means that the tree has no effect on practical decisions.

**b.** Prioritize trees by simulation using labeled and already classified cases.

Instead of removing trees, we propose prioritizing them. The prioritization can be an empirical study of the historical use and effectiveness in true/false decisions. Another way is to run a Monte-Carlo intensive simulation and prioritize trees accordingly.

### 3.7. Prioritize trees by its threat level

We define several security levels: low, normal, high, and emergency. For each level, we associate the most effective trees and the order of the trees to be visited. For each network, we designate a security manager device, which collects messages from its network devices, assesses it, and determines the network security level. When the network is initiated, the designated level is low. As time passes, messages arrive at the security manager device, which analyzes the input and decides to change the security level. Then, a message is distributed requesting a security level change. Once the level is changed, the local system activates the new tree search schedule.

### 3.8. Messaging assisted, best trees selection

MQTT is a lightweight messaging protocol, over TCP, adjusted to the IoT domain. Given MQTT, we can utilize the IoT network itself to improve performance. We use it to transfer messages and data from one device to another. For example, in case of a suspicious occasion detected by one of the sensors, using the protocol, the device sends alert messages to other members. The messages include data strings and unique data patterns that receivers should expect to receive and thus detect a malicious situation. The message may also include the most effective trees that may cope with the suspected threat.

When suspicious data reach a sensor, it is analyzed locally, and the best tree sequence is identified. This device sends a message to the security manager, containing the data with the detected anomaly and the sequence of trees to visit and act accordingly. The messaging protocol is an adjustment of HTTP.

### 3.9. Experiment using the random forest in an IoT

In this section, we describe a comprehensive test, simulating the building of various random forests and then runs several classification cycles for a given set of anonymous records. We used a computer with eight processors running the random forest PMI platform with 10–1000 trees per forest. It contained a random forest builder, an anonymous records classification process, and a configuration tool. We sought the best configuration, suitable for the optimized performance and accuracy of a random forest simulation. A configuration in this context is measured by the combination of the number of processors and the number of trees in a forest. For the simulation, we used 500 anonymous records and 3350 already classified samples, where each sample has 95 attributes. We ran 30 test cycles where each cycle represented a unique configuration—number of processors: 2, 4, 6, 8, and 16 and the number of trees per forest: 10, 100, 250, 500, 750, and 1000. For comparison, all test cycles used the same data set. In cases of similar trees, we ran a process that removes similar trees. The performance of the entire 30 test cycles is evaluated by its accuracy and processing time.

**Figures 5** and **6** show that accuracy, performance of each of the processes and combined are best achieved when using 10 trees per forest and 8 processors. Based on the above simulations, it seems that for the example at hand, using a relatively small number of trees per forest and multi-core processors is recommended for optimal performance and high accuracy. However, this may not be the common case. Therefore, prior to implementing RF-based anomaly detection, it is recommended that a simulation test be run with the main data. In addition, we propose a prototype of an IoT environment. The prototype is composed of one server and six Arduino OS devices. We built two configurations, A and B. In configuration A, all the devices are connected via WIFI 14 to the server, where the data transmission between two devices is done through the server. The entire RF is loaded in the server while the devices have one tree installed in them. The data flow of an incoming event in configuration A can be one of the following: 1. An event arrives at a device, the device forwards it to the server, which then runs the RF and classifies the event. 2. An event arrives at a device and the device forwards it to the server. The server forwards it to all devices. Each device checks the event against the appropriate local tree and sends the result to the server. The server then counts the results and sends the reply to the sender, which acts accordingly. The flow in configuration B is as follows: An event arrives at a device, the device propagates it to other devices, checks it against its own tree, and propagates the results back to the sender. The sender classifies the event and acts accordingly. To test the feasibility of the prototype, we used the trees built by the simulation tool and loaded it to the server and devices accordingly. We transmitted 500 events to the devices in a round robin schedule. The resulting accuracy level was similar to the level we found in the previous simulation. Performance was out of the scope of the prototype stage. Nonetheless, we did not notice streaming interruptions or delays. In future work, we intend to design and perform consistent and comprehensive tests of the device and other similar devices. Based on the results, we will be better able to determine which rules are to be executed in real time and which are to be executed online or in batch mode.

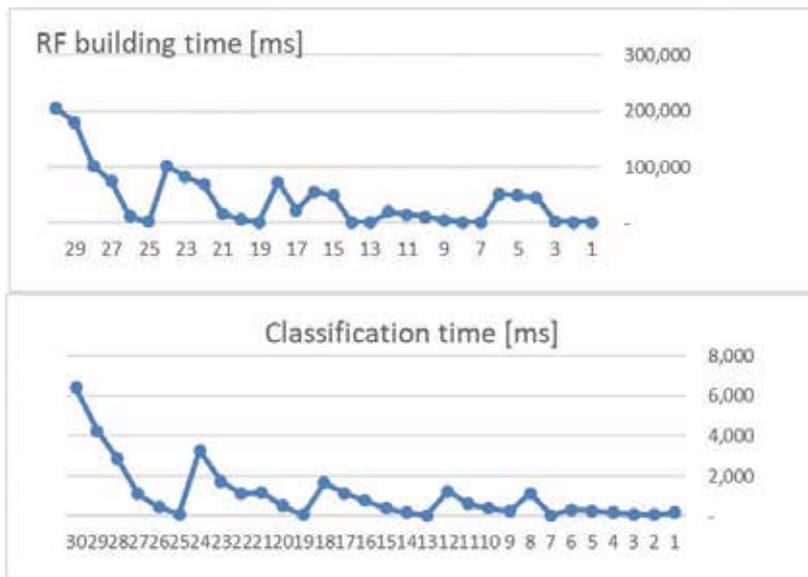

**Figure 5.** Results of running the 30 classification processes.

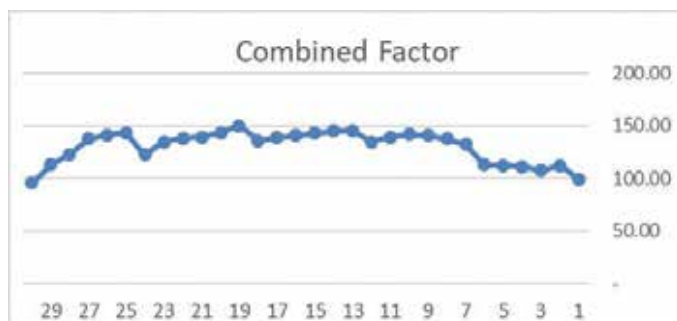

**Figure 6.** Accuracy and combined results of running the 30 classification processes.

## 4. Lightweight public key cryptographic processor suited for IoT

Due to the vast number of IoT devices and high transmission volumes, a robust and adaptive cryptography system is required. However, since IoT devices have limited memory and computation power, they are unable to execute public key cryptographic systems. To cope with this limitation, we propose a lightweight RSA process. A combination of symmetric and asymmetric encryption systems is commonly used by the industry. Symmetric encryption systems require moderate computation resources and consequently are already used in IOT. However, asymmetric public key encryption requires vast computation resources, and as a result cannot be executed by most IOT devices. In this section, we describe a lightweight RSA encryption, where three improvements are incorporated: acceleration of modular exponentiation calculation, parallel and distributed multi-core processing, and splitting the original message if the message length is very long. After each part is completed, the system collects the intermediate results and loads them into a consolidation and integration process, which generates the result. We ran comprehensive encryption and decryption processes on messages of various lengths. The results prove that lightweight RSA is ready to be incorporated in IoT devices.

The rest of this section outlines the relevant literature review. Then, we describe an example of smart modular exponential calculation, which runs efficiently in an IoT architecture.

### 4.1. Literature review

Lin et al. [24] proposed the execution in parallel on CPU/GPU hybrids, of the Montgomery algorithm, to improve RSA performance and security. Fadhil and Younis [25] proposed a hybrid system, running RSA on multi-core CPU and multi GPU cores. For comparison purposes, they implemented variants of RSA, Crypto++, and the sequential counterpart. Multi-thread CPU improved performance by 6, over the sequential CPU implementation, and with GPU, it improved 23 times over the sequential implementation. The throughput gained for 1024 bits was ~1800 msg/sec, and for 2048 bits, it was ~250 msg/sec. Yanga et al. [26] suggested a parallel block Wiedemann algorithm in cloud to enhance the performance of GNFS and reduce communication costs, involved in solving large and sparse linear systems over GF.

### 4.2. Example of the acceleration of a modular exponentiation calculator

The calculation of "a factor b modulo n" is the heart of RSA cryptography and is also the most resource consuming component. Dividing this calculation into smaller parts will allow distributed and parallel processing of this calculation, where each smaller part is calculated by one sensor and later is integrated to obtain the result of "a factor b modulo n." The underlying concept is the following conceptual equation: $((a \bmod n) * (b \bmod n)) \bmod n = (a*b) \bmod n$. This concept is used by the following algorithm to calculate modular exponentiation. Step 1: Translate the input into a binary number. Step 2: Start at the rightmost digit, let k = 0, for each positive digit calculate the value of $2^k$, Step 3: Calculate mod n of the powers of two ≤ b, Step 4: Use modular multiplication properties to combine the calculated mod n values. Steps 2 and 3 can be executed in parallel by several connected sensors. The results from the sensors are then sent to the sensor requested the encryption/decryption, to execute step 4 and obtain the final result. Using a network of 7688 devices, we ran a comprehensive test, which proves the feasibility of executing RSA using parallel and distributed processing.

# 5. Conclusion

Connecting sensors to the Internet exposes the entire network to malicious penetrations. This is due to poor computation resources in standard sensors, which do not allow the execution of robust security systems. Hence, lightweight primitive systems should be implemented in IoT. To maintain current Internet security level, we adjusted implementations of known security concepts and mechanisms, which contribute to the security of the Internet of things. In this chapter, we focused on three key security elements where downsizing is feasible without compromising security: (a) Eliminating the frequent use of detailed data in the classification process. (b) Adjusted random forest machine learning to work in a distributed and parallel mode, when building the forest and during the detection process. (c) Adjust RSA cryptography calculations which are executed in parallel and distributed. The proposed solutions have smaller footprints, are efficient, and in most cases demonstrate better performance. We prove that downsizing and parallel processing are the most appropriate approaches for implementing comprehensive concepts for proper operation in constrained environments of IoT.

We are currently working on expanding current research areas. For example, additional improvements in RF implementation and exploring other machine learning technologies to check its applicability to IoT anomaly detection. We are exploring other asymmetric cryptography systems to check their applicability to IoT. In parallel, we are investigating authentication methods and technologies to discover a suitable one for IoT, or we are considering building an IoT-specific authentication.

# Author details

Menachem Domb

Address all correspondence to: dombmnc@edu.aac.ac.il

Ashkelon Academic College, Computer Science Department, Ashkelon, Israel

# References

[1] Aldosari HM, Snasel V, Abraham A. A new security layer for improving the security of internet of things (IoT). Inteernational Journal of Computer Information Systems and Industrial Management Applications. 2016;8:275-283. ISSN: 2150-7988

[2] Mairal FB, Ponce J, Sapiro G. Online dictionary learning for sparse coding. In: Proceedings of the 26th Annual International Conference on Machine Learning. Montreal, Quebec, Canada: ACM; 2009. pp. 689-696

[3] Jankov D, Sikdar S, Mukherjee R, Teymourian K, Jermaine C. Real-time high performance anomaly detection over data streams: Grand challenge. In: Proceedings of the 11th ACM International Conference on Distributed and Event-based Systems. Tokyo, Japan: ACM; 2017. pp. 292-297

[4] Vlachos M, Freris NM, Kyrillidis A. Compressive mining: Fast and optimal data mining in the compressed domain. The VLDB Journal. 2015;**24**(1):1-24

[5] Sakurada M, Yairi T. Anomaly detection using autoencoders with nonlinear dimensionality reduction. In: Proceedings of the MLSDA 2014 2nd Workshop on Machine Learning for Sensory Data Analysis. ACM; 2014. p. 4

[6] Reeves G, Liu J, Nath S, Zhao F. Managing massive time series streams with multi-scale compressed trickles. Proceedings of the VLDB Endowment. 2009;**2**(1):97-108

[7] Chilimbi TM, Hirzel M. Dynamic hot data stream prefetching for general-purpose programs. In: ACM SIG-PLAN Notices. Berlin, Germany: ACM; 2002;**37**(5):199-209

[8] Lane T, Brodley CE. Temporal sequence learning and data reduction for anomaly detection. ACM Transactions on Information and System Security (TISSEC). 1999;**2**(3):295-331

[9] Kasiviswanathan SP, Melville P, Banerjee A, Sindhwani V. Emerging topic detection using dictionary learning. In: Proceedings of the 20th ACM international conference on Information and knowledge management. ACM; 2011. pp. 745-754

[10] Aldroubi A, Cabrelli C, Molter U. Optimal non-linear models for sparsity andsampling. Journal of Fourier Analysis and Applications. 2008;**14**(5-6):793-812

[11] Mansoori EG, Zolghadri MJ, Katebi SD. SGERD: A steady-state genetic algorithm for extracting fuzzy classification rules from data. IEEE Transactions on Fuzzy Systems. Aug 2008;**16**(4):1061-1071. ISSN: 1063-6706

[12] Dubois D, Durrieu C, Prade H, Rico A, Ferro Y. Extracting decision rules from qualitative data using sugeno integral: A case-study. In: Proceedings of the 13th European Conference, ECSQARU 2015. Compiègne, France: Springer; Jul 2015;**9161**:14-24. ISBN: 978-3-319-20806-0; ISSN: 0302-9743

[13] Daniel W. Gulwani S, Hart T, Zorn B. FlashRelate: extracting relational data from semi-structured spreadsheets using examples. In: Proceedings of the 36th ACM SIGPLAN Conference on Programming Language Design and Implementation. Vol. 50, Issue. 6. New York: ACM; Jun 2015. pp. 218-228

[14] Kosina P, Gama J. Very fast decision rules for classification in data streams. Data Mining and Knowledge Discovery. Jan 2015;**29**(1):168-202. ISSN: 1384-5810

[15] Jafarzadeh H, Torkashvand R, Asgari C, Amiry A. Provide a new approach for mining fuzzy association rules using apriori algorithm. Indian Journal of Science and Technology. Apr 2015;**8**(S7):127-134. ISSN: 0974-6846

[16] Pourpanaha F, Peng Limb C, Mohamad Saleh J. A hybrid model of fuzzy ARTMAP and genetic algorithm for data classification and rule extraction. Elsevier, Expert Systems with Applications. 2016;**49**(7):4-85

[17] Mashinchi R, Selamat A, Ibrahim S, Krejcar O. Granular-rule extra action to simplify data. In: Intelligent Information and Database Systems. Vol. 9012 of the Series LNCS. Mar 2015. pp. 421-429. ISBN: 978-3-319-15704-7

[18] Yang H, Xiao C, Qiao Y. Study on anomaly detection algorithm of qar data based on attribute support of rough set. International Journal of Hybrid Information Technology. 2015;**8**(1):371-382. DOI: http://dx.doi.org/10.14257/ijhit.2015.8.1.33. ISSN: 1738-9968. IJHIT Copyright © 2015 SERSC

[19] Tang H. A simple approach of data mining in excel. In: 4th International Conference Browse Conference Publications, IEEE Xplore; 2008

[20] Tong S, Koller D. Support vector machine active learning with applications to text classification. Journal of Machine Learning Research, Leslie Pack Kaelbling. 2001:45-66

[21] Osugi T, Kun D, Scott S. Balancing exploration and exploitation: A new algorithm for active machine learning. In: Fifth IEEE International Conference on Data Mining. IEEE Xplore, 2005. DOI: 10.1109/ICDM.2005.33

[22] Lang T, Flachsenberg F, Luxburg U, Rarey M. Feasibility of Active Machine Learning for Multiclass Compound Classification. 2016. PMID: 26740007. DOI: 10.1021/acs.jcim. 5b00332

[23] Bharathidason S, Jothi Venkataeswaran C. Improving classification accuracy based on random forest model with uncorrelated high performing trees. International Journal of Computer Applications (0975-8887). Sep 2014;**101**(13)

[24] Lin C, Liu J, Li C-C, Chu P-W. Parallel modulus operations in RSA encryption by CPU/GPU hybrid computation. In: Taiwan, Conference Paper, IEEE Xplore: 29. Jan 2015. DOI: 10.1109/AsiaJCIS.2014.25

[25] Fadhil HM, Younis MI. Parallelizing RSA algorithm on multicore CPU and GPU. International Journal of Computer Applications (0975-8887). Feb 2014;**87**(6)

[26] Yanga LT, Huanga G, Jun Feng B, Xua L. Parallel GNFS algorithm integrated with parallel block Wiedemann algorithm for RSA security in cloud computing. Information Sciences, Elsevier. 2016

# IoT Standardization: The Road Ahead

Arpan Pal, Hemant Kumar Rath,
Samar Shailendra and Abhijan Bhattacharyya

Additional information is available at the end of the chapter

**Abstract**

The Internet of Things (IoT) is an emerging area of the modern technology which impacts use cases across governance, education, business, manufacturing, entertainment, transportation, infrastructures, health care, and so on. Creating a generalized framework for the IoT with heterogeneous devices and technology support requires interoperability across products, applications, and services that preclude vendor lock-in. Global standardization of the IoT is the only solution to this. Though standardization efforts in the IoT are not new with many national and international standard bodies working today, there are many open areas to debate and standardize—like reconciling country-specific efforts, empowering local solutions, etc. This chapter brings a holistic view of the existing IoT standards, discusses their interlinking, and enumerates the pain points with possible solutions. It also explains the need for country-specific standardization with the example of an Indian Standard Development Organization (SDO), vis-à-vis global initiatives, as a driver for societal uplifting and economic growth.

**Keywords:** IoT, standardization, TSDSI, ITU, ETSI, IEEE

## 1. Introduction

The **Internet of Things** (**IoT**) is the network of "things" or smart devices embedded with sensing, actuation, software, and network connectivity to sense and exchange data among the things, between the things, and with the outside world. The term IoT was coined in 1999 by British technology pioneer Kevin Ashton to describe a system in which objects in the physical world could be connected to the Internet by sensors without requiring human intervention. Though the things were initially thought as machines, today, things are synonymous with any living entity including the human beings, animals, and any other device or element on earth. The "things" should not only be addressable but also reconfigurable, reusable, locatable, uniquely identifiable, and remotely controllable.

Today, the IoT is becoming a growing topic of interest and is becoming a part of our day-to-day life. IoT applications such as remote health monitoring, disease detection and monitoring, crop monitoring, accident prediction and detection, traffic monitoring, robotic rescue operation, environment pollution monitoring, unmanned aerial vehicle (*UAV*)-based rescue operation, and so on, are some of the common applications we witness today [1, 2]. With growing number of applications and devices, the IoT is going to be the dominant technology, where any device can connect with any other device in the world. The IoT integrates ambient sensing, ubiquitous communications, intelligent analytics, and pervasive computing.

The exponential growth of the IoT is mainly attributed to (i) the massive growth of low-cost devices, (ii) advancement of wireless networks, and (iii) creation of new applications. According to a recent survey [3], 50 billion smart devices are estimated to operate by 2020 which can generate avalanche of traffic which is in the order of multiple thousand times of the current Internet traffic. In addition to this, the application requirements are also going to be stringent in terms of latency (~1–100 ms) and reliability (~99.99–99.9999%).

Most of the existing Internet standards did not have the vision to include the IoT which is relatively a newer concept. Therefore, their scope is not sufficient to support the IoT technically and economically. Moreover, IoT architecture, use cases, devices, etc., are still evolving. Today, many IoT devices have been deployed with proprietary protocols. This makes the communication between multiple IoT devices difficult. However, in the era of digital revolution, with many vendors playing in the field, with researchers and entrepreneurs working hard to develop solutions and with government agencies trying hard to reach their citizens, the world has to agree to a common standard. Not only the hardware components related to the IoT, but also the software aspects of the IoT should also be standardized, creating standardized application programming interfaces (APIs) and software services such that future applications can be deployed in a level and uniform environment, thereby enabling easy migration across systems.

Standardization is necessary to ensure (i) interoperability across products, applications, and services that preclude vendor lock-in; (ii) economy of scale, where the three sections of the society—developer (researcher), government (regulator), and the user—get benefited in a reasonable time frame; (iii) security and privacy of the data and the users; (iv) space for the researchers to take our society to another height; and (v) interoperation across physical communication systems, protocol syntax, data semantics, and domain information [2]. Though there is no single body which is responsible for making IoT standards, there are considerable efforts at national and international level, at government level, and at different organizational levels for IoT standardization. Alliances have been formed by many domestic and multinational companies to agree on common standards and technology for the IoT. However, no universal body has been formed yet. While organizations such as IEEE, Internet Engineering Task Force (IETF), ITU-T, OneM2M, 3GPP, etc., are active at international level, Telecommunication Standards Development Society, India (TSDSI), Global ICT Standardization Forum for India (GISFI), Bureau of Indian Standards (BIS), Korean Agency for Technology and Standards (KATS), and so on, are active at national level and European Telecommunications Standards Institute (ETSI) in the regional level for standardization.

This chapter brings a holistic view of the existing IoT standards and their interlinking and enumerates the pain points with possible solutions. It also explains the need for country-specific standardization with the example of an Indian Standard Development Organization (SDO), vis-à-vis global initiatives, as a driver for societal uplifting and economic growth. Section 2 details about the deployment issues of the IoT, whereas Section 3 brings out the standardization effort visible today in both national and international levels. Section 4 discusses the role of local SDOs in IoT standardization. While we discuss the economics of IoT standardization in India in Section 5, we explain the open areas of IoT standardization in Section 6. Finally, we conclude this chapter in Section 7.

## 2. The IoT framework: Deployment Issues

Though the IoT as a term is relatively new, it is quite old as a concept. The main idea of the IoT is to control and monitor "things" through the computing devices connected over a packet switched network. Today, the IoT has become a new paradigm for the Internet through the confluence of technological advancements and easy availability of devices leading to hitherto unexplored applications. The major technology drives for the IoT are [4]:

- Improvement in connectivity in terms of data rate, availability, and cost

- Wide adoption of Internet Protocol (IP) as the basic addressing mechanism for "things"

- Miniaturization of computing and communication devices along with lowered cost

- Advancement in data analytics

- Rise of cloud computing along with cost reduction in storage systems

Depending on the settings of the exact applications, there can be several patterns of interaction among the heterogeneous entities in an IoT system [1, 5]. In this section, we intend to discuss about communication models used for typical IoT systems and perform a comparative analysis. We then plan to discuss the challenges we face while we use the state-of-the-art solutions for practical deployments.

### 2.1. Communication models used in the IoT

The IoT is the network of devices which sense, generate, and transmit data to an application server that can be located either in a cloud or in a sophisticated machine. To understand the collected data and to take appropriate action, data analytics are to be used on the application server. For the IoT to become a success, communication between the devices and the application server is the core, and the models used in practice are as follows [6]:

- **Direct communication between devices (D-D):** Under this model, two end devices can directly communicate without using any intermediary as illustrated in **Figure 1(a)**. The devices can connect over a Local/Personal/Wide Area Network (LAN/PAN/WAN).

- **Communication between device and an application server in cloud (D-C):** In this model, the device communicates with an application server in the cloud. If the device is a consumer (that needs some control information to execute some functions), then it receives the required information from the concerned application in the cloud server. The model is shown in **Figure 1(b)**. A typical example of such communication model is the offering around TCS Connected Universe Platform (TCUP) [7, 8].

- **Communication through the Edge Gateway (D-E-C):** Under this model, the end devices use a local gateway as a conduit to connect to the application server in the cloud as shown **Figure 1(c)**. This deployment has a greater scope of heterogeneity at the users' end and is highly scalable. It is useful when the devices do not use generic protocols to provide the local services but need to communicate with an application server at the cloud with generic protocols (Hypertext Transfer Protocol (HTTP), Constrained Application Protocol (CoAP), etc.) [9]. Cloud service around Microsoft Azure IoT Edge [10] is a very popular example for this kind of exchange model.

### 2.1.1. Considerations in choosing a communication model

All the models mentioned above accomplish the fundamental objective of exchanging information among "things." With the advent of lightweight protocols like Message Queue Telemetry Transport (MQTT), CoAP [9], etc., which enable web service like transactions in constrained devices, the "things" have become more like the web citizens of the conventional Internet. While the direct communication model helps in quick control and actuation, it suffers from non-scalability, interoperability, and heterogeneity. IoT functionalities are also restricted as no additional service analytics is possible. In the DC model, a typical publish-subscribe or "observe" [9] relationship can achieve one-to-many communication. This provides a possibility of application services based on the analytics/intelligence incorporated in the cloud application. However, in this case, the end devices have to be IP enabled and should use generic standard protocols to remain interoperable.

The communication model through the Edge Gateway provides design flexibilities in terms of scalability, heterogeneity, and interoperability. The edge may itself be equipped with several local intelligence/analytics which may lead to reduction in the amount of network traffic exchanged with the cloud. It enables design decisions like data aggregation at the gateway and traffic optimization while communicating with the cloud with an extra cost due to additional infrastructure at the user premise along with the cloud service. **Figure 2** summarizes the above discussion on several deployment-specific attributes.

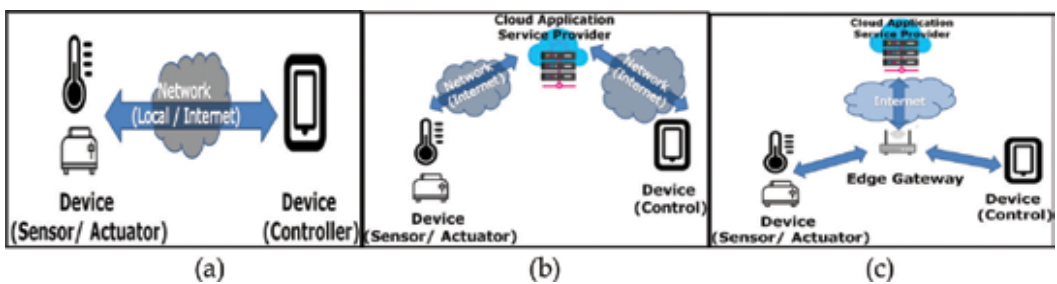

**Figure 1.** Communication model in the IoT: (a) D-D, (b) D-C, and (c) D-E-C.

**Figure 2.** Attribute-specific considerations for different IoT communication models.

## 2.2. Challenges: technical, deployment, business, and societal

Taking a thread from our earlier discussion, we now discuss the challenges we face for IoT deployment [5]. We categorize the challenges as follows:

- **Connectivity:** Connecting billions of devices or things is a major challenge. Connectivity impacts the scale of business, profit margin, and societal impact of the operation. Though cloud-based deployments rule the IoT world, edge-based deployments are picking up due to (i) low latency, (ii) ease of deployment, (iii) better security and privacy, and (iv) high data aggregation.

- **Interoperability:** The IoT is growing in various directions, and different technologies are playing different roles. Today, Wireless Fidelity (WiFi), Zigbee, Long-Term Evolution (LTE), LTE Advanced (LTE-A), Low-Power Wide Area Network (LPWAN), Bluetooth, etc., are some of the major communication technologies rule the IoT world. Seamless connectivity with different devices operating in different technologies is a major challenge. Interoperation at higher layers of the network protocol stack involving semantics, and domain-specific operations is another challenge.

- **IoT analytics:** The basic nature of the IoT is to obtain and to act on information. Therefore, IoT analytics play a major role. For practical deployment, placing the analytics platform in the IoT architecture is the major issue. Since information is generated or gathered at the devices and is communicated to the cloud with/without the support of edge, decision has to be taken such that parts of the analytics platform have to be deployed in appropriate places of the framework, i.e. whether at edge/fog or at the cloud. Factors such as delay, regulatory issues, cost, scale and ease of operation, etc., play significant roles on this.

- **Security and privacy:** It has been observed that IoT deployments are prone to security and privacy issues at device, edge, and cloud platform level. Therefore, security and privacy of the data, device, application, and the server are to be considered while deciding appropriate deployment architecture. Instead of considering security and privacy as afterthoughts of deployment, today, these are the prime concerns for any kind of deployment.

- **Business or return on investment (RoI):** Deployment decision can impact the vertical, horizontal, and consumer markets of IoT industry while struggling with the regulatory and legal aspects of the society. Based on the deployment usage and client base, IoT can be divided into (i) consumer IoT, which impacts the mass (like wellness, education, etc.) and the governance in the society; (ii) industrial IoT, which governs the communication framework of Industry 3.0 or Industry 4.0 scenarios; and (iii) commercial IoT, which includes retail and warehouse inventory controls, device tracking, health services, and so on.

- **Societal**: Societal challenges also play a major role in IoT deployment as IoT has to satisfy the customer, developer, and regulator needs of the society. This includes the mode of usage, the energy consumption, environmental impact, societal impact, etc.

Today various industries and academia have proprietary solutions (CISCO, TCS, Microsoft, IBM, etc.) to address some of the above challenges. However, the standard bodies across the world are attempting to collaborate to bring out a unified solution for seamless IoT deployment. The security and privacy which were the afterthoughts for earlier deployments are becoming the front seat candidates.

While the above challenges rule the deployment decision, standardization effort can play a significant role for the above issues. Taking it forward, we now discuss the standardization efforts we see for the IoT in the following sections.

## 3. Standardization efforts for the IoT

To maintain seamless operation of the IoT, it is essential that the "things" or devices follow a common standard with well-defined protocols and interoperable interfaces. There are several ongoing efforts in different Standard Development Organizations (SDOs) across the world to build standard platforms, protocols, and technologies to ensure seamless operation of these devices. From the perspective of technological offering, different SDOs can be broadly categorized into two classes: (i) generic and (ii) application specific.

In the first category, SDOs such as ITU, IEEE, IETF, 3GPP, and oneM2M, have traditionally performed a pivotal role in defining technology standards to cover the overall problem space. They have specified either policies or generic reference architectures or have offered a standard protocol to carry out the communication. These SDOs also specify technology domain. We shall discuss this later while discussing IETF's efforts specific to Low-Power Wide Area Network (LPWAN). These SDOs are generally open in a sense that anyone can go through the specifications from these SDOs without being a member of the same. However, to contribute one needs to be a member. IETF is an exception to this. It is indeed open in true sense. In theory, any individual can contribute to IETF standardization, and the contribution is valued in a meritocratic manner.

On the other hand, there are SDOs or alliances created in the interest of standardizing technologies for some specific domain of applications. These SDOs fundamentally use the existing architectures and protocol offerings with generic approach to create the communication model. They create specific standards for specific exchange models to fill up typical gaps in the available standard offerings. Fairhair Alliance [11], powered by the THREAD group [12], is one such example. These SDOs are generally closed within its member organizations. We further discuss how IETF plays a pivotal role in becoming the nodal entity for all the SDOs.

### 3.1. Standardization efforts for overall IoT network stack

#### 3.1.1. IoT standardization with International Telecommunication Union (ITU)

Study Group 20 (SG20) in ITU has been in charge of "IoT and its applications, including smart cities and communities." Some of the topics of the ongoing studies include semantics aspects; big data aspects; detailed requirements of networks supporting IoT applications; accounting and charging aspects; identification, security, and privacy; openness; etc.

ITU has also defined the reference architectures for different applications including smart manufacturing and Industrial IoT, e-health and e-agriculture, wearable device and services, cooperative applications and transportation safety services, monitoring and study of global processes of the earth for disaster preparedness, and so on. **Figure 3(a)** illustrates how ITU defines the component-based reference model for IoT/M2M communication. Devices are networked with or without the help of the gateways, i.e., it is a combination of **D-C** and **D-E-C** architectures explained in Section 2. **Figure 3(b)** shows an exemplary protocol stack of the reference model. It uses the standards created by open SDOs like IETF and IEEE. **Figure 4(a)** and **(b)** show how ITU defines the application-level architectures with two specific examples of e-health protocol stack. The first one follows a local gateway centric architecture. The devices connect directly to the application server in the second example.

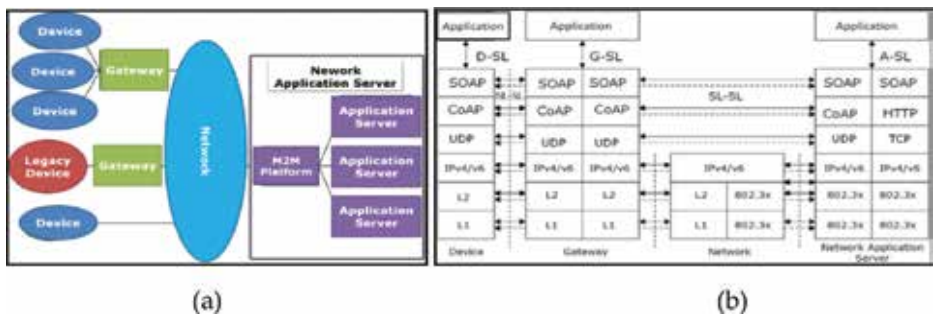

**Figure 3.** (a) Component and (b) protocol stacks in M2M ref. model [13].

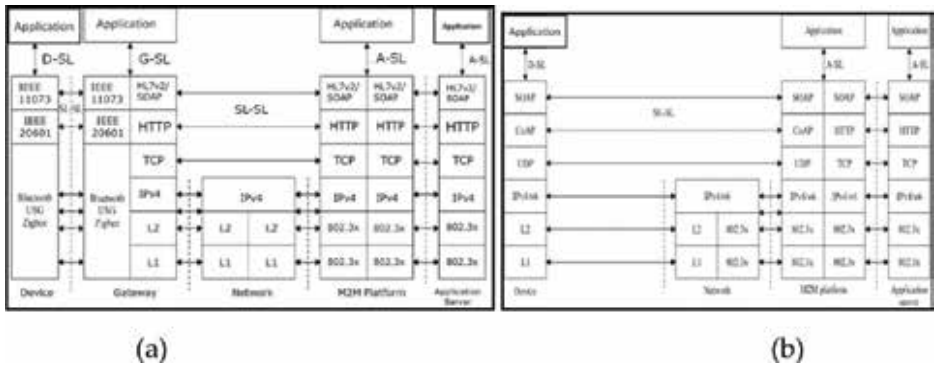

**Figure 4.** Protocol stacks for e-health (a) with and (b) without gateways [14].

### 3.1.1.1. Handling IoT deployment challenges

Though ITU has not defined any particular technology for the IoT, it has taken a key role in defining the radio spectrum. Also, as evident from the previous discussions, ITU has provided a reference architecture which can be adopted as a common platform for producing solutions for future smart city and similar IoT applications. That way ITU is taking an important role in ensuring standardization in connectivity and interoperability.

| ITU | Current activities | Roadmap | Comments |
|---|---|---|---|
| Connectivity | Defines all the layers and protocols | Spectrum allocation aspects for different future technologies such as 5G | ITU is the nodal point for defining any standard |
| Interoperability | Ensures interoperability of all standards from all SDOs | IoT framework standardization | |
| Security and privacy | Based upon the corresponding SDO solutions | | |

### 3.1.2. IoT standardization with IEEE

IEEE has been producing standards for local/personal area connectivity while playing a key role in forming the physical and Medium Access Control (MAC) layer standards. It has produced new specifications keeping the typical requirements for IoT applications in mind. The IoT standardization is being undertaken by the IEEE Standards Association (SA) under the project P2413 [14, 15], which aims to come up with an architectural framework that covers the needs for different applications and to provide necessary technological solutions by leveraging the existing body of work as much as possible. IEEE P2413 considers the IoT as a simple three-tier architecture with applications, networking and data communication, and sensing, which are essential for the IoT communication.

Today wireless LAN (IEEE 802.11 family) is still a practical MAC standard for many IoT applications. However, for the low-power operation of constrained devices in IoT applications, IEEE has come up with an access mechanism for personal area network of low-power sensing devices with low rate transmissions. The technology is standardized under IEEE 802.15.4 and termed as LowPAN. It is also made IP compatible through the standardization efforts from IETF. IEEE is putting effort in defining several futuristic technology standards covering the lower layer specifications as well as application layer APIs in the specific domains of Wireless Access in Vehicular Environment (WAVE), short range communication using visible lights, and so on [16].

### 3.1.2.1. Handling IoT deployment challenges

IEEE is taking an important role in defining the physical and data link layers to ensure low-level interoperability among devices. With IETF collaborations, it ensures compatibility of devices across the Internet. IEEE has been instrumental in standardizing the security, authentication, and authorization mechanisms for the data-link layer.

| IEEE | Current activities | Roadmap | Comments |
|---|---|---|---|
| Connectivity | Handles the MAC and physical layer aspects | To ensure interoperability with upcoming technologies for 5G and beyond along with defining technologies for low-latency/ tactile Internet | IEEE's primary focus is on the user and application-related standardization |
| Interoperability | Works with other SDOs | | |
| Security and privacy | Addresses security and authentication issues | | |
| Societal | Addresses various aspects of energy consumption at devices | | |

### 3.1.3. IoT standardization with 3GPP

3GPP unites telecommunication SDOs to produce reports and specifications for cellular communication through NarrowBand IoT (NB-IoT) [17–19].

### 3.1.3.1. NarrowBand IoT (NB-IoT)

In June 2016, 3GPP completed its first set of specification on NarrowBand IoT (NB-IoT). It is a radio standard developed for Low-Power Wide Area Network (LPWAN) to support IoT technologies. NB-IoT is designed for indoor coverage using large number of connected devices with low cost and long battery life. NB-IoT standards are not backward compatible and support three operation modes as illustrated in **Figure 5** and are as follows: (i) In-band operations utilize a resource block within the LTE carrier, (ii) guard band operations utilize the guard band within the LTE carrier, and (iii) standalone operations utilize the bandwidth of 200 kHz traditionally used by Global System for Mobile (GSM) carriers. It targets both LTE and GSM-dominant geographies. In the case of the latter, it uses GSM carrier bands, though it can still continue to have the standard guard band between the GSM carriers. In the case of the former, it uses in-band or in the guard band of LTE carriers. Apart from the physical layer, NB-IoT uses the same protocol stack as that of LTE. NB-IoT targets massive IoT deployments.

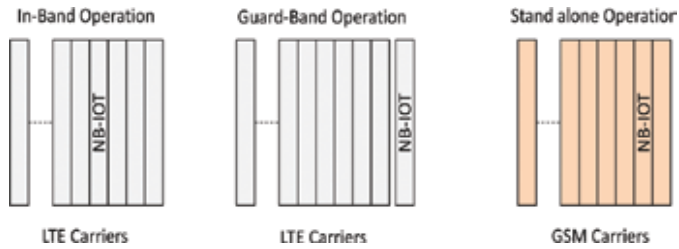

**Figure 5.** Operation mode of NB-IoT.

However, the nonconformity of NB-IoT standard with LTE standard may pose significant deployment challenges. Some of the salient features of NB-IoT are (i) less operational power consumption, (ii) reduced component cost, (iii) low data rate, and so on.

### 3.1.3.2. Handling IoT deployment challenges

3GPP's primary focus is low-power small data transfers. The issues of licensing, spectrum, interference, and so on are still need to be resolved.

| NB-IoT | Current activities | Roadmap | Comments |
|---|---|---|---|
| Connectivity | MAC and physical layer | Multicasting, mobility, and service continuity enhancements | Low power small data transfer |
| Interoperability | Backward compatible and interoperable with other standards | | |
| Security and privacy | Handles various aspects of security; privacy still needs to be addressed | | |
| Societal | Low energy consumption for the IoT | | |

### 3.1.4. IoT standardization with Internet Engineering Task Force (IETF)

IETF is a leading organization in standardizing protocols for the Internet at different levels of the network stack. It has limited its scope "above the wire and below the application". IETF is a complementing organization to IEEE, 3GPP, and ITU by creating the enabling protocols that actually connect the constrained nodes in a constrained environment in an efficient manner on top of the available physical and data-link layer technologies available from other SDOs working in that area. As evident from **Figure 6**, IETF has IoT-specific protocol offerings for every layer within its purview.

Security consideration is an integral part of any IETF document. IETF uses standardized transport layer security protocols like Transport Layer Security (TLS) and Datagram Transport Layer Security (DTLS) depending on whether Transport Control Protocol (TCP) or User Datagram Protocol (UDP) is used, respectively. The security mode (pre-shared key, certificate-based security, etc.) needs to be chosen depending on the device and network capability. However, security protocol solution optimized for constrained devices is still an open issue as TLS and DTLS are primarily not designed for constrained environments. It is an open area of research, and the question mark in **Figure 6** indicates this. In recent times, IETF has been active in creating specific standards for wide area of technologies for the IoT known as LPWAN. The IPv6 over LPWAN Working Group [20] has been formed to optimize the IETF protocol offerings for the different lower layer offerings on low-power wide area network from SigFox, LoRA [21] Alliance, 3GPP, etc., as well as to define the upper layer exchanges and signaling using the existing protocol offerings. The objective of such initiatives is to tailor the existing IETF offerings in order to cater the specific requirements to enable IP compatibility for specific access technology. LPWAN working group is yet to produce any RFCs. We need to watch out for the progress. **Figure 7** illustrates a representative network topology from LoRA alliance which is a key contributor to LPWAN-specific radio access technologies parallel to NB-IoT from 3GPP.

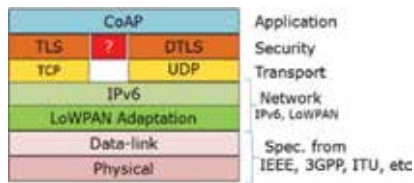

**Figure 6.** IoT offerings from IETF.

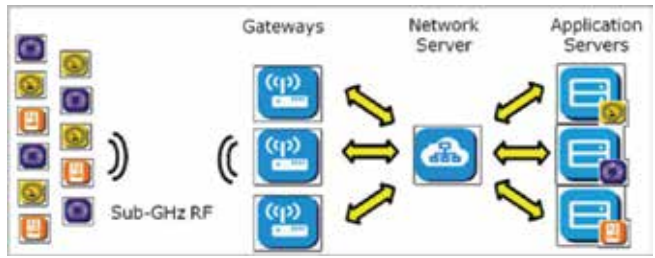

**Figure 7.** LoRA network topology [22].

### 3.1.4.1. Handling IoT deployment challenges

It is needless to say that IETF is playing a major role in defining the core standards that enable the interoperability and connectivity of billions of devices across the Internet. IETF is the key in defining the security features for future IoT/M2M devices.

### 3.1.5. IoT standardization with Organization for the Advancement of Structured Information Standard (OASIS)

IBM has developed a pair of protocols called MQTT and MQTT for sensors (MQTT-S) designed to be operated over TCP/IP except some highly real-time low-power MQTT-S mode for local exchange which operates on UDP. The protocol works in publish/subscribe mode and relies on the TCP layer for ensuring reliability and security. MQTT and MQTT-S have been there in practice for quite some time. Few years back IBM brought MQTT/MQTT-S under the umbrella of OASIS open standard community. However, it cannot ignore IETF as the pivotal entity, and there are recent efforts to augment IETF standardization with MQTT considerations [23].

### 3.1.5.1. Handling IoT deployment challenges

MQTT provides a standardized publish-subscribe mechanism to connect devices. It allows cloud-based architectures to be developed with common protocol semantics and thus helps in interconnectivity. It adopts the available security solutions from IETF to allow a common security feature.

| IETF | Current activities | Roadmap | Comments |
|---|---|---|---|
| Connectivity | Specifications for network, transport, and applications | Specify technologies for low-latency real-time Internet communication for the future. Define lightweight security solutions and privacy aware protocols | Handles Internet and core network-related standardization |
| Interoperability | Ensures the Interoperability with other protocols and applications and technologies | | |
| Security and privacy | Provides security standards. Privacy is not the mainline charter | | |

### 3.1.6. IoT standardization with oneM2M

oneM2M was formed in 2012 as a global organization with an objective to consolidate global requirements and create global standards for IoT/M2M technologies. It provides specifications for architecture, APIs, security, and interoperability guidelines and certification for M2M/IoT devices and applications. oneM2M came up with the first formal release of specification in Jan 2015, which were dated by the second release of specifications in the late 2016; the work for the third release of the specification is in progress.

As part of these specifications, it has published service layer architecture for all M2M/IoT devices to interact and exchange data seamlessly. oneM2M specification considers the IoT network layered into three service layers: application, common services, and network service layer. oneM2M provides a service layer specification for M2M services so that they can interoperate and exchange messages seamlessly. It relies on the service providers' network for message communication. Any primitive of oneM2M service layer can be mapped over IP network or other networks. **Figure 8(a)** represents oneM2M functional architecture, whereas **Figure 8(b)** explains the communication model. The interactions and protocols binding with application protocols like HTTP, CoAP, and MQTT are also being defined by oneM2M specifications.

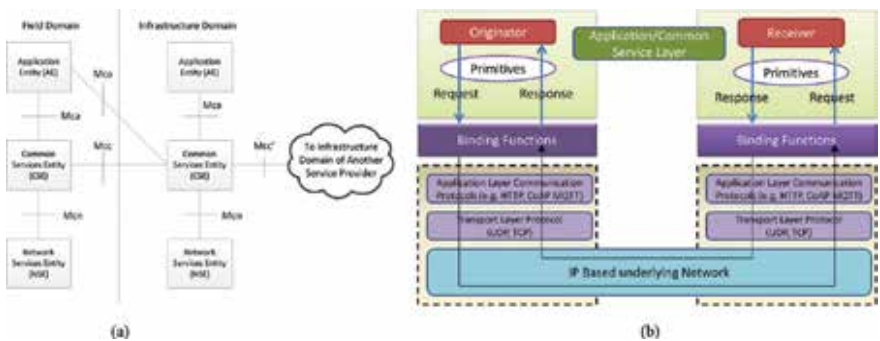

**Figure 8.** oneM2M (a) functional architecture [24] and (b) communication model [25].

### 3.1.6.1. Handling IoT deployment challenges

oneM2M is providing a universal service layer architecture which can ensure the interoperability of various IoT devices. It provides a rich set of guidelines, addressing format, APIs, and bindings with most popular IoT protocols. It also provides mechanism for non-oneM2M devices to operate with oneM2M network. This makes oneM2M a unique platform that provides a unified framework for message exchange through variety of devices and networks. However, this has significant deployment challenges such as handling heterogeneity of devices, geography-specific use cases, and interaction with variety of communication protocols. oneM2M has undertaken some pilots in Korea; the learning from which needs to be incorporated in oneM2M. oneM2M also needs to standardize the security and semantics framework as well before it is widely and ubiquitously deployed. Finally, oneM2M is also discussing on providing an open implementation of its specification which can increase the deployability of oneM2M specifications.

| oneM2M | Current activities | Roadmap | Comments |
|---|---|---|---|
| Connectivity | LoRA and NB-IoT at the south side. CoAP, HTTP, and MQTT at the north side | SigFox and any other physical transport. Any other application protocol at the north side | Working with ITU for IoT framework. Many SDOs including TSDSI have also adopted oneM2M as their local standard |
| Interoperability | Focus on device interoperability | Plan to interoperate with any other service layer specification | |
| Security and privacy | Provides basic security architecture | Security and privacy aspects are under further development | |

## 3.2. Application-specific efforts

As mentioned earlier, there are alliances and SDOs with a specific task to fill up certain gaps while using the standard offerings for a specific technology. One example is the Fairhair Alliance which is dedicated towards standardizing the technologies for lighting control and building automation [23]. The core technologies and protocols are based on the generic IoT-specific offerings from IETF, IEEE, 3GPP, and so on. Fairhair tries to fill the specific technological gaps (specific security handshakes, typical supports for multicast, exclusive protocol level optimizations, etc.) related to the applications in the concerned business domain. It is being driven by the significant players in the lighting control and home/building automation business domain like Philips, Seimens, etc. Another important participant in this alliance is the "THREAD Group" which is developing standard technologies behind home automation/smart home solutions and is largely driven by Google [12]. **Figure 9(a)** shows the protocol stack for THREAD and the relationship on IETF and IEEE specifications. It uses a mesh topology contrary to the star topology of LoRA. THREAD specification defines a Border Gateway entity to maintain connectivity between THREAD and non-THREAD networks. The topology is illustrated in **Figure 9(b)**. It defines the necessary handshakes to establish and maintain secure connection between THREAD and non-THREAD entities [12].

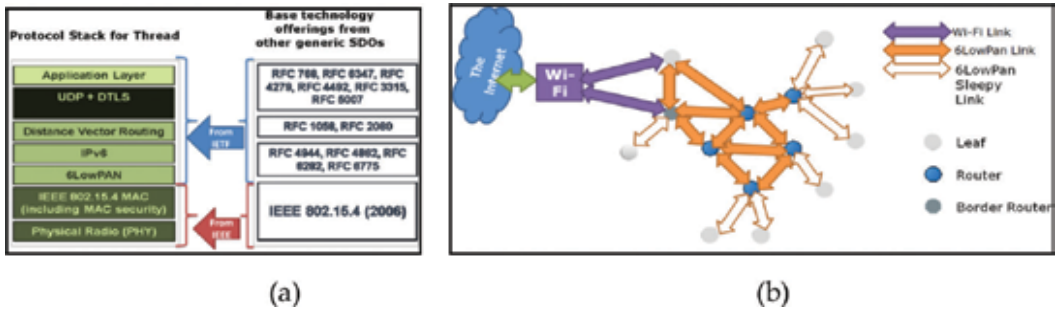

**Figure 9.** THREAD (a) protocol stack and (b) network topology [12].

### 3.2.1. Handling IoT deployment challenges

These alliances are bridging important application-specific gaps for interoperability of edge devices in smart homes.

### 3.3. IETF as a nodal entity

When the Internet was migrated from a research project to a common communication mechanism to connect computers across the globe, IETF, which has been producing standards for the Internet since 1986, became a pivotal entity. Different modes of telecommunication mechanisms considered the Internet as the conduit to reach peers globally. The offerings from different SDOs started to lean toward more and more IP-centric approach. IETF impacted the activities of the other SDOs as well. The collaboration between IETF and other important SDOs, such as ITU-T and 3GPP are started in the early 1990s. There have been several RFCs describing IETF's relationship with respective SDOs. For example, RFC7241 formulates the modes of collaboration between IETF and IEEE. RFC3113 provides the set of guidelines and principles for collaboration between IETF and 3GPP. RFC6756 does the same for collaboration between IETF and ITU-T. All these guidelines are defined by the Internet Architecture Board (IAB) which acts as an advisory body to the Internet Society (ISOC). With the new paradigm of the IoT, this collaboration approach has even more strengthened. However, as evident from our earlier discussions, the wide variety of IoT applications have given rise to application-specific alliances which have created application-specific standards. While there are specific modalities of operation between IETF and other SDOs, such formal arrangement may not be specified for all the efforts sprawling for different applications. However, all of them have to depend on IETF for the core Internet protocols, and the interaction happens as voluntary efforts from people with common interest in both IETF and the respective alliance/SDO.

Sometimes, the Work Group (WG) charter is enhanced with specific requirements from such SDOs if the sought-after solution has a large enough impact to cover several application domains to justify it as a work item in IETF. Sometimes the interested people in the community form a new WG if the initiative gets a significant support from the communities around IETF. The LPWAN WG is such an example. Sometimes, the individual SDOs create bridging specifications to fill in the required gaps on top of the relevant IETF offerings if the sought-after solution is too application specific. THREAD group and Fairhair Alliance are typical examples for such activities. The interaction can be modeled as shown in **Figure 10**.

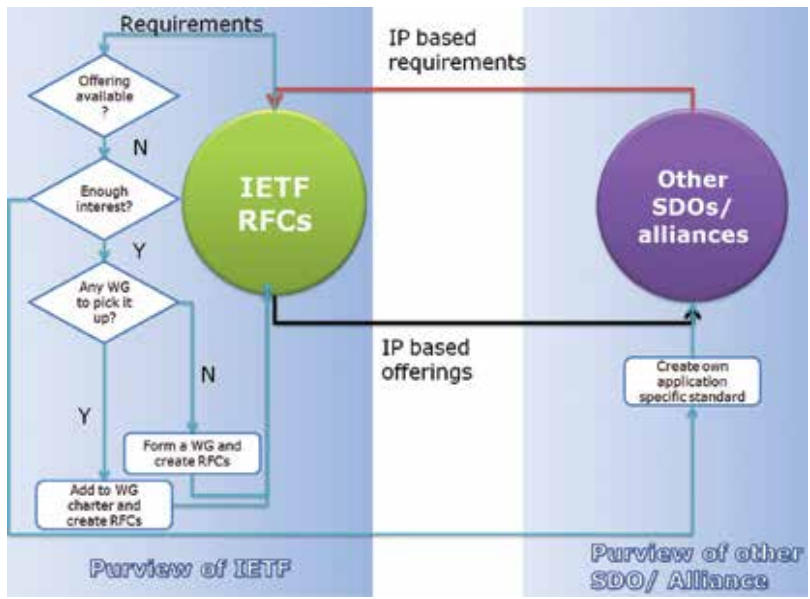

**Figure 10.** Interaction between IETF and other SDOs/alliances.

## 4. Role of local SDOs in IoT/M2M standardization

Most of the leading countries such as India, China, Korea, Japan, Europe, the USA, and so on [26–30] have their local SDOs to cater their local needs. While the global SDOs, such as ITU, IETF, and oneM2M, provide a uniform platform for the entire world, many times different geographies have conflicting requirements. Hence, these local SDOs play a major role into the success of IoT/M2M. The local SDOs are expected to adapt to the global recommendations and tailor them suitably to their requirements. Hence these local SDOs should play a dual role; (i) they should be able to provide globally interoperable ecosystem for seamless connectivity; (ii) they should also provide an equal opportunity to their local players, such as start-ups, small-scale industries, and academia, to compete in the local as well as the global IoT/M2M market.

Every country or geography in the world has very different scenarios and problems that need to be solved. Each local SDO focus is to provide requirements and standards to address the unique problems faced in the corresponding geography. However, at the same time, they should ensure that the technology used in providing solution to these use cases is not developed and deployed in isolation creating a risk of isolation of the very deployment from the rest of the IoT/M2M ecosystem.

## 4.1. India-specific efforts

Being one of the largest democracies in the world, India is expected to be the biggest consumer for the IoT. However, it must be noted that India is a very unique geography unlike the USA and Europe. India has more than 1.3 billion population which is approx. three times of that of the USA and equal to that of China, while the population density is among the highest of the developed countries. Moreover, there is a huge requirement to have affordable and low-cost solutions for any technology to be successful in India. Hence, the IoT use cases for India are also significantly different. This brings India-specific efforts for standardizations.

The economic condition of a developing country like India is very different from Europe or the USA. The economic ecosystem needs a lot of support from local SDOs and the government to create an impact on the IoT standardization. In the absence of that, there is a huge risk of getting obsoleted of local players in the IoT arena. The IoT requires to interoperable globally, and there is potential risk that large companies are likely to drive the entire standardization process, product development for IoT ecosystem. This puts the local manufacturers and start-ups into a great risk. Hence, the local SDOs should provide them an equal opportunity to contribute and adapt to the global standards and make a mark on the face of it.

### 4.1.1. IoT standardization with TSDSI

Telecommunication Standards Development Society, India (TSDSI) is an Indian SDO formed by the government of India to promote telecom standards in Indian geography. TSDSI is one of the eight organization partners of 3GPP and oneM2M for building cellular and IoT-related standards. TSDSI is an Indian counterpart of other SDOs such as ETSI [31] in Europe and ATIS in the USA. Currently TSDSI has transposed the 3GPP and onem2M specifications as TSDSI technical specifications. TSDSI also represents India in ITU-R and ITU-T for consolidating international efforts in the area of the IoT and telecommunications. TSDSI has studied various verticals important for India and consolidated all these use cases and requirements in technical reports, published in the public domain [32].

TSDSI has contributed to Low Mobility Large Cell (LMLC) standard requirements to ITU-R. LMLC is a very unique requirement of developing geographies like India with large rural populations where a vast majority of people do not even have basic networking infrastructure available. Unlike urban geography, rural areas have relative low mobility; however, they are spread over large geographic area and hence require the larger cells to cover that entire region. The members of TSDSI have provided several other key contributions to 3GPP such as TDD-based scheduling standard in future 5G networks [33].

### 4.1.1.1. Handling IoT deployment challenges

TSDSI has transposed 3GPP specifications including NB-IoT as TSDSI specifications. TSDSI is also considering transposing and adopting oneM2M specifications as one of the IoT deployment recommendation. However, there are significant ongoing efforts to study the usefulness of oneM2M specifications and tailor it according to suit Indian subcontinent requirements. Indian companies like TATA Communication are also working on creating one of the largest IoT deployments. However, it is essential that the government of India provides uniform policy to avoid silos of IoT deployment and ensure the interoperability of all the deployments in India within as well as globally.

| TSDSI | Current activities | Roadmap | Comments |
|---|---|---|---|
| Connectivity and interoperability | Same as oneM2M and NB-IoT | Planning to include India-specific requirements into NB-IoT and oneM2M | TSDSI is Type 1 organization partner of 3GPP and oneM2M |

### 4.1.2. IoT standardization with GISFI

The Global ICT Standardization Forum for India (GISFI) is an Indian standardization body active in the area of Information and Communication Technology (ICT) and related application areas, such as energy, telemedicine, wireless robotics, and biotechnology. It has been actively involved in defining various use cases related to IoT and defining a generic architecture keeping India-specific requirements into consideration. It has liaison agreements with ITU, ETSI, 3GPP, and other international SDOs in the field of the IoT and 5G communications. The IoT reference architecture under GISFI is explained in [34]. It defines the following layers as a part of its generic architecture: (i) **IoT device layer** includes individual sensors, network-enabled objects, and capillary networks consisting of data sources that are near to the physical environment. (ii) **IoT gateway layer** consists of IoT gateways and connects to the IoT service platform layer through the core network; device and gateway layer functionality can coexist in a single device. (iii) **IoT service platform layer** defines different IoT service abstractions that can be used by multiple applications. (iv) **IoT core network** is envisaged to be predominantly an IP-based network. IP connectivity could be supported over multitudes of telecommunication infrastructures such as DSL, cellular networks (2G, 3G, 4G), and so on.

**GISFI also identifies three reference points at the interfaces of these layers as follows:** (i) I1, interface from device layer to gateway layer; (ii) I2, interface from gateway layer to service platform layer through IoT core network; and (iii) I3, interface from service platform to layer-specific vertical applications.

### 4.1.2.1. Handling IoT deployment challenges

GISFI's aim is to harmonize the standardization effort within the Indian market and work closely with government or regulators, users, network providers, manufacturers, and academia and research communities. GISFI is closely working with telecom operators to decide the communication framework in addition to the frequency of operation and other communication aspects. With a generic IoT architecture proposal, it is ensuring the interoperability aspects to a certain level. IoT security and privacy framework is being framed through a separate work group, and the findings of this group are being updated with the industries and the government. With the definition of new use cases which are India specific in nature, business aspects in addition to societal aspects are also being covered to certain extent. However, the major problems we see with GISFI are as follows: (i) lack of a concrete architecture which is binding to industries and the government and (ii) difficulty in translating the India-specific requirement to standardization.

### 4.2. Specific challenges for the local SDOs

In this section, we discuss the country-specific challenges and analyze from India's perspective. India is a very unique geography in terms of population and population density. This poses unique challenges for any technology to be successful in India. People in India generally use their smart devices for longer duration of time than other parts of the world. Moreover, operators face a tough call on RoI. Therefore, for the benefit of both the parties, i.e., the user community and the operators, SDOs need to emphasize on backward compatibility while creating a new standard or adapting any existing standard. This can help in improving penetration in both rural and urban areas.

Another major concern for India is to promote the use of green and renewable energy. The pollution in India is in an alarming point. Apart from these, healthcare and education are other major areas where IoT can play a major role. The SDOs and government need to work together and build policies to ensure the maximum possible use of green energy keeping environmental issues in consideration and at the same time support various use cases. At the same time, the overall cost of the technology and devices must be kept under check. India has the advantage of the scale which makes it possible for the operators and providers to recover their RoI even with small average revenue per user (ARPU). Indian government should also keep the interest of local start-ups and manufacturers under consideration.

## 5. Economics of IoT standardization in India

The Internet has become the core of the connectivity among heterogeneous devices across the globe. It is important to create and adhere to global standard implementation of the outcomes of the research for advancing the telecommunication technologies. This helps in overall growth of the economy as standardization helps business through easy interoperability, ensuring interconnectivity, compatibility, quick time to market, etc. Thus, standardization helps the economy of scale which leads to overall economic growth as depicted from **Figure 11(a)**. So far, India has remained as a large consumer of technologies rather than a contributor. The stray contributions have been largely from the Indian counterparts of foreign corporates. India has been deploying the readily available global standards which are not created with India-specific requirements. However, the advent of IoT creates a renewed opportunity for Indian organizations, irrespective of public or private, to take lead in the standardization arena. It is important that at this juncture, India takes up the lead in identifying the problems to solve for a better living.

This can further enable the value chain behind a self-sustaining ecosystem of local indigenous innovations, productizing of the innovation outcomes, and standardization of the same. Though standardization can happen at the local level, it must impact the global SDOs to maintain compatibility at a global level, create a global business value, and also uphold

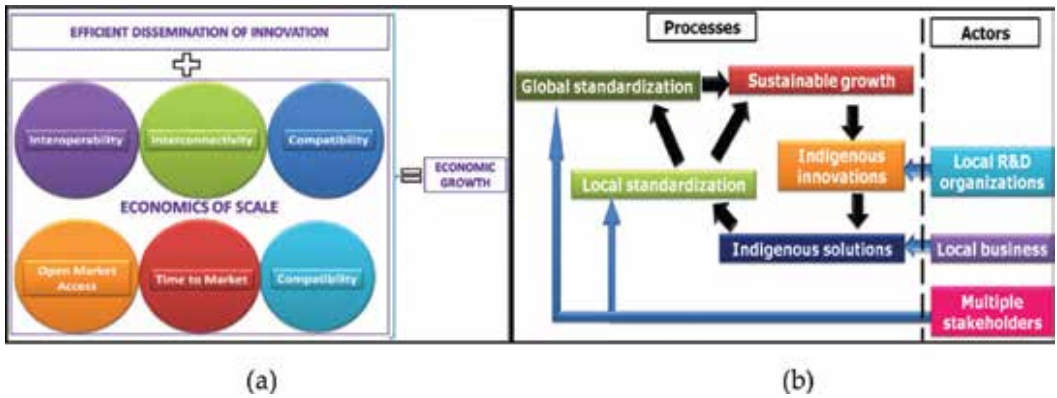

**Figure 11.** (a) Standardization and economic growth and (b) conceptual ecosystem for sustainable growth in India through IoT standardization.

India's requirements in global arena. This ecosystem should get all the stakeholders with common area of interest to complement their national peers with collaborative standardization and finally add to economic growth of the country through indigenous intellectual properties. The expected chain of activities in the conceived ecosystem is depicted in **Figure 11(b)**. It is encouraging to see regional standard bodies like GISFI and TSDSI being formed. Such initiatives create a platform for Indian stakeholders to join hands.

## 6. Open areas for standardization in the IoT

As we discussed in previous sections, there are multiple ongoing efforts for IoT standardizations. Different standard bodies and various independent alliances are targeting different areas of the IoT ecosystem, e.g., IETF is focusing on Layer 3 to Layer 5 protocols and applications. 3GPP and ITU are focusing on the radio and MAC aspects of the ecosystem. 3GPP has proposed NB-IoT standard in its release-13 for small data transfer for IoT devices. oneM2M is focusing on the service layer aspect of IoT/M2M with a vision that all the M2M devices can interoperate seamlessly. Since IoT is a completely heterogeneous system both in terms of applications and technologies, there are several challenges which need to be addressed before we can have seamlessly deployable IoT ecosystem.

In addition to the standard bodies, Indian and western academia are involved in various state-of-the-art solutions specific to lightweight protocols for IoT data and device security, user and data privacy, green energy along energy harvesting, multimedia multicasting and broadcasting, adaptive coding for multimedia communication, SDNization of application and networks, and so on. The IT service industry is also focusing on the application API standardization for seamless access across heterogeneous device and networks.

**Security and privacy:** IoT systems are able to gather sensitive data about the consumers, and companies are already using lot of Machine Learning (ML) and Artificial Intelligence (AI) tools to extract information about their consumers for their marketing purposes. Elaborate systems and policies need to be formed to provide guidance about the exposure and use of private information along with the technology enhancement to ensure that such data are not compromised and mishandled by the malicious users.

**Interoperability:** There are several efforts that are going in parallel to capture the multibillion market of IoT. This has a risk of creating an ecosystem which is fragmented and developed in silos and is not able to interoperate with each other. We should have learning from the way the Internet has been developed, and unlike the fragmented development and patching of the Internet, we should provide elaborate policy guidance to curtail such fragmented development of IoT ecosystem.

**Reliability:** With the advent of IoT and 5G, society is emerging into an always connected society. The services such as healthcare, education, and connected cars are made available through the technology. This requires that underlying technology and the applications are utmost reliable, i.e., 99.9999% or better reliability is required.

**Agility and scalability:** The future applications and network need to be both agile and scalable to user demands and operations. Operators must be able to scale up and down dynamically without sacrificing the QoS, security, and reliability. The service providers must be able to deploy applications and services which can adjust themselves to the changing network conditions and use case requirements. Moreover, this all should be possible in a cost-effective way. Hence, it is expected that virtualization of resources and machine learning and AI-based predictions are used. SDNization of network and application can be used to predict the ever-increasing demand of massive data volumes.

IoT is same or more heterogeneous than the Internet is; hence it is not a hyperbole to call the IoT as "network of network of devices." We have witnessed in the past that the Internet has faced tremendous challenges due to unbounded, unplanned, and unregulated growth. This leads to significant inefficiency and underutilization of resources in the Internet deployment. Hence, it is imperative that all deployment of IoT should be well coordinated, supervised, and bound with the proper policy from the government and standards from the SDOs of the world. Such a coordinated effort only is able to ensure that the future deployment is efficient, interoperable, reliable, as well as seamlessly connectable to any other technology.

## 7. Conclusion

With exponential increase in the number and types of smart devices over the coming years, IoT poses a major challenge for the world in general and regulators in particular. One of the biggest challenges, upon which the eventual success of IoT depends, is the development of interoperable global standards. However, IoT standards today are still wide open—in device, protocol, and software level as there are no existing global validated standardization frameworks. Without enforcement of standards, the value and commercial viability of IoT will have difficulty to reach its full potential.

This chapter has highlighted various ongoing global standardization efforts along with India's contribution to these efforts besides the unique aspects of Indian geography. To make IoT and 5G, the lifeline of IoT networks, successful in India, it is important to identify the right use cases along with the right policies of deployment while keeping the cost of the technology affordable to rural Indian population along with requirement drivers for massively large-scale deployment. This is very different from the other developed geographies like the USA and Europe where only improved quality of experience may be enough for the success of the technology. This requires that India must increase its participation in global standardization process and push India-specific requirements into standard building processes so that the Indian use cases and need of Indian's are addressed. Needless to say, similar standardization efforts in other emerging market economies also need to be synergized at a global level in addition to efforts in the developed economies.

## Author details

Arpan Pal, Hemant Kumar Rath*, Samar Shailendra and Abhijan Bhattacharyya

*Address all correspondence to: hemant.rath@tcs.com

Embedded Systems and Robotics, TCS Research and Innovation, India

## References

[1] Bandyopadhyay S, Balamuralidhar P, Pal A. Interoperation among IoT standards. Journal of ICT Standardization. 2013;**1**(2):253-270. DOI: 10.13052/jicts2245-800X.12a9

[2] Pal A, Balamuralidhar P. IoT Technical Challenges and Solutions. Artech House; 2016. ISBN-13: 978-1630811112

[3] Ericsson. White paper "More Than 50 billion connected devices". 2011

[4] Rose K, Eldridge S, Chapin L. The Internet of Things: An Overview, Internet Society Document; October 2015

[5] Pal A. Internet-of-Things, making the hype a reality. IT Professional Magazine. IEEE Computer Society. 2015. pp. 2-4. DOI: 10.1109/MITP.2015.36

[6] Tschofenig H et al. Architectural Considerations in Smart Object Networking. Tech. no. RFC 7452. Internet Architecture Board (IAB); Mar. 2015. Available from: https://www.rfc-editor.org/rfc/rfc7452.txt. DOI: 10.17487/RFC7452

[7] Balamuralidhar P, Misra P, Pal A. Software platforms for Internet of Things. Journal of the Indian Institute of Science. 2013;**93**(3):487-498

[8] TCS Connected Universe Platform (TCUP) [Internet]. Available from: https://www.tcs.com/tcs-connected-universe-platform

[9] Hartke K. RFC 7641: Observing Resources in the Constrained Application Protocol (CoAP) [Internet]. September 2015

[10] https://azure.microsoft.com/en-in/services/iot-edge/

[11]  The Fairhair Alliance [Internet]. Available from: https://www.fairhair-alliance.org/

[12]  Thread Group [Internet]. Available from: https://threadgroup.org/

[13]  ITU-T. Overview of application programming interfaces and protocols for the machine-to-machine service layer (Y.4411/Q.3052). February 2016

[14]  IEEE Standards Association – Internet of Things [Internet]. Available from: http://standards.ieee.org/innovate/iot/

[15]  IEEE-IoT. Towards a definition of the Internet of Things (IoT), revision 1. May 2015

[16]  IEEE Standards Association. Internet of Things related standards

[17]  LTE Evolved Universal Terrestrial Radio Access (E-UTRA). General Physical Description

[18]  AnIntroductiontoNB-IoT.https://www.link-labs.com/blog/overview-of-narrowband-iot

[19]  Narrow Band IoT (NB-IoT). https://cdn.rohde-schwarz.com/pws/dl_downloads/dl_application/application_notes/1ma266/1MA266_0e_NB_IoT.pdf

[20]  IPv6 over Low Power Wide-Area Networks (lpwan) [Internet]. Available from: https://datatracker.ietf.org/wg/lpwan/about/

[21]  LoRa Alliance [Internet]. Available from: https://www.lora-alliance.org/

[22]  LoRA Alliance: Technology [Internet]. Available from: https://www.lora-alliance.org/technology

[23]  Sengul C, Kirby A, MQTT-TLS profile of ACE [Internet]. January 2017. Available from: https://tools.ietf.org/html/draft-sengul-ace-mqtt-tls-profile-00

[24]  TS 0001. oneM2M Functional Architecture. http://www.onem2m.org/images/files/deliverables/Release2/TS-0001-%20Functional_Architecture-V2_10_0.pdf

[25]  TS 0004. oneM2M Service Layer Core Protocol. http://www.onem2m.org/images/files/deliverables/Release2/TS-0004_Service_Layer_Core_Protocol_V2_7_1.zip

[26]  Telecommunications Technology Association of Korea (South Korea). www.tta.or.kr/English/

[27]  Association of Radio Industries and Businesses, Japan. https://www.arib.or.jp/english/

[28]  Telecommunication Technology Committee, Japan. www.ttc.or.jp/e/

[29]  Alliance for Telecommunications Industry Solutions. USA. www.atis.org/

[30]  China Communications Standards Association, China. www.ccsa.org.cn/english/

[31]  European Telecommunications Standards Institute. www.etsi.org

[32]  Technical Reports and use case documents. http://www.tsdsi.org/main/tr/

[33]  5G India forum. https://coai.com/5g_india_forum

[34]  Technical specification—IoT platform. Dec 2013. http://www.gisfi.org/wg_documents/GISFI_IoT_201312438.doc

# Cooperative Human-Centric Sensing Connectivity

Albena Mihovska and Mahasweta Sarkar

Additional information is available at the end of the chapter

**Abstract**

Human-centric sensing (HCS) is a new concept relevant to Internet of Things (IoT). HCS connectivity, referred to as "smart connectivity," enables applications that are highly personalized and often time-critical. In a typical HCS scenario, there may be many hundreds of sensor stream connections, centered around the human, who would be the determining factor for the number, the purpose, the direction, and the frequency of the sensor streams. This chapter examines the concepts of HCS communications, outlines the challenges, and defines a roadmap for solutions for realizing HCS networks. This chapter is organized as follows. Section 1 introduces the concept of cooperation in information and communications technologies (ICT), and in the context of IoT. Section 2 discusses cooperation in the context of the personal and extra-personal user space and identifies the remaining open challenges and requirements for realizing the benefits of this approach to enabling more resources and services in a hyper-connected society. Section 3 defines a roadmap toward realizing simple, efficient, and trustable systems based on advanced technologies combining security, cloud, and IoT/big data technologies and outlines the challenges related to this vision. Section 4 concludes the chapter.

**Keywords:** human-centric sensing, smart connectivity, multisensory communications

## 1. Introduction

Information and communications technologies (ICT) have progressed rapidly in this millennium for people to communicate and exchange information using multimedia (speech, video/image, text), and the same has extended to Internet of Things (IoT) and machine-to-machine (M2M) and machine-to-human communication (M2H). Propelled by the explosive growth in IoT, this trend is only going to accelerate in the years to come with new inventions in the area of human-IoT interaction technologies to deliver powerful engaging and intuitive experiences. The cyberspace of the future would rely on multisensory communications for a virtually rich personalized communication experience involving human-to-human (H2H), human-to-machine (H2M), and M2M interactions. Concepts like artificial intelligence (AI), machine learning, crowd and cloud computing, virtualization, and quantum computing (QC) will involve the human as a critical computing and interpreting node in a distributed computing architecture, pushing the concept of IoT toward a vision of Internet of Beings (IoB).

Cooperation between the human nodes is a powerful generator of streams of data that with the advances in multisensory communication will evolve to become complex and diverse in terms of the type of content, size, context, value, purpose, and so forth.

## 1.1. The concept of cooperation

Cooperation in communication networks is not a new concept, and its applicability and essence have evolved jointly with advances in ICT. Cooperation has been widely researched in the context of cellular and cognitive communications [1–6] for the purpose of optimizing network performance, planning, and deployment, as a way to enhance network capacity, release extra resources, and enable the flexible use of the frequency spectrum. Cooperation in the context of sensor network management has been studied as a way for energy-efficient data transmission [7], for improving the tolerance and dependability of the network [8], for optimal path discovery [9], and for enabling trustworthy node relationships [10]. Most recently, cooperation has been explored to enable that diverse IoT applications share resources among each other in order to enhance their functionalities and improve the level of services they deliver [11, 12]. There are an ever-growing number of sensors that automatically communicate to the Internet without human intervention, and these are essential in allowing people to acquire knowledge about their environment and determine how to use it for their needs. The data generated by human-wearable/mobile devices is another key source of information about the person, enabling the personalization of the IoT applications. However, there are still the challenges of the limited spectrum availability and energy limitations to deploy practical applications fully. With the increased complexity of data, the requirements for more processing power and longer battery life have stimulated the research in the direction of energy harvesting based on cooperative transmission as a way to increase the reliability and battery power of wireless devices [13, 14].

**Figure 1** shows an example of interconnected devices in a home environment cooperating for the generation of data that would be streamed in support of eHealth applications [15]. All the devices are connected to the home PC for collecting and processing the home signals to extract the care recipient's context. The primary user interface (UI) device is a large touch screen. An optional smartphone is the secondary UI device, also facilitating the collection of data from consumer devices, such as a wearable activity/sleep monitor (i.e., wearable device) and a sleep monitor. Two other consumer devices may form a Zigbee mesh network for socket sensing and controlling, and a gateway and lamps would be used for lighting controlling. Audiovisual sensing in the living room may be facilitated by the motion sensing input devices as in **Figure 1**. Medical sensors would be used for monitoring vital body signs, which are connected via Bluetooth in the scenario shown in **Figure 1**. These are meant to provide data to the patient at given times every day or during exercising at home. The environmental monitoring sensors would communicate via Zigbee radio with custom-packaged sensing modules.

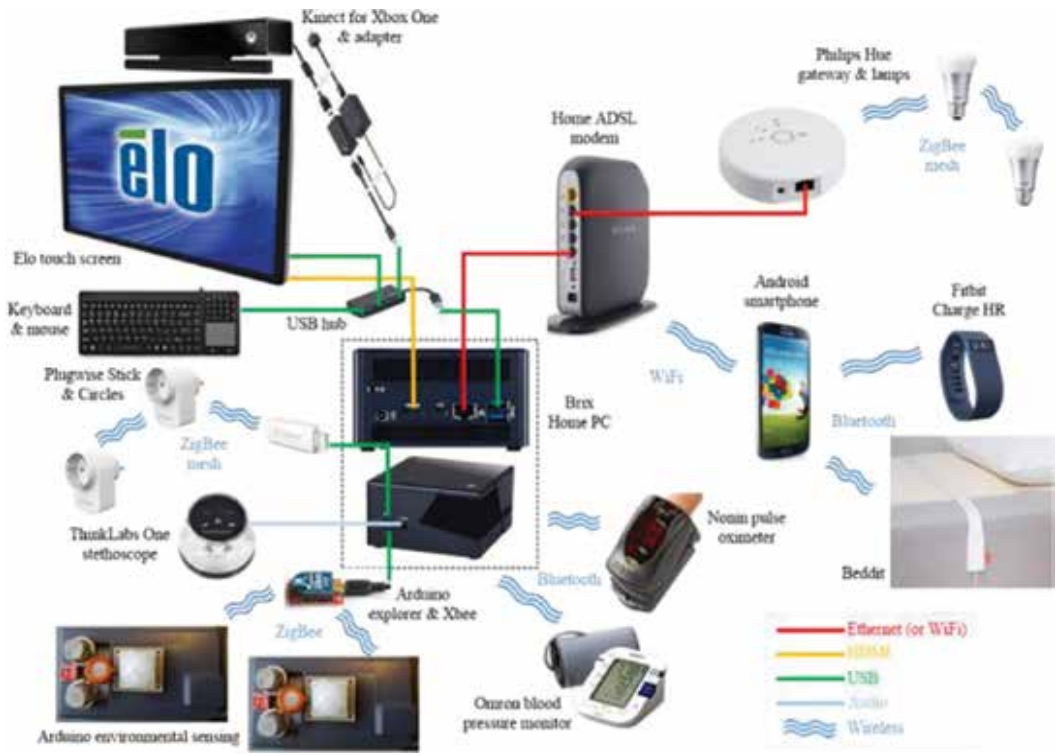

**Figure 1.** Cooperation approach to sensing within a home environment.

The cooperation scenario shown in **Figure 1** has been implemented as part of an eHealth platform developed in [15], and the data collected by the devices shown would be quite diverse requiring different approaches to their processing. On the one hand, there are the data extracted from the environmental signals to model the expected sensor values and understand deviations from those. Much more elaborate processing would be needed to distinguish the heart and lung signals obtained through the stethoscope and to measure the metadata related to these two organs. On the other hand, there are data associated with the voice commands (handled by the Kinect software) and face tracking and analytics. An enhanced face tracking system based on Kalman filters would also boost the processing speed performance, regardless of the visual complexity of the scenario, which could be critical in an eHealth-related scenario [16].

All the sensors in the scenario of **Figure 1** are cooperating in order to transfer their data to the cloud. Because of the bulkiness of some of the collected data, they would be processed locally by different algorithms implemented in a device gateway [17]. The resulting metadata would be indexed in a local database, from which they would be streamed via a remote proxy to the cloud. This process is shown in **Figure 2**.

In order to deliver personalized applications, cooperation is required between the IoT sensing environment and the data management environment (i.e., the cloud) to guarantee the quality, security, capacity, and reliability of information exchange, which would be sent via proper interfaces to the cloud environment. Within the cloud, algorithms will cooperate in order to extract meaningful information from the collected data and relate it to a particular user and use case. An example of a key requirement to a successful cooperation in the above context relates to trust when collecting and storing the data.

## 1.2. Cooperation and interoperability

An interconnected IoT world implies diversity and innovation, both in terms of devices, applications, technologies, and user needs. Different from the single-purpose wireless system standards, the wireless technologies delivering human-centric services have the hard task of operating an ever-growing number of heterogeneous networked devices that can communicate with each other or with people or robots for satisfying very dynamic and high-level user expectations. There are still no interoperability standards in place, although international standardization bodies, such as the International Telecommunication Union (ITU) [17] and the European Telecommunications Standards Institute (ETSI) [18], have dedicated an effort within the areas of the various study and working groups. Due to the multitude of stakeholders within an IoT scenario, to reach the potential for advancements in the area would only be possible through standards that facilitate interoperability among systems and devices, provide unqualified privacy and security, address the unique needs of the developing world, and leverage existing ubiquitous technologies such as social media applications and mobile devices.

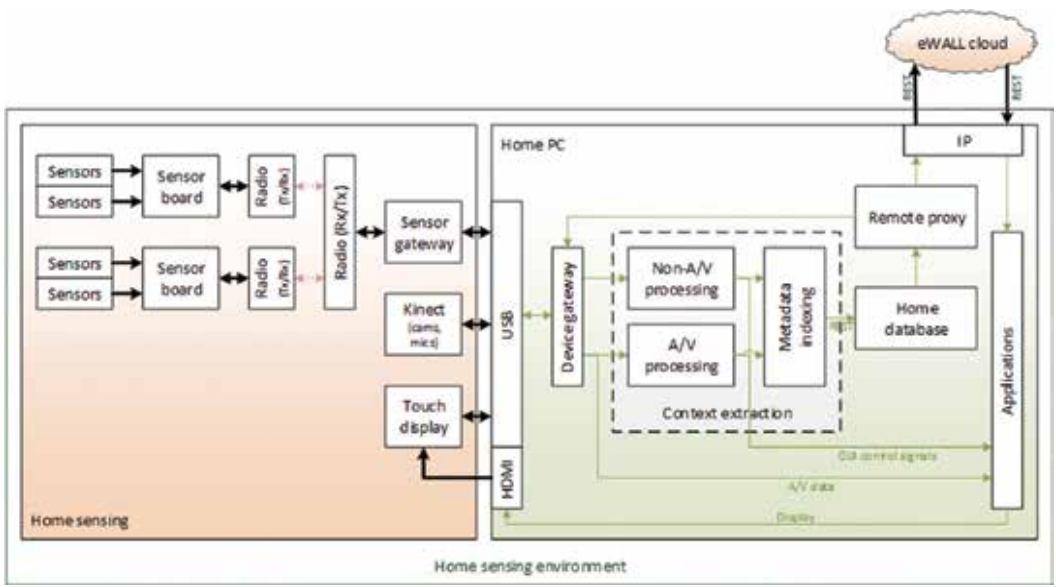

**Figure 2.** Cooperation of sensing devices for information processing.

Establishing ways to put in place cooperative behavior in a heterogeneous IoT landscape opens up a new road to deal with the challenge of interoperability.

## 2. Cooperation in a human-centric context

Cooperation in a human-centric context provides the unique opportunity to exploit the dynamics of dependencies between the individual users, which are heterogeneous and time-varying and thus create possibilities to address multiple scenarios simultaneously.

### 2.1. The human node capabilities

In a smart connectivity scenario, the human becomes a critical computing and interpreting node represented by the human-centric sensing network (HCS-N) that can enable the required scalability. The HCS-N is built around the user to enable information about the personal and extra-personal space and, thus, not only provides access to resources, services, and applications [19] but also promotes binding of the person to these two spaces, thus releasing an additional information to support many tasks of daily life. Such awareness gives the possibilities for the extraction of data on demand for the purpose of releasing additional resources or creating a trustworthy network for a particular application need. The HCS-N would be able to cooperate with other human nodes (i.e., HCS-Ns), within itself and/or with the environment depending on the need. The rapid formation of the HCS-N around the human sensing node would eventually evolve into a smart ubiquitous wireless network, relying mostly on short-range technologies (e.g., Zigbee, Z-Wave, IEEE 802.22 standard, etc.) and low-consumption node devices to enable multi-hop connections between self-configurable nodes.

Many approaches have been proposed for modeling the relationships between nodes in various contexts.

In [10], a social mathematical model was developed to find suitable node partners in large-scale wireless environments of dynamic topologies and resource-constrained nodes. Building the model, a set of parameters may be designed, taking into account the characteristics of each node, alongside with its capabilities. Thus, making it possible for the nodes to present themselves, to share their specifications and services, and evaluate to the benefits of forming a temporary ad hoc network. This concept may be easily extended to model the ad hoc networks composed of human nodes with the strict requirement of complying with minimum levels of security, privacy, and trust. In addition to choosing nodes with the right capabilities and functionalities, those nodes should also be reliable and trustworthy.

Game theory has been studied extensively as an approach to enabling cooperation. Craciunescu et al. [6] proposed a novel set of functions to model the node selection process in a scenario of cooperative wireless communications. A utility function would reflect the behavior and influence that a selected node may have on the quality of the cooperation to be established. The utility function could be adjusted to reflect on the parameters defining the cooperation scenario with the overall goal to maximize the overall network performance. This approach has a strong potential for HCS-N cooperation because of the ability to also assess security and reliability of the selection. The general block chart of establishing a cooperation in an HCS-N context is shown in **Figure 3**.

Reliability in the context of HCS becomes a multidimensional concept when we consider the evolving complexity of data content to be delivered through such cooperation. Immersive multisensory communications have been gaining momentum because of their strong potential to improve the quality of our life more than ever before. However, the existing technology still lacks solutions allowing for the detection, sensory analysis and evaluation methodology, coding/decoding, synchronization, transmission, and reconstruction over the ICT infrastructure of complex data associated with the olfactory, gustatory, and tactile experiences of a user. Regardless, it is an enabling trigger for ongoing groundbreaking research in preventive medical diagnostics (e.g., certain types of cancer, diabetes, etc., can be diagnosed early by means of smell) and in industrial process surveillance, waste reduction, and prognosis of critical process states. Enhancing the sense of presence/immersion, is of great benefit to various military and first responders training applications, can boost the efficacy of tele-surgery and numerous active ambient assisted living applications, and is well aligned with the cutting-edge research in augmented reality, smart mobile personal devices, sensors, and wearable devices. Replicating the sense of touch is essential in robotic technologies where certain forms of surgical operations may be needed. Touch has already become an essential feature of the success of a number of everyday devices from mobile phones to PC and tablets. The integration of the physical and chemical senses will enhance user experience and open to a large variety of application contexts ranging from education to environmental monitoring and from gaming and entertainments to healthcare. The applications would range from reliable detection of hazardous materials and low-cost and efficient environmental monitoring and food control up to advanced and noninvasive medical examinations and easily deployable threat detection systems.

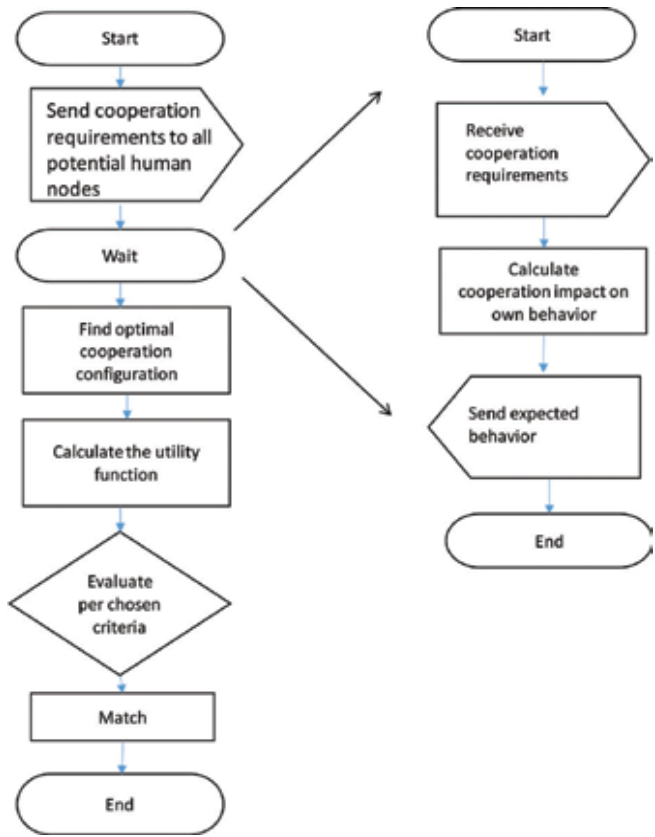

**Figure 3.** General steps during HCS-N cooperation establishment.

Such an immersive scenario defines very strict requirements for the levels of reliability, privacy, and security that are current approaches to cooperation lack and should be performed in parallel with research on cybersecurity.

The following key requirements for cooperation in a human-centric context can be summarized:

- Self-organization at the network level
  - Security, neighbor discovery, path optimization, authentication
  - Edge management and processing
- HCS-N aware data and service management
- Automatic, controlled establishment of HCS-N cooperation
- HCS-N-aware context management
- User satisfaction

## 2.2. Cooperation in the intrapersonal space

A good example of an intrapersonal space cooperation is the smart body area network (S-BAN), introduced as the smallest unit of the HCS-N in [19]. An S-BAN can be built by placing sensors on the human body and/or implantable devices within the human body as part of very advanced health monitoring and stimulating systems. A communication link between an implanted transmitter inside the brain with different body parts can be established via a brain-computer interface (BCI). The BCI provides a real-time artificial communication channel between the brain and external devices such as smart phones and wearable devices.

The scenario is shown in **Figure 4**. Multiple transmitters are implanted within the brain matter, and the signals are transmitted wirelessly to a receiver placed externally. By using multiple implantable transmitters in the brain, one would enhance the bandwidth though the transmission data rate is reduced [20]. Multiple tags would increase the quality and quantity of the electrocorticographic (EcoG) signals, but the probability of collisions would increase; therefore, special anti-collision algorithms should also be applied. Multiple transmitters can be used for implementing many channels responsible for moving or connecting many different body parts. Various access control schemes may be applied, but TDMA- or FDMA-based approach generally is preferred. Implanted transmitters do not transmit data all the time but respond to the neural signal generations and the distribution of the signals. Therefore, an enhanced MAC protocol considering the low power consumption of transmitters, bandwidth utilization, throughput enhancement, and minimizing the transmission delay is recommended to BCI applications.

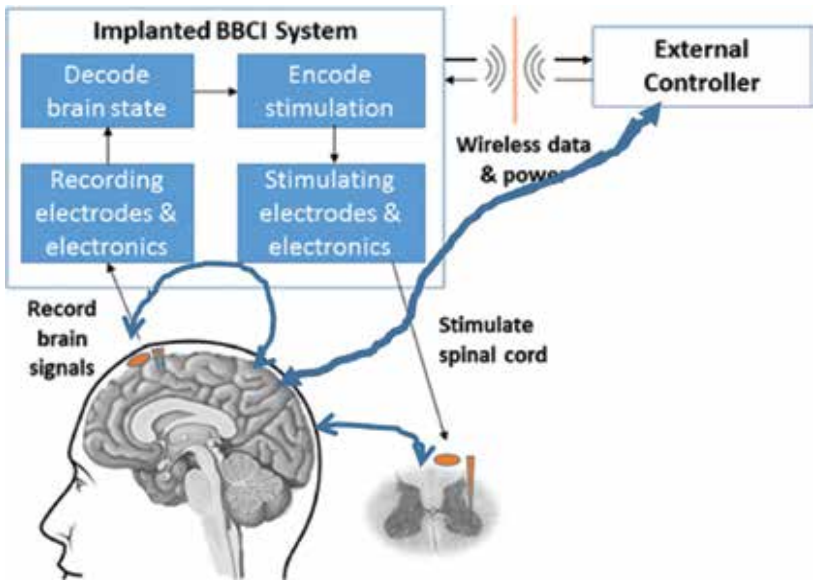

**Figure 4.** Intrapersonal cooperation by BCI.

BCI application transmissions need to be wireless, low power, and energy-efficient. Data to be perceived are usually about 200 Mbps to be within the constraints of human safety and tolerance. The signal should be received and interpreted in exactly the same way as it has been transmitted, and the full understanding of the radio signal transmission through the human tissues, blood, and other matters, which is still an open challenge, is essential to guarantee the patient's safety and signal quality. Both UHF passive RFID and UWB system could efficiently and accurately transmit up to millions of independent signals, accommodating all possible future BCI needs.

There have been substantial efforts for developing various signal propagation models for radio frequency waves from wearable and implantable devices for both narrowband and wideband communication systems. However, knowledge of the biological channel characteristics such as path loss, received signal strength, channel capacity, impulse response, and delay spread for networks of implantable devices transmitting data are very much open issues at the time of writing of this chapter despite the extensive research results available on the radio communication channel models in mobile communication networks. The general channel model for body area networks described by IEEE 802.15.6 working group has no focus on the particular part of the human body or human tissue [2]. Recently, ETSI has launched a study group on SmartBan that defined and specified a low-power physical and medium access control layers for SmartBans and studied the related coexistence issues of the radio environment in this scenario. In April 2015, ETSI TC SmartBan released its first two standard publications, i.e., technical specification (TS) 103,326 for an ultralow power PHY [21] and TS 103325 for a low complexity MAC [22]. Despite that the specifications are considered externally placed on the human body devices rather than implantable ones, it was concluded that more research is required on robustness in high-interference environment [23].

Due to the ethical difficulties arising with implantable devices, research in this area has focused on experimenting with simulated environments. An example of such an experimental framework based on the use of a wireless identification and sensing platform (WISP), which has been developed within the Wireless Laboratory at the Department of Electrical and Computer Engineering at San Diego State University is shown in **Figure 5**.

As it is not conventionally plausible to provide power to implantable sensors by batteries for longer duration, the setup used programmable passive RFID implantable tag (WISP), which is battery-free and uses power transfer mechanism to excite its circuitry, in contrast to active RFIDs which contain batteries. Separated Ziploc bag has been used for each WISP and placed close to each other over a chemical solution of a glycerin and saltwater (emulating the human tissue and blood) barely touching it, mimicking implantable electrodes. To replicate real-world scenarios, the WISP would be implanted inside the human brain surrounded by blood and tissue fluid, and the RFID antenna is sitting below the beaker as the RFID antenna is placed outside the brain located on top of the skull. **Figure 5** also shows how the RFID antenna is connected to the Impinj RFID reader through cable and how the controller (laptop) is connected to the RFID reader through an Ethernet cable. The plastic beaker with various titivations of glycerin and salt water from 1 to 4 cm allows for performing the analysis of the depth requirement of the sensor implantation. UHF RFID signals from multiple tags from various implant depths would be captured and analyzed in MATLAB in terms of received signal strength (RSSI) and signal-to-noise ratio, channel capacity, and path loss.

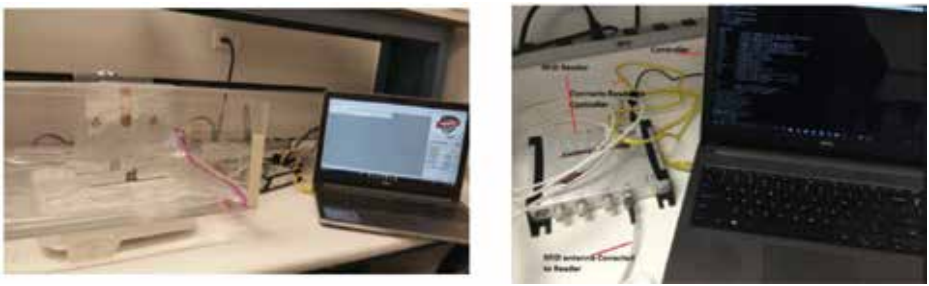

**Figure 5.** Simulated laboratory environment for signal transmission for BCI applications.

Important requirements to be considered during such scenario setup are the size of the sensor devices, which should be as small as possible, lightweight, and low maintenance. The BCI technology has a great potential in helping patients and hospitals by monitoring critical physiological signals and also for implantable camera-based diagnosis. Patient comfort can be greatly increased if the implantable and wearable biomedical devices are small, wireless, and with batteries that last long. The UWB and RFID radio technology can also be used in wireless endoscopy with higher-resolution images transmitted at lower power by the UHF passive RFID and UWB transmitter, also, for the diagnosis of Crohn's disease, celiac disease, benign and cancerous tumors, ulcerative colitis, gastrointestinal diseases, Barrett's esophagus, and several others.

## 3. Simple, efficient, and trustworthy convergence

Future development of applications and technologies is greatly pushed by the explosive growth of data and the growing number of interconnected devices. Because of the scale and complexity of the data to be generated, the concept of hyper-data (i.e., big) has been introduced.

In order to capitalize on the enormous business potential presented by such rapid digitalization, new interoperable architectural and platform solutions are required. The value created by collecting, communicating, coordinating, and leveraging the data from connected devices depends on evolving the key IoT technologies that relate to identification, sensors, localization, wireless and information exchange protocols, data storage and security, and their seamless convergence with advanced cloud computing concepts.

In a scenario of hyper-connectivity, supporting networks would need to meet the challenge of the generation of massive sets of streaming digital data, defined by volume, variety, veracity, velocity, and value [24]. This would require distributed application logic cloud infrastructures, able to handle the diversity of data sources and formats and to support the continuous nature of the data acquisition. Security becomes more difficult to address as it is difficult to develop a generic security strategy or model [25], also in view of the emerging "openness" of the networks. Streaming data, in addition, demands ultrafast response times from security and privacy solutions [26]. To realize a comprehensive hyper-data platform, a set of sophisticated and scalable analytic functions should be implemented at infrastructure, platform, and software level, while some of the key requirements to consider relate to response time, reliability, accuracy, and so on.

The following functionalities are minimum requirements for the support of the reliable gathering, exchange, and processing of hyper-data in order to take an intelligent real-time decision-making in relation to a given application:

- Support of 24/7 continuous collection of data from various sources and environments by means of an elastic wireless system based on a converged low power consumption of local area mesh-based communication network (e.g., based on IEEE 802.14.5) complemented by a wide-area mesh-based communication network (based on IEEE 802.22 or 802.16.n network) [27]

- Optimized single APIs capable to expose the collected data to a sophisticated in terms of prediction accuracy, sensitivity, and speed of response data processing platform

- Transformation of the exchanged data into an active decision related to a personalized user application by means of novel parallel/distributed data mining algorithms able to handle multidimensional datasets

- Protection of the collected, exchanged, and transformed data by means of on-the-fly deployments/positioning of security/privacy functions that may be enabled by a combination of software-defined networking (SDN) and network function virtualization (NFV)

- Trustworthiness of decision-making enabled by distributed ledger technology (DLT)

- Protection of data based on privacy-by-design approach

A conceptual vision for a platform able to handle the real-time analysis of large diverse and unstructured datasets acquired in a continuous manner is shown in **Figure 6**.

At the user equipment (UE) level, data would be collected from multiple (in the order of thousands and more) nodes located indoors at the user's home, office, car, or public spaces. These nodes can be battery-operated sensor devices or Internet-enabled personal devices. Functionalities enabling lower layer security to the IoT device or as protection of the physical infrastructure should be put into place at the UE level. The collected data may be processed at the edge level for real-time decision-making while keeping the data mining components at the remote level and distinct locations (e.g., at the cloud). This approach allows for isolation of the individual microservices, and this simplifies the security-related design. The edge component may implement distributed ledger technology (DLT) to enable trustworthiness between the various platform entities. The edge component will also implement some storage capabilities and will expose the API to the collected metastreams of macro-data. Other functionalities at the edge level are the processing and indexing of the micro-data, configurations, and exchange of control data, and decisions that would trigger alarms. The API should support both message-based and pull and push communication.

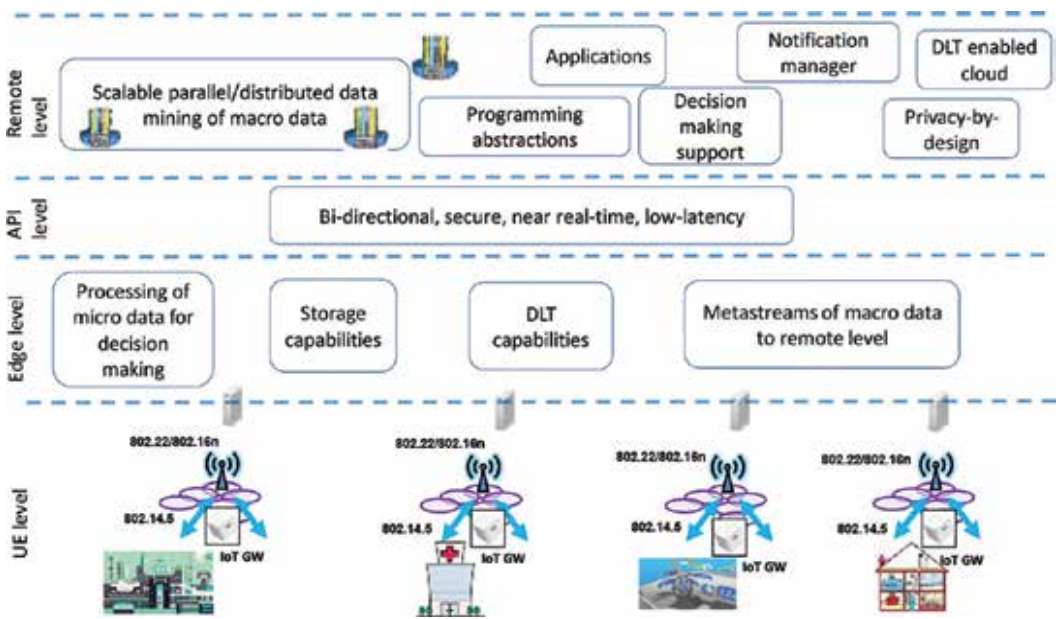

**Figure 6.** Conceptual vision of a converged IoT cloud platform for hyper-data.

By introducing an edge level, one can enable certain tasks, such as giving a fast response to a query sent by a user or securing the data transmissions between sensors and concentrators by defining a zero trust security at the entering points. Security can be implemented as a service by means of distributed networking solutions based on technologies such as software-defined networking (SDN), network function virtualization (NFV), network slicing, and evolved cloud computing paradigms. Blockchain and other distributed ledger technologies like Tangle and Hashgraph have recently drawn much attention due to their distributed nature and the possibility of not needing a Central Authority (CA) for establishing trust. Such technology can enable to implement a distributed ledger SDN that can reduce attack window times by allowing IoT forwarding devices to quickly check and download the latest flow-table rules.

The degree to enable privacy and protection of the data is expressed through the capability of a system to anonymize and pseudonymize data, which may be enabled by the privacy-by-design approach currently under standardization within ETSI and within the European Union Agency for Network and Information Security (ENISA).

An emerging trend is the increasing complexity of the data to be collected, prompted by research on how to involve all five human senses for an immersive experience. Such data can be enabled by new types of digital sensors able to replicate the human touch, smell, and taste and their synchronous integration with traditional sensors, which requires also a novel approach to extracting the contextual information carried by the multisensory data and technologies for reproducing the experience at the receiver's end.

There are a number of open critical challenges in order to be able to deploy the above technological advances as a converged and operational platform. The main difficulty of enabling convergence comes forth from the conflicting properties of the IoT and cloud environments. The IoT environment resides at the UE level and is typically resource constrained and location specific. The remote-level functionalities have plentiful of resources and are typically location independent.

A key number of open challenges have been summarized below:

- Collecting, exchanging, and processing of hyper-data without distorting the quality of the collected data and without compromising the personal aspects of the processed data

- Processing of hyper-data in a user- and application-centered manner, without overwhelming the user with affluence of information, which requires sophisticated data mining algorithms

- Supporting real-time processes

- Enabling resources from the cloud for the IoT applications and data residing at the physical infrastructure level, which requires novel functionalities and capabilities to be deployed at the edge level

- Enabling a unison of performance response/trade-offs between the wireless and cloud infrastructure, which requires flexible software solutions

- Interoperability of solutions

Standards, regulation, and open-platform solutions are key to the deployment and commercialization of any of the above research solutions toward achieving the vision of HCS-N cooperation and realizing its social potential. Common standards are essential to achieving interoperability, which in turn is essential for reliable and smooth operation of technological solutions across various deployment scenarios and for stimulating further innovation.

Currently, there is an effort within the ITU-T to standardize Blockchain and DLT, which is also closely linked to the standardization effort on ITU-T SG16 (multimedia), ITU-T SG17 (security), and ITU-T SG20 (IoT, smart cities, and communities). Another standardization effort is within the IEEE with the objective to evolve the mesh function in wireless networks as 802.16n, 802.15.4g, and 802.15.4e, as an integral part of a converged IoT/cloud/big data scenario.

## 4. Conclusions

This chapter introduced and explored the concept of HCS-N cooperation on various levels and in the context of a converged IoT-cloud-hyper-data scenario. This vision is of global importance for releasing the potential of robust interoperable technologies to deliver business and/ or societal applications. A scenario of hyper-connectivity requires a supporting platform that is open in nature and allows for its deployment under any type of legislative framework to realize in full the visions of smart cities and digital single market. The role of the human user, an ability to deal with an ever-increasing amount of sensors, smart objects and data, enhanced security and privacy, and trust are some of the key open research challenges to be resolved.

The emerging hyper-connectivity trend implies pervasive and exponentially increasing complex type of data traffic that pushes against the boundary of the power and design of current communication and information processing networks. The intensely high streams of wireless traffic and immersive data necessitate scalably to their continued growth of wireless and data management architectures, offering the required efficient processing techniques and capacity without an additional infrastructure expansion.

## Acknowledgements

Some of the research work on BCI presented in this chapter was supported by award number EEC-1028725 from the National Science Foundation (NSF). Authors would like to thank

Dr. Joshua R. Smith and Dr. Matt Reynolds, professors of electrical engineering at the University of Washington, Seattle, for their continuous support and assistance in the research.

## Acronyms and abbreviations

ICT             Information and communications technologies

IoT             Internet of Things

M2M             machine to machine

M2H             machine to human

H2H             human to human

SDN             software-defined networking

NVF             network virtualization function

HCS-N           human-centric sensing network

BCI             brain computer interfaces

## Author details

Albena Mihovska[1]* and Mahasweta Sarkar[2]

*Address all correspondence to: amihovska@btech.au.dk

1 Aarhus University-BTECH, Herning, Denmark

2 San Diego State University, San Diego, CA, USA

## References

[1]  David Alvarez Corrales L, Giovanidis A, Martins P. Coverage gains from the static cooperation of mutually nearest neighbours. In: Proceedings of IEEE Global Conference on Communications (GLOBECOM'16); December 2016. Washington, DC/New York: IEEE; 2016. pp. 4-6. ISBN: 978-1-5090-1328-9/16/

[2]  Baccelli F, Giovanidis A. A stochastic geometry framework for analyzing pairwise-cooperative cellular networks. IEEE Transactions on Wireless Communications. 2015; **14**(2):794-808. DOI: 10.1109/TWC.2014.2360196

[3]  Tanbourgi R, Singh S, Andrews JG, Jondral FK. A tractable model for noncoherent joint-transmission base station cooperation. IEEE Transactions on Wireless Communications. 2014;**13**(9):4959-4973. DOI: 10.1109/TWC.2014.2340860

[4]   Semov P, Poulkov V, Mihovska A, Prasad R. Self-resource allocation and scheduling challenges for heterogeneous networks deployment. Springer International Journal on Wireless Personal Communications. 2016;**87**(3):759-777. DOI: 10.1007/s11277-015-2640-7

[5]   Mihovska A, Luo J, Mino E, Tragos E, Mensing C, et al. Requirements and algorithms for cooperation of heterogeneous networks. Springer International Journal on Wireless Personal Communications. 2009;**50**(2):207-245. DOI: 10.1007/s11277-008-9586-y

[6]   Craciunescu R, Mihovska A, et al. Use of behavior and influence functions for relay selection in cooperative communications. In: Proceedings of IEEE Global Conference on Communications (GLOBECOM'15); December 2015. San Diego/New York: IEEE; 2015. pp. 6-10. DOI: 10.1109/GLOCOMW.2015.7414145

[7]   Weiwei F, Feng L, Fangnang Y, Lei S, Sojiro N. Energy-efficient cooperative communication for data transmission in wireless sensor networks. IEEE Transactions on Consumer Electronics. 2010;**56**(4):2185-2192. DOI: 10.1109/TCE.2010.5681089

[8]   Nikodem J. Autonomy and cooperation as factors of dependability in wireless sensor network. In: Proceedings of Third International Conference on Dependability of Computer Systems 2008 (DepCos-RELCOMEX '08); 26-28 June 2008. Szklarska Poreba, Poland, New York: IEEE; 2008. DOI:10.1109/DepCoS-RELCOMEX.2008.50

[9]   Vasilev V, Mihovska A, Poulkov V, Iliev G. Optimization of wireless node discovery in an IoT network. In: Proceedings of IEEE Global Conference on Communications (GLOBECOM'15); December 2015; San Diego; California. New York: IEEE; 2015. pp. 6-10. DOI:10.1109/GLOCOMW.2015.7414153

[10]  Kasabova S, Gechev M, Vasilev V, Mihovska A, Poulkov V, Prasad R. On modeling the psychology of wireless node interactions in the context of internet of things. Springer International Journal on Wireless Personal Communications. 2015;**85**(1):101-136. DOI: 10.1007/s11277-015-2730-6

[11]  Liang C-W, Hsu JY-J, Lin K-J. Auction-based resource access protocols in IoT service systems. In: Proceedings of 2014 IEEE 7th International Conference on Service-Oriented Computing and Applications (SOCA '14); 17-19 November 2014; Matsue, Japan. New York: IEEE; 2014. pp. 49-56. DOI:10.1109/SOCA.2014.50

[12]  Ranjibaran S, Mohammadi A, Manshaei MH. On cooperation mechanism in internet of things with multiple sponsors. In: Proceedings of IEEE Global Conference on Communications (GLOBECOM'16). 4-6 December 2016; Washington DC, New York: IEEE; 2016. DOI:10.1109/GLOCOMW.2016.7848973

[13]  Mulatu MA. Energy cooperation in communication of energy harvesting tags. AEU—International Journal of Electronics and Communications. 2017;**71**:145-151. DOI: 10.1016/j.aeue.2016.10.016

[14]  Zou Y, Peng J, Liu K, Jiang F, Lu H. Energy-efficient cooperative spectrum sensing for cognitive sensor networks with energy harvesting. In: Proceedings of Control and Decision Conference (CCDC); 28-30 May 2016. Yinchuan, China, New York: IEEE; 2016. DOI:10.1109/CCDC.2016.7531382

[15] Kyriazakos S et al. eWALL: An open-source cloud-based eHealth platform for creating home caring environments for older adults living with chronic diseases or frailty. Springer International Journal on Wireless Personal Communications. 2017;**97**(2):1835-1875. DOI: 10.1007/s11277-017-4656-7

[16] Bardas G, Astaras S, Diamantas S, Pnevmatikakis A. 3D tracking and classification system using a monocular camera. Springer International Journal on Wireless Personal Communications. 2017;**92**(1):63-85. DOI: 10.1007/s11277-016-3839-y

[17] International Telecommunication Union (ITU). http://www.itu.int

[18] European Telecommunication Standardization Institute (ETSI). http://etsi.org

[19] Mihovska A, Pejanovic M, Prasad R. Human-centric IoT networks. In: Dixit S, Prasad R, editors. Human Bond Communication: The Holy Grail of Holistic Communication and Immersive Experience. 1st ed. Hoboken: Wiley; 2017. DOI: 10.1002/9781119341451.ch5

[20] Al Ameen M, Ullah N, Chowdhury SM, Islam R, Kwak K. A power efficient MAC protocol for wireless body area networks. EURASIP Journal on Wireless Communications and Networking. 2012;**33**. DOI: 10.1186/1687-1499-2012-33

[21] Smart Body Area Network (SmartBAN). Enhanced ultra-low power physical layer, document ETSI TS 103 326 V1.1.1, April 2015. Available from: http://www.etsi.org/deliver/etsi_ts/103300_103399/103326/01.01.01_60/ts_103326v010101p.pdf

[22] Smart Body Area Network (SmartBAN). Low complexity medium access control (MAC) for SmartBAN, document ETSI TS 103 325 V1.1.1, April 2015. Available from: www.etsi.org/deliver/etsi_ts/103300_103399/103325/…/ts_103325v010101p.pdf

[23] Viittala H et al. ETSI SmartBAN system performance and coexistence verification for healthcare. IEEE Access. 2017;**5**:8175-8182. DOI: 10.1109/ACCESS.2017.2697502

[24] Bellazzi R. Big data and biomedical informatics: A challenging opportunity. Yearbook of Medical Informatics. 2014;**9**:8-13. DOI: 10.15265/IY-2014-0024

[25] Sfar AR, Natalizio E, Challal Y, Chtourou Z. A roadmap for security challenges in the Internet of Things. Elsevier Journal on Digital Communications and Networks. 2017. DOI: 10.1016/j.dcan.2017.04.003

[26] Cloud Security Alliance. Top ten big data security and privacy challenges, report, November 2012. Available from: https://www.isaca.org/Groups/Professional-English/big-data/GroupDocuments/Big_Data_Top_Ten_v1.pdf

[27] Harada H, Mizutani K, Fujiwara J, Mochizuki K, Obata K, Okumura R. IEEE 802.15.4g based Wi-SUN communication systems. IEICE Transactions on Communications. 2017;**EB100-B**(7):1032-1043. DOI: 10.1587/transcom.2016SCI0002

# The Internet of Things in a Smart Connected World

Hyun Jung Lee and Myungho Kim

Additional information is available at the end of the chapter

**Abstract**

The internet of things (IoT) constitutes a network of embedded devices that incorporate sensors and communication functions. The IoT is becoming one of the core technologies of the Fourth Industrial Revolution. This is because the IoT creates new values in the connected smart world by collecting big data, uploading data into clouds, and processing data in intelligent systems. The newly created values in intelligent systems differ from previously generated values that were based on the simple automated systems of the Third Industrial Revolution. In this chapter, we present a brief introduction of the IoT, which connects to the Internet through incorporating sensors and communication functions in various smart objects. In the IoT era, it is possible to create a networked smart world with powerful new services and products that create new values. As applications of the IoT, we introduce smart homes, smart electronics, smart connected cars, smart grids, smart healthcare, smart wearable devices, etc. In addition, we illustrate a specific IoT complex in a smart city as one of the smart connected applications of the IoT. Finally, we describe the predicted hyper-connected smart world that will be achieved through the IoT.

**Keywords:** internet of things, IoT, big data, cloud, intelligent systems, hyper-connected, smart world

## 1. Introduction

The internet of things (IoT) can connect the enormous offline world with people through the Internet. To achieve this, developed sensors are used to collect data from connected smart objects in the physical world. The gathered data are then uploaded into the cloud and become big data. These data are then integrated and utilized for the development of intelligent systems. Therefore, the IoT is one of the core technologies that is driving the Fourth Industrial

Revolution. Moreover, intelligent systems are continually being developed to process big data through the IoT. One of the special characteristics of these intelligent processing-based services and products is the capacity for customization and personalization. Consequently, new and potent values can now be created in smart systems using smart technologies, including the IoT, for a dynamic smart world.

In the early 2000s, during the advent of the IoT, radio frequency identification (RFID) technology was developed for logistic and inventory management applications. It was mainly applied to reduce product distribution and factory production costs. It was also utilized to trace the locations of products being delivered using location-based information systems. RFID technology then continuously evolved and developed into machine-to-machine (M2M) applications, which enable direct communications, monitoring, and controls between devices with a remote application infrastructure using communication channels. More recent M2M communication has expanded into the Internet. Specifically, utilizing wired or wireless communication channels between IP networks, it transmits data between humans and things and between things and other things, such as between household appliances. The Internet itself has also evolved into the IoT as the third generation of the Internet. The first generation of the Internet was developed to be enterprise oriented as the Internet of Computers (IoC), and the second generation of the Internet focused on customers as the Internet of People (IoP) [1]. Eventually, the internet of things (IoT) became an advanced form of M2M.

In 1999, the term "the Internet of Things" was first coined by Kevin Ashton [2]. Initially, the term referred to a type of computer network that can gather a lot, and a wide variety, of data from all of the physical things in the offline world. In order to obtain these data, these things have embedded sensors that record data and transmit them through connections to the Internet using IP networks. According to Kevin Ashton [2], the unique importance of the IoT comprises the following factors. First, the IoT was introduced as a new and powerful method to gather information that was not possible to be gathered in the past. From these tremendous amounts of collected data, the IoT enables the discovery of an almost infinite amount of previously inaccessible facts. Consequently, many manufacturing companies are now attempting to transform themselves from manufacturing to service-based companies, such as the General Electric (GE) Company. Predix is GE's cloud-based platform (PaaS) for industrial Internet applications that combine people, machines, big data, and analytics [3]. Applications of IoT technology are manifold and diverse. For example, government organizations can use discovered data extracted from the IoT to discover and prevent terrorist attacks. It is also easy to extend IoT-based systems due to good scalability and flexibility. In fact, IoT-based systems can be extended as much as the Internet itself has been extended. For example, new services that are based on IoT applications, such as IoT-based new car-sharing services or parking lot searching services, can be added to previously built systems, leading to the possibility of an infinite extension of the services. Indeed, with relatively little effort, systems and services can be expanded to create new and massively powerful values and opportunities. It is anticipated that the world will witness exponential expansion of diverse applications of the IoT in the near future.

Using big data collected, processed, and integrated through the IoT, intelligent systems have been developed to connect intelligent things. The IoT is thus closely related to intelligent systems because its development is based on the enormous amount of collected data. Generated data in the online and offline world can be propagated and shared in real time by anyone who needs or wants it. They can be also used and analyzed to provide products or services in business and public sectors. Using the IoT, it is possible to collect personalized data, such as at what time someone came to a physical location, in what he or she is interested in, and how long he or she remained in that place. After data analysis, customized and personalized services can be generated that are dynamically developed depending on users' analyzed characteristics, as well as requirements. To analyze data and generate a relevant service, it is also necessary to utilize intelligent applications and systems. For instance, Amazon's Dash Button device uses Wi-Fi and Bluetooth technology. It is enabled by a mobile phone, collects personalized data, and provides a corresponding customized service. To do this, the Button technology must be connected to the IoT, intelligent systems, big data, cloud, etc. It processes different contents each time that the button is pushed through the use of smartphone apps to send and receive information. In this way, it provides valuable customized content. In this system, big data technology is also requisite because data are accumulated each time that the button is pushed. By using this button system, it is possible to connect the online to the offline world by gathering so much data from the offline world. Therefore, the IoT is clearly different from previously developed electronics and technologies because it can create new opportunities, services, businesses, and platforms by connections and communications with the online to the offline world.

This chapter is organized as follows. In Section 2, we introduce the global growth of the IoT, trends in the global markets, and current and potential uses of the IoT in government and business sectors. Section 3 introduces sensors, networks, and service interfaces of IoT-based technologies and created services. In Section 4, we discuss IoT-based service applications, such as smart workplaces, smart factories, smart healthcare systems, etc., as well as an example of a smart city application and potential hazards of the IoT. Finally, we present some conclusions.

## 2. Global growth and trends of the IoT

### 2.1. Global market growth

In the global market, a variety of expectations exist regarding the internet of things (IoT). These expectations are related to how IoT devices will be connected, what are the services and values that will be created, how it can be used to increase a company's market share, etc. Although forecasts may vary slightly regarding the ubiquity of the IoT, it is obvious that it is growing dramatically. This rapid growth is attributable to the creation of new service markets, the expansion of the IoT devices, and the ease with which the IoT can be applied to industry, governments, products, and services. It is also clear that the growth of the service market, in particular, will comprise a major portion of the IoT market.

Concerning this IoT device market, Gartner predicts that, by 2020, the number of connected things will reach 25 billion and the service market will grow to USD $ 300 billion by the same year [4]. In Directions 2016 [5], the IDC forecasts that the number of terminals connected to the Internet will reach approximately 80 billion units in 2025. In addition, Cisco expects that, by 2030, there will be over 37 billion Internet units, the number of IoT devices will reach 50 billion, and the IoT will develop into the Internet of Everything (IoE) [6]. Gartner also predicts that China, North America, and Western Europe will be most active in adopting IoT devices, which will account for 67% of all Internet devices in 2017 [1].

In addition, the service market is also expected to occupy a large proportion of the IoT market. According to Gartner, in 2020, more than half of all existing Internet devices will connect with regular customers. Moreover, the number of customers using home automation systems and entertainment information will amount to 13 billion [7]. Cisco also predicts that 250 million people will be connected to the Internet by 2020. According to IDC, the expected IoT market will be USD $ 1.46 trillion by that same year [8]. These forecasts are based on the development of IoT-related products and the increase of related software and applications. Business and labor markets associated with data centers and management infrastructures will also be expanded to manage increasing data traffic. The consumer segment is predicted to comprise 5.2 billion units, accounting for 63% of the total installed capacity, leading to the ubiquitous use of IoT devices. Moreover, the business sector is anticipated to reach 3.1 billion connected units by 2017 [9]. To leverage the IoT, Mckinsey [10] defined nine key relevant environments: factories, cities, healthcare, retail stores, workplaces, logistics, transportation, housing, and offices. Economic effects range from USD $ 3.9 trillion to USD $ 11.1 trillion, depending on the availability of the IoT [10]. Machina Research (2015) predicts that the global market for the IoT will reach USD $ 1.2 trillion by 2022 [11]. In 2013, the market was USD $ 200 billion, but Machina Research forecasts that the market will grow 22% annually [11]. In addition, market size is expected to increase in the order of terminal, platform, and service by 2022. The average annual growth rate of service and platforms from 2013 to 2022 is expected to be 90.0 and 66.1%, respectively.

It should be noted that the growth of the service market is intimately related to semiconductor chipsets, communication modules, terminals, platforms including systems and solutions, and communication and service applications for device markets that support the IoT. From 2013 to 2022, each of these markets is forecast to have 19.2, 18.7, 8.8, 66.1, 17.0, and 90.0% of the compound average annual growth rate (CAGR). Global consulting firms, Gartner and IDC, forecast that the global IoT market will grow at a CAGR of 31.4 and 17.5% in 2013 and 2020, respectively. According to Cisco, the market value created by IoT corporations is expected to be USD $ 14.4 trillion over the next 10 years, and the public sector will be approximately USD $ 4.6 trillion. IDC expects that the IoT market will increase from approximately USD $ 2 trillion in 2013 to USD $ 7 trillion in 2020. Demands related to software applications, services, and devices for the IoT will also continue to increase. Consequently, in accordance with this demand, service markets from smart factories, smart healthcare systems, connected services, etc. [12] will also grow.

## 2.2. Trends in governments around the world

Currently, in order to realize economic and social innovations, governments and public sectors are also focusing on the internet of things (IoT) as a means of announcing policies that they want to promote. Through this, national governments around the world are rapidly establishing public goals, such as strengthening national competitiveness, improving people's quality of life, and taking actions that will catalyze major economic development. Certain large countries, based on developed information and communications technology (ICT), are strongly supporting the development of the IoT as a national project, including the USA, Japan, China, Europe, and South Korea. China, for example, established the Sensor Network Information Center in 2009 and the Intelligent Things Communications Center in 2010. Through these two institutions and others, China is announcing, establishing, and promoting various national projects. One of them is the "12-5 Plan for Development of the IoT" as part of the twelfth 5-year plan from 2011 to 2015 in 2011. It is building IoT pilot complexes targeted at facilitating the use of the IoT and the cloud as strategic measures [13]. The EU has also announced an implementation plan, including the 2009 IoT Detailed Treatment Plan. The UK is increasing IoT development funds and has announced that it is planning to invest $ 100 billion in the development of IoT technology by 2025. In 2008, the USA focused on building a hyper-connected network infrastructure to extend its existing communication infrastructure to the IoT. In early 2000, Japan accelerated national projects related to the IoT. In 2013, Japan implemented major ICT strategies, such as building smart towns, smart grids, and remote monitoring capabilities. In 2013, South Korea announced a comprehensive IoT plan for the development of technology and related market creation.

## 2.3. Global business trends

Many large global companies are actively participating in technology development and building ecosystems of technology focused on the internet of things (IoT) market. For example, Google has announced an ambitious plan to include the smartphone operating system "Android" on all major devices, such as televisions, automobiles, and watches. The company is also continuing strategic mergers and acquisitions (M&As) with related companies, e.g., the Nest company, which provides control services for room temperature, and Dropcam, an Internet surveillance camera manufacturer. Cisco has also led the IoT platform with IOx as an environment for the execution of IoT applications. In addition, Cisco recently announced that it had acquired Tail-f Systems, as a provider of network management solutions, and will acquire Assemblage, a real-time collaboration solution provider. Cisco, as the global market leader in networking equipment, has built an "Interloud" for the entire Internet of Everything (IoE) and is actively pursuing the IoT business through its "Smart Connected Communities" project. In addition, Qualcomm leads the open-source object Internet framework to connect devices with AllJoyn. General Electric (GE), as a leading equipment manufacturer, has announced that it will create new value with the "Industrial Internet Consortium" in connection with the IoT. In GE, the adopted IoT is available to provide new types of services or events. For example, GE's Predix collects data to monitor factories or systems, estimate possible faults during factory or system operations, and provide appropriate solutions for these faults [3]. AT&T is also working with Cisco, GE, IBM, Intel, and IoT network providers that connect all devices [12]. In recent years, M&As have also been increasing in global IT companies, such as Cisco and Google. This has been identified as a major activity that is preparing for the dominance of the IoT era. Therefore, it is important to ensure competitiveness in each service industry, including distribution, healthcare, security, and finance. It is also essential to possess the capabilities of an IoT value chain, such as content, platform, networks, and devices. For example, platform vendors, such as Microsoft and Oracle, are working to take advantage of their platforms, Microsoft Azure (Azure) and Java ME (Java Platform, Micro Edition), respectively, to prepare for a strong position in the IoT platform market. Moreover, Qualcomm, Intel, and other chipset vendors have focused their devices on the IoT network through AllJoyn and Quark. They are specifically focused on wearable devices and smart homes in the IoT market [12].

## 3. IoT technology and service

Sensors play a critical role in the internet of things (IoT). Sensors collect data on the Internet by smart devices, which are then used to upload information to the cloud. To achieve this, sensors are embedded in physical devices or exist in the form of external devices. Sensing technology is utilized to acquire a broad range of information, such as position, motion, images, etc. They can also collect surrounding environmental data, including temperature, humidity, heat, atmosphere composition, light, and sound. The IoT is also used to remotely control air conditioning, heating, and lighting. It is important to note that many physical sensors are also evolving into smart sensors with built-in standard interfaces for improving information-processing capabilities and applicable functions. Sensors can also include virtual sensing functions that extract specific information from the sensed and accumulated data. Moreover, virtual sensing technology can be implemented in the actual IoT service interface. Using multidisciplinary sensor technology, which is one-dimensional higher than existing independent sensors, it is also possible to extract more intelligent and high-dimensional information.

For the connection of sensors, the network interface plays the role of connecting physical network devices. For wired and wireless IoT networks, physical devices include wireless personal area networks (WPAN), Wi-Fi, 3G, 4G, LTE, Bluetooth, Ethernet, broadband convergence network (BcN), satellite communication, microwaves, serial communication, and PLC. These and other advanced communications systems enable the possibility for people, things, and services to become closely and rapidly connected.

The devices, such as sensors and network modules, are fixed on terminal devices for the collection of data. In other words, the development of sensor technology is essential to collect and extract data from objects. In addition, it is obviously necessary for network modules to communicate with these sensors, constituting an interworking of Internet communication, an application system, and an embedded system for providing user interfaces (UI). For activating the IoT, optimization and evolution of network technology are very important. The IoT can be connected to a network in a variety of ways. For example, things can be directly connected to a wireless network or connected to a smartphone through communication systems, such as Bluetooth. In the case of non-portable products, it can be connected to a protocol such as Wi-Fi, which is fixed in a certain place, such as a smart home or Industry 4.0.

It is important to note that the IoT service interface differs from traditional network interfaces. The primary aim of the IoT service interface is to offer value-added services through transformation, processing, extraction, and accumulation of sensed data. Additionally, it must make it possible to judge, contextualize, recognize, protect privacy, ensure security, authenticate, allow, discover, shape, etc., for the creation of services. The IoT service interface interlocks three major components: people, things, and services. For the application services to perform specific functions, the IoT must provide some interfaces for accumulating, processing, and transforming data for services, such as ontology-based semantics, open-sensor APIs, augmentation, virtualization, location identification, process management, open platform technology, etc.

The new types of value chains can be created based on the sensor devices, networks, and services in the IoT environment. This means that it can create new types of services that are based on different types of value chains on a data platform that is based on the particular device's sensing technology. The IoT contributes greatly to the derivation and creation of services based on connections between devices, things, and people. Ultimately, the created services, operations, and products will be based on convergence between data and services using data collected through sensors.

The processed data can also be accumulated in a cloud computing environment as big data. It is obviously critical to integrate data collected from distributed things through the IoT for the creation of advanced services. To achieve this, a data platform that can integrate distributed, collected, and aggregated data is requisite. This platform enables the creation of services that can generate value from different types of data. Service applications on such a data platform are introduced in the next section.

## 4. IoT service applications

The internet of things (IoT) is expanding the service market that is focused on public safety and distribution through merging with various industries. It is anticipated to be expanded to intelligent transportation services; social infrastructure, such as buildings and bridges; remote management services, existing healthcare, and smart energy-related fields. If the IoT becomes firmly established, its influence is expected to include everyday life, as well as all industries, due to the development and increased use of certain technologies, such as wireless networks, communication modules, sensors, and smart terminals. Furthermore, medical, transportation, manufacturing, distribution, education, and other fields will bring significant changes to existing processes and services.

### 4.1. Smart workplace

The smart workplace constitutes a new paradigm for working that will greatly increase collaboration, communication, and intelligent decision-making. It is based on connected, knowledge-based, integrated, and intelligent work facilities that depend on the new technology platform. One of the core technologies involved in creating smart work places is the IoT [7, 14]. Software applications that will be supported by the IoT have also been developed to support smart workplace environments, such as videoconferencing, new knowledge-sharing capabilities, and tracking the location of key mobile business assets.

## 4.2. Smart factory

The smart factory is not the automation-based factory system that existed in the Third Industrial Revolution, but is rather an intelligent system to support customization according to customers' requirements. This results in greatly increased production efficiency, more accurate and less expensive inventory systems, etc. Smart factories are developed by intelligent systems that are based on collected data from intelligent devices, integration of the collected data for the creation of services, and uploading the data to the cloud. In factories, it is important to interconnect facilities, such as overall systems, processes, and machines, in order to enable advanced services, such as innovation of production processes and cost reduction in supply chains. The IoT has also assumed a role in monitoring and maintaining infrastructure in smart factories.

## 4.3. Smart health

For smart health, hospital information systems usually use the internet of things (IoT) to monitor and connect patients, doctors, medical devices, and application systems, such as X-rays, using sensors. Some healthcare systems, such as IBM Watson, possess partnerships between people and systems. For example, instead of always requiring the presence of a medical doctor, in some cases, IBM Watson can treat patients by itself because it possesses expert knowledge and constitutes an intelligent system. In this type of case, the IoT is used to track, collect, and integrate remote data and the location of mobile assets in order to create and provide intelligent and advanced medical services. It is also applied to greatly increase the efficiency of healthcare infrastructure and resource usage. It is important to note that the developed applications can also substantially increase profits. Consequently, the more resources that can be saved, the greater the likelihood that new services will be developed. In fact, eight out of ten healthcare leaders (80%) stated that innovation has expanded since the advent of IoT use [7, 14].

## 4.4. Smart connected retailers

Nearly half of retailers worldwide allow network access on individual mobile devices to build the internet of things (IoT). This can create many new experiences and services for customers. For example, such applications of the IoT use a store's location service to provide customized information about products. It also assists in obtaining and retaining customers due to customization systems based on collected, accumulated, and processed data concerning individual customers. Currently, the retailing process is changing from a supplier-based value chain to a value-added value chain that is based on customer-centric services. Through the IoT, it is now possible to collect customers' personalized information, and the accumulated data can be applied to develop new types of services that can be based on intelligent systems. Since the IoT can facilitate more beneficial and customized services for individual customers, developing such services is currently very popular.

## 4.5. Smart farm

Recently, with smart farms, many countries and farmers are actively attempting to utilize the Internet, nano-based devices, and robot technology. In 2014, the National Weather Service and the Department of Agriculture established an open data policy and developed various smart agricultural services [15]. For example, Fujitsu grows hydroponic lettuce using its Internet technology platform (Akisai) and is developing it as a new type of farm. In agriculture, food seeds, seedlings, and information about them can be sent directly to consumers, allowing people to grow agricultural products themselves at home. Of course, commercial farmers can also use such services supported by the information provided by the IoT. In addition, by using the IoT, it is now possible to remotely monitor and control conditions for crops and farms. It can monitor and control essential factors, such as humidity, sunshine, temperature, etc.

### 4.6. Smart connected car

Unlike in the past, automobiles can be now viewed as a digital mobile software system and not as a machine with an engine. Accordingly, such modern cars are often termed "connected cars." In fact, advanced cars have more than 100 million lines of source code, which supports autonomous operation, self-parking, control, infotainment, safety, performance monitoring with built-in sensors, and inter-vehicle communication. Gartner predicts that, by 2020, connected cars will deliver a new in-vehicle maintenance service and autonomous navigation capability. It is further expected that there will be more than 250 million such units, and one out of five vehicles globally will be connected to a wireless network through the internet of things (IoT) [16]. This rapid increase in vehicle connectivity will affect the overall functionality of telematics, autonomous navigation, infotainment, as well as mobile services, such as mobile banking and remote offices. Over the next 5 years, the proportion of new vehicles with these features is anticipated to increase at a truly dramatic rate, and connected cars will constitute a major part of the IoT [17].

### 4.7. Smart city

Hall [18] defines a smart city as a city that "monitors and integrates conditions of all of its critical infrastructures, including roads, bridges, tunnels, rails, subways, airports, seaports, communications, water, power, even major buildings, can better optimize its resources, plan its preventive maintenance activities, and monitor security aspects while maximizing services to its citizens." According to Harrison et al. [19], the smart city is defined by "connecting the physical infrastructure, the IT infrastructure, the social infrastructure, and the business infrastructure to leverage the collective intelligence of the city." Recently, the definition of the smart city has been expanded to include not only physical aspects, such as city infrastructure, but also concepts that comprise nonphysical factors, such as the environment and governance. The United Nations Conference on Trade and Development (UNCTAD) [20] defines the smart city as smart mobility, smart economy, smart living, smart governance, smart people, and smart environment. Data for smart cities originate from all infrastructure and things in the city based on internet of things (IoT) technology. Services are then developed to enable citizens to have greatly expanded and personalized options in their lives by using the collected data. The IoT overall was developed for the purposes of connecting various things to exchange information and realize value-added information services. Consequently, if the IoT is intelligently applied to cities' facilities, management, and security, city functions could be performed much faster and more efficiently than was previously the case. If a hyper-connected society that connects things and cities becomes a reality in the near future, we will experience truly smart cities that can integrate city management systems that were previously operated individually.

As progress has been made in IoT uses and applications, public sectors are linking building security systems (57%), street lighting (32%), and automobiles (20%) to create an organic technological environment that will support the smart city of the future. The most widely deployed IoT applications in this sector comprises remote monitoring and control of urban devices (27% responded that this is the main application) and constitutes an essential step toward actualizing the smart city's integrated infrastructure.

Paul Manwaring [21], cofounder of the IoT Living Laboratory in Amsterdam, stated that "we need to empower communities to solve their own problems." Certainly, problems still exist that need to be solved to achieve sustainable development. These problems are mainly due to industrialization activities that are based on digital technology.

### 4.8. IoT demonstration complex

The internet of things (IoT) has been identified as a core technology for building smart cities. Therefore, many countries around the world are promoting smart cities to obtain various benefits. As one of the efforts to solve the abovementioned problems, we focus now on trash cans equipped with IoT sensors to assess load quantity in real time. In early 2016, 76 IoT sensors were attached to trash cans in major commercial districts in Seoul, Korea. In June 2017, Goyang city built a smart collection management system based on the IoT [22] as the IoT demonstration complex. The IoT sensors are installed in the trash cans in various locations along city streets and in resident public trash cans to manage loads in real time. A load detection sensor, a solar compression device, and a garbage collection tracker and system are installed in the trash cans. The IoT trash can with the load-sensing control is equipped with a sensor inside of the trash can's lid to measure the load in the trash can in real time, and the compression trash can is automatically compressed to prevent trash can overflow when too much garbage accumulates. In addition, the sensor is powered by solar energy. In garbage collection vehicles, a tracker is installed, and the vehicle position and collection routes are displayed in real time. The amount of garbage collected by each vehicle in the landfill can also be quantified and systematically managed. The measured data in the smart trash can are transmitted to the Goyang city demonstration center server and to environment-friendly smartphones. Finally, garbage-loading information can be checked and managed in real time. This is an example of using the IoT to successfully solve a generally occurring problem in most cities.

### 4.9. IoT threats

It is certain that the internet of things (IoT) will provide tremendous opportunities in manifold regions and industries. However, a fundamental gap still exists between understanding and preparing for the anticipated ubiquity of the IoT. For example, although 98% of organizations that have adopted the IoT claim to be able to analyze data, almost all respondents (97%) stated that it is still difficult to generate value from these data. In fact, more than one-third of companies are not extracting and analyzing corporate network data and using these insights to improve business decisions. One of the biggest limitations is security of data and information to protect IoT-based systems from external threats.

# 5. Conclusion

In this chapter, we introduced the internet of things (IoT), which is a new type of a network that connects device to device, device to people, device to place, etc. The network communications are based on an Internet protocol (IP), such as that used for the Internet. The communications are conducted using embedding or external sensors in devices or objects. Through these communications, tremendous amounts of data are generated. These data are termed big data and are uploaded to a cloud system. This enormous amount of data can then be utilized to create valuable new services and products. In addition, through using the accumulated data, some systems and markets provide powerful intelligent services and applications, such as smart workplaces, smart factories, etc.

We are already living in a hyper-connected world where people and intangible things are networked through the IoT. Indeed, the IoT is leading the era of superfusion that is creating multifaceted economic, social, and ethical values that converge with various industries and expressed as productive business models. In the era of the IoT, most devices use gathered information and network connectivity that actively exploit collected data through a variety of sensors to drive opportunities for new products and services. From this perspective, the IoT integrates intelligent networks which can be systematically linked with humans, things, and services for distributed sensing, networking, and processing.

As one of the IoT applications, the smart city was introduced in this chapter. The smart city can be understood as a kind of hyper-connected world comprising the overall society, business platforms, the environment, etc., with newly developed technologies, such as big data, cloud, and artificial intelligence. Smart cities can also embed these applications and innovations, such as in connected vehicles, smart homes, etc.

Initially, the IoT was developed for simple communications between devices and objects through RFID and M2M technology. However, the IoT is creating a new type of hyper-connected world that comprises connected societies, connected environments, etc. It also creates entirely new types of services, products, and businesses that were not even envisioned in the past. For example, when the Internet first appeared, it was not expected that it would revolutionize the world, but it did. This time, the IoT is changing the world and to no less of an extent.

In near the future, in our hyper-connected world, we will be able to experience a truly smart world which integrates systems that were previously operated individually and create powerful new values and opportunities that we have never experienced.

## Author details

Hyun Jung Lee* and Myungho Kim

*Address all correspondence to: hjlee5249@gmail.com

Department of Economic and Social Research, Goyang Research Institute, Goyang-si, Gyeonggi-do, South Korea

## References

[1] Information Research Service Global. IOT·AI-based Smart Home (Home IoT) related Innovation Technology Trend and Future Prospects. Available from: http://www.irs-global.com [Accessed: December 04, 2017]

[2] Ashton K. Making Sense of IoT. 1999. Available from: http://www.google.co.kr/url?sa=t&rct=j&q=&esrc=s&source=web&cd=1&ved=0ahUKEwjz3vj-uIXYAhVIgrwKHYGzC6cQFggrMAA&url=http%3A%2F%2Fwww.arubanetworks.com%2Fassets%2Feo%2FHPE_Aruba_IoT_eBook.pdf&usg=AOvVaw2eucTstKelqZE5OGGRCBRC [Accessed: December 04, 2017]

[3] Predix. Available from: https://www.ge.com/digital/predix [Accessed: December 04, 2017]

[4] Gartner. Forecast Analysis: Semiconductor Assembly and Test Services, Worldwide, 3Q14 Update in Direction 2016. 2014. Available from: https://www.gartner.com, https://www.gartner.com/doc/2907217?ref=SiteSearch&sthkw=2014%20November%20IoT&fnl=search&srcId=1-3478922254) [Accessed: December 04, 2017]

[5] IDC Directions. Digital Transformation at Scale, Innovation in a Changed World. 2016. Available from: http://www.cvent.com/events/idc-directions-2016/custom-19-d98c-2b4263944a4cb8c785d4bdbec196.aspx [Accessed: December 04, 2017]

[6] Cisco white paper. Internet of Everything: A $4.6 Trillion Public-Sector Opportunity. 2014. Available from: http://www.cisco.com/web/services/portfolio/consulting-services/documents/internet-of-everything-public-sector-white-paper.pdf [Accessed: December 04, 2017]

[7] Barker C. 25 Billion Connected Devices by 2020 to Build the Internet of Things. ZDNet. 2014. Available from: http://www.zdnet.com/article/25-billion-connected-devices-by-2020-to-build-the-internet-of-things/ [Accessed: December 04, 2017]

[8] Framingham M. Worldwide Spending on the Internet of Things Forecast to Reach Nearly $1.4 Trillion in 2021, According to New IDC Spending Guide. IDC. 2017. Available from: https://www.idc.com/getdoc.jsp?containerId=prUS42799917 [Accessed: December 04, 2017]

[9]  Gartner. 2017 Will See 8.4 Billion Connected 'Things' Will Be in Use in 2017, Up 31 Percent From 2016. 2017. Available from: https://www.gartner.com/newsroom/id/3598917 [Accessed: December 04, 2017]

[10]  McKinsey. The Internet of Things: Mapping the Value beyond the Hype. 2015. Available from: https://www.mckinsey.com/~/media/McKinsey/Business%20Functions/ McKinsey%20Digital/Our%20Insights/The%20Internet%20of%20Things%20The%20 value%20of%20digitizing%20the%20physical%20world/The-Internet-of-things-Mapping-the-value-beyond-the-hype.ashx [Accessed: December 04, 2017]

[11]  Machina Research. The Global IoT Market Opportunity Will Reach USD4.3 Trillion by 2024. 2015. Available from: https://machinaresearch.com/news/the-global-iot-market-opportunity-will-reach-usd43-trillion-by-2024/ [Accessed: December 04, 2017]

[12]  IITP, 2014. IoT Status and Major Issues, ICT Insight 04

[13]  Ministry of Science, ICT and Future Planning Ministries of the Republic of Korea. Master Plan for Building the Internet of Things (IoT) that leads the hyper-connected, digital revolution. 2014. Available from: http://www.google.co.kr/url?sa=t&rct=j&q=& esrc=s&source=web&cd=1&ved=0ahUKEwjepe_EmNfYAhUBvrwKHZwYAF0QFggm MAA&url=http%3A%2F%2Fwww.kiot.or.kr%2FuploadFiles%2Fboard%2FKOREA-IoT%2520Master%2520Plan.pdf&usg=AOvVaw2Z5CjmAItMtT-rM6S51i-K    [Accessed: January 12, 2018]

[14]  Hewlett Packard Enterprise. Internet of Things, Today and Tomorrow. 2016. Available from: http://www.arubanetworks.com/assets/eo/HPE_Aruba_IoT_Research_Report.pdf [Accessed: December 04, 2017]

[15]  Economic Review. 2016. Available from: https://kerala.gov.in/documents/10180/ ad430667-ade5-4c62-8cb8-a89d27d396f1 [Accessed: December 04, 2017]

[16]  Gartner. The Internet of Things Challenges Smart City Processes and Culture. 2013. Available from: https://www.gartner.com/doc/2619517?ref=SiteSearch&sthkw=2014%20 November%20IoT&fnl=search&srcId=1-3478922254 [Accessed: December 04, 2017]

[17]  IT World. Connected Car Occupies Major Part of Internet of Things for Next 5 Years. 2015. Available from: http://www.itworld.co.kr/news/91595#csidxae88edf9540bff89177c f9bba41bbce [Accessed: December 04, 2017]

[18]  Hall RE. Thee vision of a smart city. In: Proceedings of the 2nd International Life Extension Technology Workshop, Paris, France, September 28, 2000. Available from: https://www.osti.gov/scitech/servlets/purl/773961 [Accessed: December 04, 2017]

[19]  Harrison C, Eckman B, Hamilton R, Hartswick P, Kalagnanam J, Paraszczak J, Williams P. Foundations for smarter cities. IBM Journal of Research and Development. 2010;**54**(4):1-16. Available from: http://fumblog.um.ac.ir/gallery/902/Foundations%20 for%20Smarter%20Cities.pdf [Accessed: December 04, 2017]

[20] UNCTAD. Issues Paper On Smart Cities and Infrastructure. 2016. Available from: http://unctad.org/meetings/en/SessionalDocuments/CSTD_2015_Issuespaper_Theme1_SmartCitiesandInfra_en.pdf [Accessed: December 04, 2017]

[21] Paul Manwaring. The IoT Living Lab at The Smart City Innovation Summit Asia—SCIS Asia 2017. 2017. Available from: https://www.linkedin.com/pulse/iot-living-lab-smart-city-innovation-summit-asia-scis-manwaring [Accessed: December 04, 2017]

[22] Kim JH. Goyang City, IoT Garbage Can Appear. May 17, 2017. e4ds. Available from: http://www.e4ds.com/sub_view.asp?ch=30&t=1&idx=6203 [Accessed: December 04, 2017]

# A Reference Architecture for Digital Ecosystems

Alexandru Averian

Additional information is available at the end of the chapter

**Abstract**

Digital ecosystems are a new type of application based on a "universal digital environment" populated by digital entities that form communities that evolve and interact with information exchange and who trade digital objects that are produced through the system. Entities that participate and form the ecosystem can be applications running not only on simple devices: wearable, sensors, actuators, but also on complex services executed on smartphones, tablets, personal computers, company servers, etc. A reference architecture for digital ecosystems is a step toward standardization, as it defines a set of guidelines in designing and implementing a digital ecosystem. Often such architectures are very abstract, difficult to understand and implement. In this chapter, we introduce a vendor- and technology-neutral reference architecture for digital ecosystems and apply this architecture to an actual use case.

**Keywords:** digital ecosystems, reference architecture, adaptive, context-aware

## 1. Introduction

A series of architectures have been taken into consideration in the construction of digital ecosystems. Briscoe presents in [1] the first neural calculus applications that bring together, in a new approach, elements of theory with service-oriented architectures, multiagent systems and distributed computing components, the proposed digital ecosystems to deliver business support. In [2], the principles and semantics used in digital ecosystems are formulated, as shown in [3], Chang et al. continue to research digital ecosystems by broadening their scope in areas such as transport, education and health. Article [4] presents the implementation steps toward an agent-oriented architecture of the digital healthcare ecosystem. All examples assume the existence of an environment of communication and intelligent agents or context-conscious applications that respond to changes that occur in the environment. In this chapter, we introduce a new, high-level, reference architecture for the development of digital ecosystems. We propose a set of steps toward a more specific reference architecture for digital ecosystems, and introduce Reference Architecture for Digital Ecosystems (RADE), a six-layer architecture comprising: environment, context management, interaction, adaptation to goals, species management, user integration, and finally, we apply proposed architecture to an actual use case. This chapter is structured as follows. Section 2 presents the related work and current results in the field of ecosystem representation. Section 3 presents a new reference architecture for digital ecosystems. Section 4 describes a possible implementation of proposed reference model. Last section presents the conclusions and hints for future work.

## 2. Related work

As introduced in [5], the ecosystem oriented architecture (EOA) model, named architectural style, defines a digital business ecosystem as being an open ensemble of distributed services. This differs from a service-based architecture (SOA) because it needs to address new issues such as: decentralization, managing a distributed knowledge base, self-organization, and self-rebuilding. EOA is not a larger SOA or other SOA [6], it addresses a new type of evolutionary, dynamic, knowledge-sharing, self-organizing, self-controlling, self-reliant architectural model as it occurs in natural ecosystems [7]. The EOA architecture applied to a digital business ecosystem presupposes the following components:

- Services described from the perspective of the problem (business), the computational description (interface) is not sufficient, and a business specification is needed.

- Service registry—intelligent discovery mechanisms based on the business specification are required because the services will not be known in the construction of the system.

- Model repository—it is separated from the service register and must allow for a model-driven development, so that models can be categorized by users (folksonomy) and improved.

Minimal support services that facilitate communication between services and ecosystem development help participants integrate and publish new services.

An SOA-based application is owned and managed by a large organization, operates within its network, and interconnection (B2B) with other organizations is usually required. The user community must be the owner of the ecosystem, a P2P network is more appropriate in this case being more "democratic." On the other hand, in a digital ecosystem, there must be no hierarchical topology based on interest, it must not be a single point of administration, and the system must be self-configuring and adaptive.

## 3. Reference architecture for digital ecosystems

In this section, the RADE model is presented—a reference architecture for digital ecosystems. After a brief introduction, the section 3.1 presents the anatomy of the digital species participating in ecosystem. The following sections describe each layer in the species structure. The section concludes with a series of considerations on security, identity, and trust.

In order to present the components of a digital ecosystem, we start from the structure of a natural ecosystem. A natural ecosystem is made up of biotic (living entities) and abiotic components, whose interaction creates an ecosystem that perpetuates itself. More specifically, it consists of one or more communities of organisms of different species that manifest themselves in their habitat. Of each species, there may be several populations that activate in their microhabitat [8]. A community is formed by groups of populations of various species that operate in the same habitat. A habitat is a distinct part of the environment. An individual organism or a population can migrate from one habitat to another in search of resources, thus being able to compete with other organisms. A microhabitat is a subdivision of a habitat with its own specific properties. A population that operates in a microhabitat will tend to occupy a niche. This is a functional relationship between a population and the environment it occupies. Niche emerges as a strong adaptation of a population to the occupying microhabitat [9]. One can find more details about the habitats in articles [1, 10], which presents the first applications in the neural calculus area that bring together, in a new approach, elements of theory with service-oriented architectures, multiagent systems and element-distributed computing, and the digital business ecosystems.

Ecosystems are described as complex adaptive systems (CAS) consisting of diverse components that interact locally and are subject to the process of natural evolution and selection. Digital ecosystems are composed of populations of agents evolving (through natural selection) in the distributed environment. These fall within the definition of complex adaptive systems, having a nonlinear evolution and always aiming at a dynamic balance. We present the following components of a digital ecosystem:

- the communication medium is a P2P communications network or a real-time data distribution system (such as DDS—Data Distribution Service), TCP/IP- or UDP-based networks;

- habitat is a key element in the functioning of an ecosystem, it is a network node running the ecosystem services in which support services and optimization services (EVE) are run [5, 6]. Each user has a habitat that connects and launches queries in the ecosystem, users can define and activate new services. Habitats communicate with each other and can form clusters of habitats based on data exchanges between them.

- the entities of a digital agent ecosystem, context-conscious applications, or local or remote services, sensors or intelligent objects (IOTs) that can be connected to the habitat to provide context information; these are the species that populate a digital ecosystem.

- the context defines the environment in which a particular entity is in the ecosystem at a given time, it contains primary data relevant to the application, a high-level context obtained by aggregating primary data, and context contexts.

Reference architecture for digital ecosystems represents a step forward for standardization, because this defines a set of guidelines in designing and implementing a digital ecosystem. Usually, these types of architectures are very abstract, hard to understand and implement.

RADE architecture integrates devices and species, and it takes a device or a thing and upgrades it with context-awareness, adaptability, and autonomicity and produces a digital species populating a digital ecosystem. The anatomy of a species is formed from six main layers and a security component, which can also be found on every layer, as can be seen in the next section (**Figures 1–6**).

### 3.1. Anatomy of species in ecosystems

From an ecosystem perspective, every actor participating in a supply chain ecosystem is represented by its digital counterpart in digital ecosystem. An entity is part of a species if it is designed and programmed to behave in a certain specific way, to use a certain type of resources and to act according to a specific context. In this section, we introduce the reference model of species participating in a digital ecosystem. The introduced model proposes a set of guidelines in designing and implementing of digital counterpart of players taking part in a supply chain ecosystem. Anatomy of species is comprised of six main layers and a security component, which can also be found on every layer, as can be seen in the figure below.

A reference architecture operates with components with a high degree of similarity, so they can be assembled correctly and safely, resulting in complex yet scalable solutions, providing flexibility for various application scenarios. In this respect, the following guiding principles can be found in different areas of architecture:

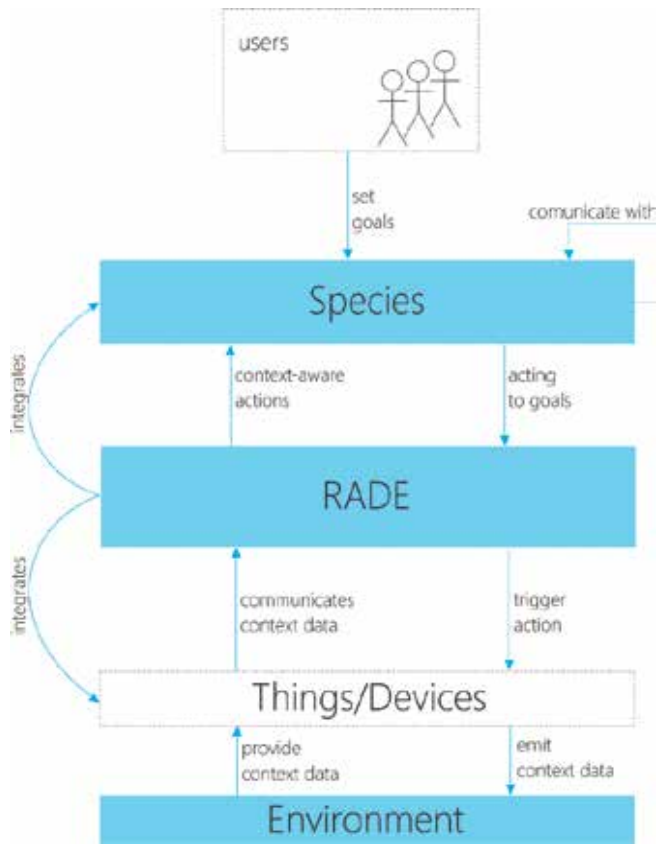

**Figure 1.** RADE overall architecture.

- Heterogeneity—ecosystems are open systems, a reference architecture must cover a wide variety of logical, physical and virtual entities, processing patterns and standards, and it must be able to use a wide variety of hardware and software platforms.

- Flexibility—this assumes that the system is easily changed, permits easy assembly of heterogenic components, and the easy assembly of a varied set of components and services.

- Weak coupling—this involves the use of poorly coupled processing modules and a communication medium that allows the digital entities involved to be decoupled in time and space.

- Scalability—this requires the system to admit a large number of connected entities (theoretically unlimited); the system being open, it must admit the addition of new participants in a flexible way.

- Security—certain areas of application of digital ecosystems (e-health) will imply a strong connection between the physical world and the digital world, for the realization of secure systems, the model should include multilevel security measures including identification and authorization of digital entities and users, data protection, and authentication.

### 3.2. Environment

The environment level is the mode of communication between species, and it extracts the information from other entities and helps to communicate data/orders in the environment. The digital environment can be a peer-to-peer (P2P) system that has a number of advantages over a centralized (client/server) model that is not resilient, error tolerant, scalable, and vulnerable to attacks. These advantages result from the network definition mode, P2P is defined as a network in which the nodes are equivalent to each other in the sense that all nodes (in principle) can execute the same set of functions needed for network to work. The most important features of a P2P network are as follows:

- resource sharing through direct transfer with no centralized servers, however, sometimes centralized servers that can be used for setting up the network, node management, etc.;

- no centralized nodes, no central fall points, no central attack points;

- nodes actively participate in operations such as handling information, finding resources, and storing and managing data;

- the network has the ability to adapt to changes in connectivity, to changes in typology, and the ability to reconfigure itself after finding an error;

- the typology of a P2P network is tolerant to defects, having the ability of auto-organization in order to keep functioning;

- the network can be structured or not, and physical proximity between nodes is not important;

- IP, the links between nodes are TCP connections but can be represented as pointers to the IP address.

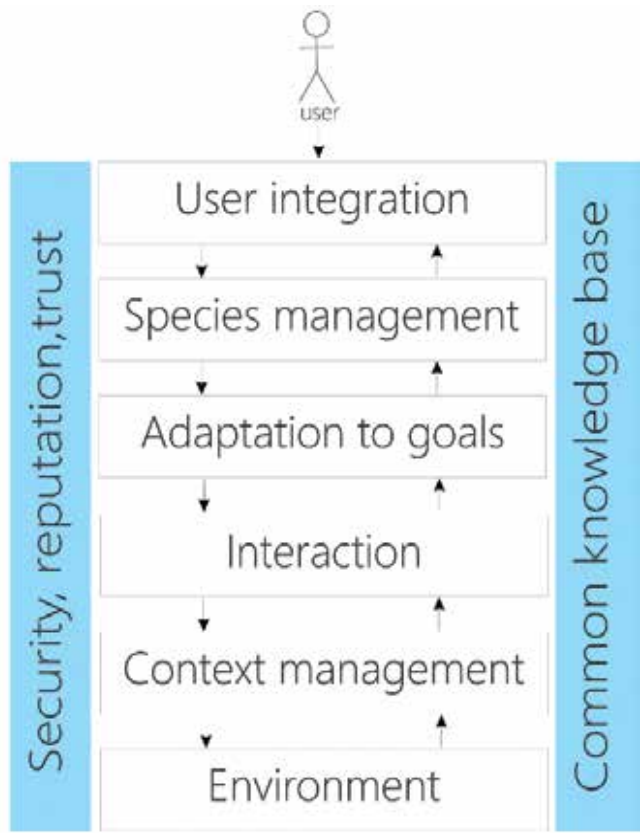

**Figure 2.** Anatomy of species in digital ecosystems.

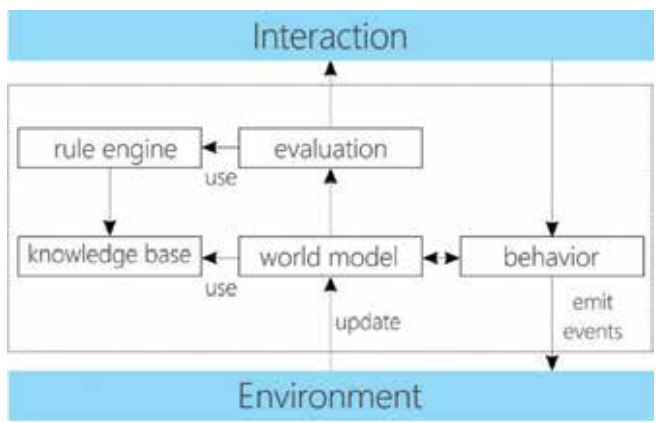

**Figure 3.** Context management.

Depending on the type of application, the level of access to the environment can be implemented through a messaging-oriented machine-to-machine type such as Data Distribution Service (DDS), Extensible Messaging and Presence Protocol (XMPP), Advanced Message

Queuing Protocol (AMQP), Message Queue Telemetry Transport (MQTT), or Constrained Application Protocol (CoAP). These systems are widely used to implement IOT applications. Message-Oriented Middleware (MOM) is a middleware product family that facilitates messaging across distributed systems. A MOM system uses one of the following communication paradigms: message passing, indirect queuing, *publish/subscribe* communication (data are published through a topic, and customers receive all messages posted by the topic they are subscribed to). Of the three communication models mentioned, *publish/subscribe* is best suited for building the level of access to the environment within the RADE architecture because it provides asynchronous, scalable multi-to-many communication. In this scheme, the messaging emitters communicate with the subscribers, without prior knowledge, through a distributed P2P infrastructure. The system allows a decoupling in terms of time, space, and synchronization. Disconnection over time allows the broadcaster and receiver to communicate without having to be online at the same time and to cooperate directly. Decoupling in space refers to the fact that the transmitter and receiver are unaware of each other, and their identity and location are not relevant. Synchronization decoupling refers to the fact that the receivers and transmitters do not have to synchronize, and the communication is accomplished by asynchronous notifications implemented with a callback function system. For sending messages to the environment or for exchanging messages between entities, an RPC communication scheme will be used.

### 3.3. Context and niche

Context data can only be used in the presence of a well-defined goal. Thus, some contextual information along with a defined purpose can generate actions to be taken. In other words, if we have a formal context consisting of a series of objects and a set of attributes, the purpose of the application is defined by a set of opportunities (affordances) that invite to action. Opportunities create a relationship between subject and object, the subject being the application/entity and the object can be any context information. These define the rules by which when a context object changes its state, the subject can act. The rules define the role a species has in its environment. The totality of interactions between a species and the environment forms a niche. The goal can be defined not only at the level of an application and at the level of a user queries, but also at the level of applications that work in the background or at the level of intelligent objects or autonomous robots that work for the user but without direct intervention. Niche expands the concept of context, as defined by Dey in [11], to digital ecosystems; it encompasses any relevant information that is useful to characterize not only the situation of an entity in the ecosystem but also the mode of action of the entity over environment. In article [12], there are introduced as the stages of implementation of a multiagent health ecosystem, and finally, article [13] defines the context of e-health as avatars and virtual organisms as "a collection of information from e-health systems used by people and applications that characterize the situation of another entity (usually a person but may also be an application) in its environment used to interpret the state or behavior of the entity concerned." A wide range of context approaches have been presented in the literature, especially in the field of ubiquitous computing. These are key-value, object-based, logic-based models based on ontologies, using graphical representation or markup. A classification of context patterns can be found in [14]. All examples involve the existence of a communication medium and intelligent agents or context conscious applications that respond to changes that occur in the environment. The context programming model for digital ecosystems introduced by the author in [15] and presented extensively in [16] is an objectual model where the context is represented as a set of communicating streams; the streams can be connected to various filtering/processing operations, finally a series of reactive variables in the application are updated. Generally, the context management level contains logic for extracting and processing context data. At this level, the following operations are executed:

- extraction and primary processing of data,

- aggregation of primary data to obtain more complex contexts,

- assessing the context situation and issuing signals to the upper layers,

- reception at the upper level of orders, actions, behavior adaptation,

- issuing events, actions, and context data to the environment describing the state of the entity.

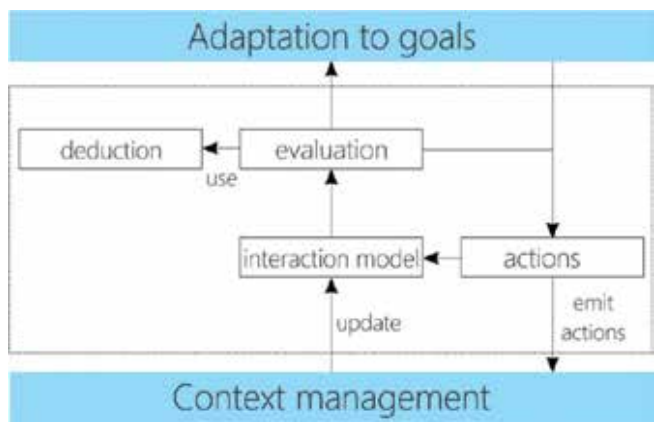

**Figure 4.** Interaction level.

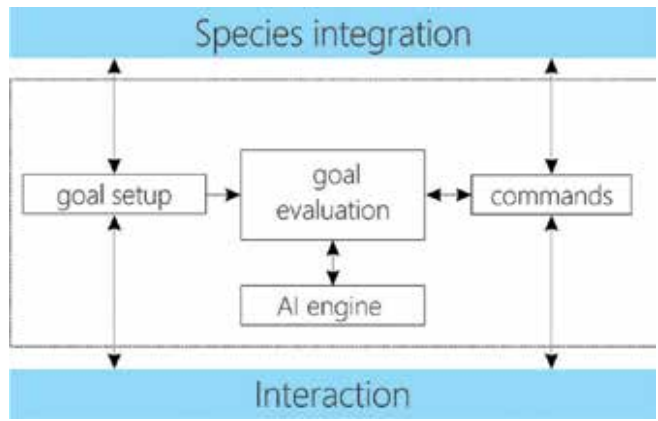

**Figure 5.** Adaptation to goals (optimization level).

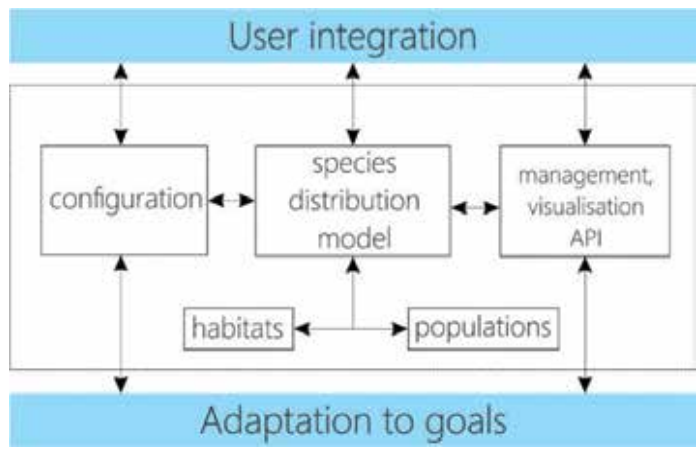

**Figure 6.** Species integration.

The context management level consists of five components as can be seen in the following figure.

The *knowledge base* component can be considered as a database where all the data describing the entity context are stored; they can include information about other objects or entities relevant to the application and their context data. In some cases, it will be implemented through a database management system (SGBD), and in other cases, it could just be a simple collection of information.

The *world model* component is a data model that reflects the real world situation in which the entity in question is running—the context situation. The component also retains the contexts' context of other entities and integrates the changes that occur in these situations. The data are stored in the knowledge base. The rule engine component defines and manages context rules and mechanisms. Generally, this component consists of a set of evaluation rules that also contain actions that can be taken to change the context situation. Concrete implementation can be varied, depending on the specificity of the application; in some cases, it can be a simple set of rules that evaluates data in the form of key-value pairs, and in other cases, it can be given by sophisticated systems for processing its ontology data system mining. The evaluation component evaluates the context situation reflected in the world model component and the changes that occur in it. Evaluation uses the assessment engine and produces status changes and actions that it communicates to the level of interaction. The behavioral component receives commands from the interaction level and emits events in the environment that signal the behavior of the entity. This component may add changes to the context model (world model).

### 3.4. Interaction

The interaction level receives information about changes that happen in the context, events, or messages received from other entities in the environment, orders (from trusted/authorized entities) that can lead to reconfiguration of the purpose. In the case of a reconfiguration (e.g.,

"abort mission"), or if there is a situation in which urgent measures must be taken (the occurrence of threat in the species), the level of interaction emits in the environment through the context component of the context a message by which it signals the situation. It also signals changes to higher levels of adaptation to goals and species integration and can ask (after receiving confirmations) to change or reconfigure the objective/purpose of the current entity. This level maintains communication and ongoing interaction with other participants in the environment, issues actions and events that describe the entity's behavior and intentions in the environment. The interaction level consists of the following components as shown in the below

- model interaction is a model that defines the application's action mode, how to relate the entity to the environment, reflecting the entity's behavior in the real world. The interaction model describes how users understand the application. Defining a pattern of interaction is essential. Once defined and understood, users can understand and track the way an entity operates. This is a fundamental pattern that describes how certain elements relate to each other, may contain sub-modules for various subcomponents, and together they constitute the general pattern.

- evaluation permanently assesses the current situation and events occurring in the environment, applies the rules of the interaction model, generates messages for the higher level, and also issues commands to the action component.

- deduction—a deduction engine that can be used in decision-making on interaction and can deduce new interaction rules from observations on the evolution of the site.

- Actions—receives orders from the higher level or from the assessment component on the same level, applies entity-specific actions and issues lower layer behaviors to be published in the environment.

The interaction model depends on the application, it can also be a model with simple one-way rules, the cause-effect form, or it can be a very complex interrelation pattern. For example, for business digital ecosystems, *ActionWorks* Business Interaction Model can be used to coordinate the interaction between a customer group and a group of providers through a four-step feed-back loop: preparation, negotiation, delivery, and acceptance. For other applications, the Complexity of Interaction Sequences (CIS) model introduced in [17] can be used. This model uses interaction sequences that are defined as action steps that change the status of a system, and any problem that needs to be resolved is seen as a state to be reached as a result of executing a sequence of steps.

### 3.5. Adaptation to goals

Adaptability is the ability of a system to change its behavior according to new, unexpected situations [18]. The adaptive properties of an organism are closely related to the self-organizing property [19] and the emergence phenomenon. Applications from a digital ecosystem must solve concrete problems but also be computationally efficient. It will seek to establish a balance between the freedom of a system to self-organize and the constraints that apply to obtain useful solutions. Briscoe in [20] proposes a digital ecosystem model that incorporates evolutionary and self-organizing properties specific to natural ecosystems. The model applies an EOA architecture to a distributed multiagent system, and evolving mechanisms are made on two levels within the evolutionary component (EvE). The first level is formed by a P2P network of agents (evolutionary population) that feeds a second optimization system that operates locally (at the habitat level) and exploits evolutionary algorithms to identify solutions that satisfy relevant local constraints. The local search process of the solutions is accelerated by the exchange of values (migration of individuals) between different habitats in which a calculation with similar constraints is executed. This level has the role of effectively solving complex problems so that the system is getting closer to the purpose for which both an individual system and the whole group are configured.

The goal evaluation component could implement the adaptive reference architecture defined in [21], which presents the structure of a MAPE-K Loop Adaptation Manager, comprising a series of activities to be followed in order to have complete feedback and adaptation. MAPE-K comes from monitoring, analysis, planning, execution, and knowledge. The four steps present in the loop correspond to the four activities that are also found in the medical field: observation, diagnosis, solution, and treatment. The Knowledge Base retains information about the adapted system and its context, and this information is used by all four stages of the feedback. Depending on the degree of adaptation of the component, an increasingly complex knowledge base is needed, along with the advanced deliberative mechanisms leading to the adaptation process. The system continually assesses its own state and context in which it finds and issues decisions (and internal or external commands) that adjust the state of the system toward the goal. At this level, an AI engine or a set of evolutionary algorithms such as genetic algorithms, bee colony optimization [22], and intelligence swarm will be used. In the case of digital ecosystems implemented as multiagent systems, membrane-computing models [23] can be used for specification and implementation.

### 3.6. Species integration

The concepts of species, individuals, integration and cohesion are widely debated in the literature of biology and ecology [24]. The term integration refers to the active interaction between the components of a system. Cohesion refers to cases where a component of a system behaves like the whole system, relative to a particular process. Thus, the presence and action of a part of a system does not affect the activity of another part of the system, although all parts are uniformly responsible for a certain type of stimulus and behave similarly to the same process. The level of species integration allows the integration and configuration of participating entities within a species. Also, at this level is the general purpose of a species, splitting and managing population populations to respond to queries or to solve a specific problem. A population is a part of a species that operates in a specific context. An entity becomes part of a population and a species if it is programmed to act in a certain way specific to the species, to use a certain type of resources and to act according to a specific context to the population to which it belongs.

### 3.7. User integration

The user and his applications are part of the ecosystem. The user integration level integrates the users and applications with which it interacts on the last layer, the level can be viewed as a service or as a graphical interface located on the highest level of architecture. The concrete implementation of this level is dependent on the specificity of each application, actors and usage cases. At this level, setup commands and queries or commands will be launched by the ecosystem. In some cases, this layer takes care of authentication, authorization, accounting for the use of shared resources and payment services.

### 3.8. Security, identity, and trust

In building a digital ecosystem, security issues must be considered on each level. Depending on the specificity of the application, a series of attacks can be triggered at the level of connected devices, network, operating systems, application level, or user level. A digital ecosystem is an open system, besides the "classic" security issues, there may be problems of reputation and trust. In a distributed system, such as a digital ecosystem, there is a need for trust between users and organizations. Trust is a multidimensional concept that is hard to define and difficult to measure [25].

Article [26] analyzes trust from the technological, economic, behavioral, and organizational perspective. The technological dimension of trust expresses the subjective probability of an organization to believe that a particular infrastructure can facilitate transactions in line with its expectations. The technological dimension includes security services, mechanisms that ensure the confidentiality, authenticity, nonrepudiation and integrity of transactions, as well as mechanisms that ensure identity control and access to resources. A distributed identity management system must exist in the ecosystem so that it is possible to ensure the identity of a service provider as well as consumers to control access to resources. The economic dimension involves establishing relationships of interdependence between organizations (based on a cost-benefit analysis) and the use of IT infrastructure for trading, data transfer, and know-how. In [27], a model for the management and accounting of the use of services in digital ecosystems based on an SOA architecture is presented. The behavioral dimension of trust is derived from the characteristics of interpersonal behavior, which relate to competence, predictability, honesty, and good intentions. The organizational dimension of trust results from the use of good practices, quality standards, audit, risk management strategies, and process management standards.

## 4. Supply chain ecosystem

There are a variety of supply chain management models in literature as results from a recent review [28]. Markus and Loebbecke [29] use the term ecosystem as a unit of analysis in describing groupings of suppliers and distribution chains, which are understood as loose sets of organizations engaged in the creation and delivery of products and services, the same term is used by Iansit and Levien in [30] describing strategy as an ecology. In [31], the authors present the opportunity to develop a digital ecosystem for transportation and warehousing logistics. This involves building a supply chain [32] that would facilitate the integration and collaboration of small and medium-sized enterprises (SMEs) in particular, would encourage cooperation, would be an opportunity to create synergy, facilitate incubation, increase, and would bring prosperity to the business. The "Virtual Collaborative Consortium" digital ecosystem implemented in Australia is an example, which is a collaborative environment for all those involved in the product distribution chain. In [33], an agent-based distributed supply chain model is proposed and a number of open issues are formulated. In [32], the delivery chain problem is formulated in terms of task dependency network, a mathematical model is proposed, and equilibrium and convergence issues are studied. In this section, we present an application of the digital ecosystem architecture on a section of the Amazon retailer chain as introduced in [34].

The automation of operations in a warehouse seems to be a difficult operation, but some companies have already made great strides in this direction. The orange robots, as can be seen in the next figure, are simple machines that move horizontally on a 2D grid in all directions, can enter under the shelves in the warehouse, lift them, and carry them to the desired destination. Kiva robots are generally used to transport the shelves of objects to be shipped to the selection and packaging table. After taking over the objects, they carry the shelves back to their place. In addition, they can be used for warehouse shelving operations, for more efficient use of storage space, for sorting and ordering shelves for delivery. Robots with a mobile arm operate on packing and putting packages on the conveyor. The drones' species connects the packages on the platform and takes them to their destination. The following table summarizes a case of using the architecture for digital ecosystems on a section of the Amazon retailer supply chain. It includes actors, purpose, preconditions, a correct usage scenario, and postconditions (**Table 1**).

| | |
|---|---|
| Actors | **1.** Human operator |
| | **2.** Client |
| | **3.** Kiva robot species |
| | **4.** Species of mobile handler robots |
| | **5.** Drone species—Prime Air |
| Goals | Delivery of products to recipients. Customer orders are quickly honored, delivery is done with the help of the drones in rural areas and peripheral urban areas. |
| Preconditions | There is a stock of products displayed on a website. |
| High-level success scenario | **1.** The operator picks the general role for every species and for the robot population. |
| | **2.** The client makes an order in the system through the website. |
| | **3.** The system checks the stock and sends a movement order of the product to the packing line. |
| | **4.** Kiva Robots will bring the rack with the ordered products to the packing line. |
| | **5.** Manipulating robots pack the products and place the package on the delivery line. |
| | **6.** At the end of the line, another manipulating robot extracts the package from the tape and places it on a platform. |
| | **7.** A Prime-Air drone picks up the package, reads the code extracts the address of the destination and performs the delivery flight. |
| | **8.** The delivery is made, confirmed and the drone comes back to base. |
| Postconditions | The client confirms the reception of the package online and can use the product. |

**Table 1.** Use case summary.

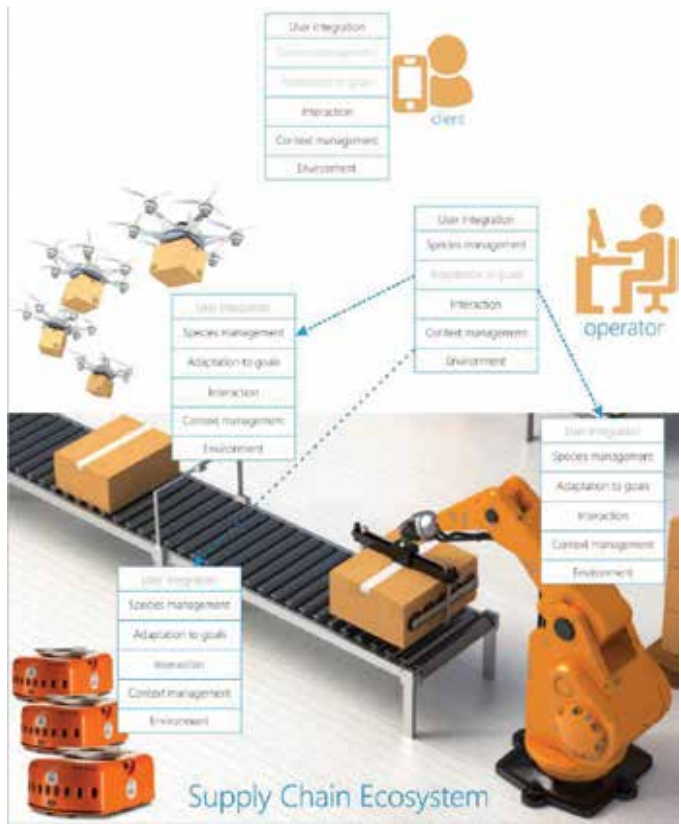

**Figure 7.** RADE architecture mapping on use case [34].

As can be seen from the **Figure 7**, the Kiva robot species, the arm robots, and the drones species do not contain the functions of the first level of user integration, the applications used by the operator species also implement the level of user integration but lack the level of adaptation to the goal. We can also see that all species integrates species management functions; at this level, each species of robots can be configured, setting the mode of work—the purpose. The operator performs this configuration through its own level of species management. Otherwise, it can be observed that all other layers of the RADE architecture are present within each species. Security features were omitted in this example for simplification. It can be considered that all digital objects are reliable, and access to the system is checked at the network level. In the following figure, we can see how to map the levels of the RADE architecture for each actor.

## 5. Conclusions

The research area of digital ecosystems is becoming more and more important. Despite a large number of relevant works in this area, the level of knowledge is still insufficient. Digital ecosystems offer many opportunities but also challenges for researchers and developers. In this chapter, we introduce a vendor and technology neutral reference architecture for digital ecosystems and a possible application of this architecture to an actual use case. The introduced architecture proposes a set of guidelines in designing and implementing a digital ecosystem. The proposed model consists of the following six layers: environment, context management, interaction, adaptation to goals, species management, and user integration. In the final part of the chapter, we presented supply chain ecosystem, an application of the RADE model for a section of an ecosystem supply network, containing four species, namely the species of human operators, the species of drones, the Kiva robot species, and robots with a mobile arm. This work follows the study conducted in [15, 16] and continue the research that was presented in respective paper. Future research will seek to refine the model, will try to integrate with blockchain technology, and adopt some algorithms from AI domain.

## Acknowledgements

The work presented in this chapter has been funded by the Sectoral Operational Programme Human Resources Development 2007–2013 of the Ministry of European Funds through the Financial Agreement POSDRU/159/1.5/S/132395.

## Author details

Alexandru Averian

Address all correspondence to: aaverian@gmail.com

Politehnica University of Bucharest, Romania

## References

[1] Briscoe G, De Wilde P. Digital Ecosystems: Evolving Service-Orientated Architectures no. 507953. 2006 1st Bio-Inspired Model. Network, Information and Computing Systems. 2006

[2] Boley H, Chang E, Digital ecosystems: Principles and semantics. In: Proceedings of 2007 Inaugural IEEE International Conference on Digital Ecosystems and Technologies; 2007. pp. 398-403. DOI: 10.1109/DEST.2007.372005.

[3] Chang E, West M. Digital ecosystems and comparison to existing collaboration environment. WSEAS Transactions on Environment and Development. 2006;**2**(11):1396-1404

[4] Vasilățeanu A, Șerbănați LD. Towards an agent-oriented architecture of the digital healthcare ecosystem. UPB Scientific Bulletin, Series C: Electrical Engineering. 2012;**74**(2):87-102

[5] Briscoe G, Sadedin S, Wilde P. Digital ecosystems: Ecosystem-oriented architectures. Natural Computing. 2011;**10**:1143-1194

[6] Ferronato P. Architecture for Digital Ecosystems, beyond Service Oriented Architecture (IEEE-DEST 2007). In: 2007 Inaug. IEEE-IES Digit. Ecosyst. Technol. Conf., 2007. DOI: 10.1109/DEST.2007.372047

[7] Levin SA. Ecosystems and the biosphere as complex adaptive systems. Ecosystems. 1998;**1**(5):431-436. DOI: 10.1007/s100219900037

[8] Begon M, Harper JL, Townsend CR, others. Ecology. Individuals, Populations and Communities. USA: Blackwell Scientific Publications; 1986

[9] Lawrence E. Henderson's Dictionary of Biology. London, UK: Pearson Education; 2005. ISBN: 1408234300

[10] Briscoe G, Sadedin S. Natural science paradigms. Nachira F, Dini P, Nicolai A, Le Louarn M, Rivera Lèon L, editors. Digital Ecosystems. Luxembourg: Office for Official Publications of the European Communities; 2007. pp. 48-55. ISBN: 92-79-01817-5

[11] Dey AK. Understanding and using context. Personal and Ubiquitous Computing. 2001;**1**(5):4-7

[12] Serbanati LD, Ricci FL, Mercurio G, Vasilateanu A. Steps towards a digital health ecosystem. Journal of Biomedical Informatics. 2011;**44**:621-636. DOI: 10.1016/j.jbi.2011.02.011

[13] Șerbănați LD, Vasilățeanu A, Niță B. Strengthening context-awareness of virtual species in digital ecosystems. Bucharest: Conference: 19th International Conference on Control Systems and Computer Science (CSCS). 2013:503-510

[14] Strang T, Linnhoff-Popien C. A context modeling survey. In: Workshop Proceedings. First International Workshop on Advanced Context Modelling, Reasoning And Management at UbiComp; Nottingham, UK. 7 September 2004. Workshop on, https://doi.org/10.1.1.2.2060

[15] Averian A. A programming model of context-aware applications in digital ecosystems. In: 17th International Multidisciplinary Scientific GeoConference SGEM 2017. Vol. 17; 2017. pp. 37-44. DOI: 10.5593/sgem2017/21

[16] Averian A. Towards More Context-Awareness in Reactive Digital Ecosystems. In: Proceedings of Creativity in Intelligent Technologies and Data Science. Second Conference, CIT&DS 2017, Volgograd, Russia, September 12-14, 2017. 2017. pp. 640-654. DOI: 10.1007/978-3-319-65551-2_46

[17] Appert C, Beaudouin-Lafon M, Mackay WE. Context matters: Evaluating interaction techniques with the CIS model. In: Fincher S, Markopoulos P, Moore D, Ruddle R, editors. People and Computers XVIII—Design for Life: Proceedings of HCI 2004. London: Springer London; 2005. pp. 279-295

[18] Bradbury JS, Cordy JR, Dingel J, Wermelinger M. A survey of self-management in dynamic software architecture specifications. In WOSS '04: Proceedings of the 1st ACM. SIGSOFT workshop on Self-managed systems, no. November, 2004. pp. 28-33. DOI: 10.1145/1075405.1075411

[19]  Serugendo GDM, Gleizes MP, Karageorgos A. Self-Organising Software: From Natural to Artificial Adaptation. Springer-Verlag, Berlin, Heidelberg: Springer Science & Business Media; 2011. ISBN: 978-3-642-17347-9

[20]  Briscoe G, Sadedin S, Paperin G. Biology of Applied Digital Ecosystems. Digital EcoSystems and Technologies Conference; DEST'07. Inaugural IEEE-IES. IEEE, 2007

[21]  Horn P. Autonomic Computing: IBM's Perspective on the State of Information Technology. USA: IBM Press; 2001

[22]  Anescu G. A fast artificial bee colony algorithm variant for continuous global optimization problems. University Politehnica of Bucharest Scientific Bulletin Series C-Electrical Engineering and Computer Science; 2017;**79**(1):83-98

[23]  Vasile C, Dumitrache I. Multi-agent membrane systems. Scientific Bulletin Series C. 2016;**78**:3-12

[24]  Lee M, Wolsan M. Integration, individuality and species concepts. Biology and Philosophy. Nov. 2002;**17**(5):651-660. DOI: 10.1023/A:1022596904397

[25]  Hosmer LT. Trust: The connecting link between organizational theory and philosophical ethics. Academy of Management Review. 1995;**20**(2):379-403. DOI: https://doi.org/105465

[26]  Ratnasingam P. Trust in inter-organizational exchanges: A case study in business to business electronic commerce. Decision Support Systems. 2005;**39**(3):525-544. DOI: 10.1016/j.dss.2003.12.005

[27]  Malone P, Jennings B. Distributed accountability model for digital ecosystems. 2nd IEEE International Conference on Digital Ecosystems and Technologies IEEE-DEST 2008. 2008. pp. 452-460. DOI: 10.1109/DEST.2008.4635163

[28]  Seuring S. A review of modeling approaches for sustainable supply chain management. Decision Support Systems. 2012;**54**(4):1-8. DOI: 10.1016/j.dss.2012.05.053

[29]  Markus ML, Loebbecke C. Commoditized digital processes and business community platforms: New opportunities and challenges for digital business strategies. MIS Quarterly. 2013;**37**(2):649-654

[30]  Iansiti M, Levien R. Strategy as ecology. Harvard Business Review. 2004;**82**(3):68-81. DOI: 10.1108/eb025570

[31]  Chang E, West M. Digital ecosystems a next generation of the collaborative environment. Eight International Conference. 2006;**214**:3-23

[32]  Walsh WE, Wellman MP. Decentralized supply-chain formation: A market protocol and competitive equilibrium analysis. Journal of Artificial Intelligence Research. 2003;**19**:513-567. DOI: 10.1613/jair.1213

[33]  Walsh WE, Wellman MP. Modeling Supply Chain formation in Multiagent Systems. International Workshop on Agent-Mediated Electronic Commerce. Springer, Berlin: Heidelberg; Vol. 1788; 1999. pp. 94-101

[34]  Averian A. Supply chain modelling as digital ecosystem. In: Proceedings of International Scientific Conference ITEMA 2017; 2017. pp. 27-35

*Edited by Jaydip Sen*

The term "Internet of Things" (IoT) refers to an ecosystem of interconnected physical objects and devices that are accessible through the Internet and can communicate with each other. The main strength of the IoT vision is the high impact it has created and will continue to do so on several aspects of the everyday life and behavior of its potential users. This book presents some of the state-of-the-art research work in the field of the IoT, especially on the issues of communication protocols, interoperability of protocols and semantics, trust security and privacy issues, reference architecture design, and standardization. It will be a valuable source of knowledge for researchers, engineers, practitioners, and graduate and doctoral students who are working in various fields of the IoT. It will also be useful for faculty members of graduate schools and universities.

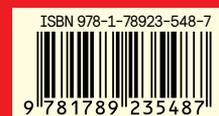